\documentclass[aps,prc,twocolumn,superscriptaddress,amsmath,showpacs,amssymb]{revtex4-1}
\usepackage{bm,physics}
\usepackage[dvipdfmx]{graphicx}
\usepackage{color}

\bmdefine{\ba}{a}
\bmdefine{\bb}{b}
\bmdefine{\bx}{x}
\bmdefine{\by}{y}
\bmdefine{\bz}{z}
\bmdefine{\bn}{n}
\bmdefine{\bp}{p}

\newcommand{\BM}{\begin{pmatrix}}
\newcommand{\EM}{\end{pmatrix}}

\renewcommand{\d}{\dagger}
\newcommand{\Tc}{\mathcal{T}}
\newcommand{\Lc}{\mathcal{L}}
\newcommand{\Mc}{\mathcal{M}}

\newcommand{\hphi}{\hat\varphi}
\newcommand{\hpsi}{\hat\psi}

\newcommand{\intx}{\int\! d^3x\;}
\newcommand{\intxd}{\int\! d^3x'\;}
\newcommand{\intxxd}{\int\! d^3x\,d^3x'\;}

\newcommand{\ex}{\mathrm{ex}}

\begin{document}
\title {Bose--Einstein condensation of alpha clusters and 
new soft mode in $^{12}$C--$^{52}$Fe $4N$
nuclei 
in field theoretical superfluid 
cluster model 
}

\author{R.~Katsuragi}
\affiliation{Department of Electronic and Physical Systems, 
Waseda University, Tokyo 169-8555, Japan} 
\author{Y.~Kazama}
\affiliation{Department of Electronic and Physical Systems, 
Waseda University, Tokyo 169-8555, Japan} 

\author{J.~Takahashi}
\email{takahashi.j@aoni.waseda.jp}
\affiliation{Department of Electronic and Physical Systems, 
Waseda University, Tokyo 169-8555, Japan} 

\author{Y.~Nakamura}
\affiliation{Department of Electronic and Physical Systems, 
Waseda University, Tokyo 169-8555, Japan} 
\affiliation{Nagano Prefectural Kiso Seiho High School, Nagano 397-8571, Japan} 
\author{Y.~Yamanaka}
\email{yamanaka@waseda.jp}
\affiliation{Department of Electronic and Physical Systems, 
Waseda University, Tokyo 169-8555, Japan} 

\author{S.~Ohkubo}
\email{shigeo.ohkubo@rcnp.osaka-u.ac.jp}
\affiliation{Research Center for Nuclear Physics,
Osaka University, Ibaraki,
Osaka 567-0047, Japan}

\date{\today}

\begin{abstract}
Bose--Einstein condensation of $\alpha$ clusters in light and medium-heavy nuclei 
is studied in the frame of the field theoretical superfluid cluster model. 
The order parameter of the phase transition from the Wigner phase to the 
Nambu--Goldstone phase is a superfluid amplitude, square of the moduli of which 
is the superfluid density distribution.  The zero mode operators 
due to the spontaneous symmetry breaking of the global phase in the {\it finite} number 
of $\alpha$ clusters are rigorously treated.    The theory is systematically 
applied to $N \alpha$ nuclei  from $^{12}$C--$^{52}$Fe at various 
condensation rates. In $^{12}$C   it is found that the energy levels 
of the gas-like  well-developed $\alpha$ cluster states  above the Hoyle state
are reproduced well in agreement with  experiment  for realistic condensation 
rates of $\alpha$ clusters. The electric E2 and E0 transitions are calculated and found 
to be sensitive to the condensation rates.  
The profound {\it raison d'\^{e}tre} of the  $\alpha$ cluster gas-like states 
above the Hoyle state, whose structure  has been  interpreted 
geometrically  in the nuclear models without  the order parameter 
such as the cluster models or {\it ab initio} calculations,  is revealed. 
It is found that in addition to the Bogoliubov--de Gennes vibrational 
mode states collective states of the 
zero mode operators appear systematically at low excitation energies
  from the $N\alpha$ threshold energy. These collective states,
new-type soft modes in nuclei due to the Bose--Einstein 
condensation of the $\alpha$ clusters,
emerge systematically  in light  and  medium-heavy  mass regions and are
also located  at high excitation energies from the ground state in contrast 
to the traditional concept of soft mode 
 in the low excitation energy region.
\end{abstract}

\maketitle
\par

\section{INTRODUCTION}

\par
The nucleus shows various aspects of collective motion such as quadrupole 
vibration in spherical nuclei and  rotation in deformed nuclei,
and pairing vibration in normal nuclei and  rotation in superfluid nuclei. 
These collective motion  and  the phase transitions from spherical to deformed and 
from normal to superfluid \cite{Bohr1969B,Ring1980}  are observed 
 in medium-heavy and heavy mass region where mean field picture works well. 
The emergence of quadrupole deformation and the superfluidity has been understood 
as a consequence of  condensation of a quadrupole boson 
with an angular momentum $J=2$ and a Cooper pair boson with $J=0$, 
respectively.  The phase transition from the Wigner phase to the Nambu--Goldstone 
(NG) phase  in nuclei composed of {\it finite} number of nucleons can be understood by 
the order parameter, the deformation $\delta$  for the quadrupole collective motion 
and the pairing gap energy $\Delta$ for the pairing collective motion.

In light nuclei, collective motion due to the emergence of cluster structure, 
especially $\alpha$ cluster structure, has been studied extensively in the last 
decades \cite{Suppl1972,Wildermuth1977,Ikeda1980,Suppl1998}. The 
 cluster model has been established as one of the three 
nuclear models \cite{Suppl1972,Wildermuth1977,Ikeda1980,Suppl1998} as well as
 the shell model  \cite{Mayer1949,Jensen1949} and the collective model \cite{Bohr1952}.
This collective motion caused by the spontaneous symmetry breaking of rotational 
invariance due to the  $\alpha$ clustering  has been  observed in a wide range of 
nuclei  throughout the periodic table. Typically, they are, for example, 
the $\alpha$+$\alpha$ cluster structure in $^8$Be in the p-shell region 
\cite{Suppl1972}, the $\alpha$+$^{12}$C structure in $^{16}$O and $\alpha$+$^{16}$O 
structure in $^{20}$Ne  around  the beginning of the sd-shell region \cite{Ikeda1980},
 and the $\alpha$+$^{36}$Ar cluster  structure 
in $^{40}$Ca  and the $\alpha$+$^{40}$Ca cluster  structure in $^{44}$Ti 
around the beginning of the fp-shell region \cite{Michel1998,Yamaya1998,Sakuda1998},
 and the $\alpha$+$^{208}$Pb cluster band in $^{212}$Po
 \cite{Buck1994,Ohkubo1995,Astie2010,Suzuki2010} in the  heavy mass region.

The collective motion related to $\alpha$ cluster  condensation or  superfluidity 
in nuclei has been paid attention in the frame of the  many-body theory in the 
last decades \cite{Eichler1972,Gambhir1983,Koh1998}. 
However, in heavy and medium-heavy mass region,  $\alpha$ cluster superfluidity
 with a characteristic collective motion has not been confirmed experimentally. 
On the other hand, in light nuclei the  Hoyle state in $^{12}$C  
with a well-developed three $\alpha$ cluster structure 
near the $\alpha$ threshold energy has been extensively studied 
theoretically \cite{Fujiwara1976,Uegaki1977,Fukushima1978,Tohsaki2001,
Matsumura2004,Kurokawa2004,Ohtsubo2013,Buck2013,Funaki2015,Nakamura2016}, 
and experimentally \cite{Freer2011,Itoh2011,Kirsebom2017,Aquila2017,Smith2017}, 
 and  attracted much attention as a candidate for a Bose--Einstein condensation 
(BEC) of $\alpha$ clusters.

Matsumura and Suzuki \cite{Matsumura2004} have shown in the cluster model 
calculations   that the occupation probability of the $\alpha$ clusters sitting 
in the lowest $0$s state is about 70$\%$ for the Hoyle state. However, 
from this occupation probability,  whatever the 0s state occupation probability 
may be,  it is impossible to know  that the  global phase 
of the system is frozen, that is,  to conclude that the BEC of $\alpha$ clusters 
is realized. It is because the {\it order parameter} is not defined 
in the traditional  cluster models such as the generator 
coordinate method \cite{Uegaki1977}, the resonating group method \cite{Fukushima1978}, 
the cluster gas model \cite{Tohsaki2001,Funaki2015}, the correlated Gaussian 
stochastic variation method \cite{Matsumura2004}, the orthogonality 
condition model \cite{Kurokawa2004,Ohtsubo2013}, the local potential 
model \cite{Buck2013} and the Faddeev theory \cite{Fujiwara1976}. 
The {\it ab initio} calculations such as the anti-symmetrized molecular 
dynamics \cite{Kanada2007}, the fermionic molecular dynamics \cite{Chernykh2007}, 
the no-core shell model \cite{Roth2011} and  the  lattice 
calculations \cite{Epelbaum2011} have also no order parameter 
in their theoretical frames.
To know whether 
the phase transition from the Wigner phase to the NG phase of BEC 
is realized, the order parameter of spontaneous symmetry breaking (SSB) should be 
defined  in the frame of  theory.
Also up to now, no decisive experimental evidence of BEC of $\alpha$ clusters 
such as superfluidity or quantum vortex has  been observed.

The very recent precise experiments \cite{Kirsebom2017}
reported that the direct three $\alpha$ decays from the Hoyle state is 
less than 0.043\% \cite{Aquila2017} and 0.047\% \cite{Smith2017}, 
respectively. Reference \cite{Smith2017} noted that this could indicate that 
the $\alpha$-condensate interpretation of the Hoyle state is less likely to 
be correct. However, the likelihood of the BEC cannot be judged by the amount 
of the 0s state occupation probability of the $\alpha$ clusters because BEC of 
$\alpha$ clusters could be realized by a smaller but significant amount of 
condensation probability. The emergence of  BEC can only be  concluded 
by investigating the  order parameter of the system. 

A cluster model, in which the order parameter that characterizes the phase 
transition from the Wigner phase of normal solid, liquid or gas-like cluster
structure to the NG phase of Bose--Einstein $\alpha$ cluster condensate,
is defined is highly needed.
In such a model, it is essential to treat the Nambu--Goldstone (NG) operators 
or zero mode operators
rigorously for the systems with a {\it finite} number of bosons like the three 
$\alpha$ bosons in the Hoyle state.
In a previous work~\cite{Nakamura2016}, we have proposed an $\alpha$ cluster model, 
in which an order parameter of  BEC of $\alpha$ clusters is defined based on 
quantum field theory with a spontaneous breakdown of the global phase symmetry, 
and applied it to  the Hoyle and excited states above it in $^{12}$C, assuming 
100\% condensation of the three $\alpha$ clusters. 
In the present paper, we show, with a realistically smaller but significant amount of 
condensation probability around 70\%, that the Hoyle  state and the excited states 
are well understood as a Bose--Einstein condensate of $\alpha$ clusters.

It has been well-known that the $\alpha$ cluster structure persists not only 
in light nuclei but also in medium-heavy nuclei as well, most typically 
in the $^{44}$Ti region \cite{Suppl1998,Michel1998,Yamaya1998,Sakuda1998}, 
where $j-j$ coupling becomes important due to the strong spin-orbit force. 
The collective motion such as superdeformation, for example, in $^{36}$Ar 
and $^{38}$Ar,
can  also be understood to be caused by clustering \cite{Sakuda2004,Sakuda2013}.
The   $\alpha$ cluster structure study  around $^{48}$Cr in the fp-shell region 
was done in Refs.\cite{Sakuda2002,Descouvemont2002,Mohr2017}. 
The $\alpha$ cluster condensation  in sd-shell region up to $^{40}$Ca
 and in heavier nuclei
including $^{52}$Fe was paid attention in Ref. \cite{Yamada2004} and  
  Refs. \cite{Kokalova2003,Kokalova2006,Itagaki2010,Oertzen2011,Itagaki2011}, 
respectively.
In the present work, we study BEC of $N\alpha$ clusters with $N= 3$--$13$, i.e.,
${}^{12}$C\,, ${}^{16}$O\,, ${}^{20}$Ne\,, ${}^{24}$Mg\, 
${}^{28}$Si\,, ${}^{32}$S\,,${}^{36}$Ar\,, $^{40}$Ca, $^{44}$Ti, $^{48}$Cr 
and $^{52}$Fe.  
We show that collective states of the zero mode operators that 
are new-type soft 
modes due to  BEC of  $\alpha$ clusters emerges systematically.  
In $^{12}$C  the observed excited $0_3^+$ and $0_4^+$ states with a well-developed 
$\alpha$ cluster structure above the Hoyle state  
can be interpreted
 as such a type of soft modes of the zero mode
operators associated with the SSB of the global phase of the Bose--Einstein 
condensate of $\alpha$ clusters. This interpretation
 is in contrast to the 
traditional  interpretations based  on the  geometrical picture  of the 
$\alpha$ clusters in the configuration space.



The organization of the paper is as follows. In Secs.~\ref{sec-ModelFormulation}
 and \ref{sec-BdG}, the formulation of the present model is given in detail. 
Section~\ref{sec-ModelFormulation} is devoted to the formulation 
of a cluster model with an order parameter
that characterizes the phase transition from the Wigner phase of the normal
 $\alpha$ cluster state to the Bose--Einstein condensate of NG phase 
of {\it finite} number of $\alpha$ clusters  in the frame of quantum
 field theory. The zero mode operators 
due to SSB of the global phase is treated  rigorously,
 keeping the canonical commutation relations.
Section \ref{sec-BdG} is devoted to the formulation of 
the Bogoliubov--de Gennes equation.
In Sec.~\ref{sec-ElecTranProb},  the electric transition probabilities 
for the Bose--Einstein condensate states are formulated.
In Sec.~\ref{sec-BEC12C70}, the BEC of three $\alpha$ clusters in $^{12}$C,
for which many experimental data are available, are studied in detail under an
assumption of realistic condensation rate of 70\%.  
In Sec.~\ref{sec-12CN0}, the energy level structure, wave functions  
and electric transitions of $^{12}$C at various condensation rates of $\alpha$ clusters,
including  condensation rate of 100\%, 
are investigated. The robustness of the energy level structure 
for the different condensation rates is shown. 
In particular, the systematic appearance  of the collective states 
of the zero mode operators
  is illustrated for different  condensation rates. 
We proceed to the investigations of the BEC of many $\alpha$ clusters 
from $^{16}$O to $^{52}$Fe in Sec.~\ref{sec-Nalpha}. 
There it is emphasized that the 
zero mode states of the Nambu--Goldstone mode due to BEC of $\alpha$ clusters 
is a new  kind of soft mode, which appears even in the highly excited energy region 
in contrast to the concept of the traditional concept of soft mode 
such as a quadrupole collective motion in the low excitation energy region. 
 We analyze the eigenequation to determine the zero mode states in 
some detail, and clarify the robustness of the spectra of 
the zero mode states for various condensation
rate and number of $\alpha$ clusters.
Summary is given in Sec.~\ref{sec-Summary}.

\section{Model AND FORMULATION}
\label{sec-ModelFormulation}

The formulation was originally presented for BEC
of trapped cold atoms in Ref.~\cite{Nakamura2016} and 
is called the interacting zero mode formulation (IZMF in short). 
The reason for IZMF is explained as follows: 
The canonical commutation relations of the field operator do not allow us to
disregard the zero mode operators in finite-size systems such 
as in the trapped cold atomic systems and nuclei, which is associated 
with the spontaneously broken symmetry, whereas they can be suppressed 
for homogeneous systems because of their point like 
contributions in continuum. Once the zero mode operators,
 denoted by ${\hat Q}$ and ${\hat P}$, are present, a naive choice of the bilinear
unperturbed Hamiltonian leads to the difficulties that no stationary vacuum 
exists and that the phase of the order parameters should diffuse. 
To avoid the difficulties, we take the nonlinear unperturbed Hamiltonian of
${\hat Q}$ and ${\hat P}$. The crucial point for
our analysis on $\alpha$ cluster states is that this nonlinear 
Hamiltonian adds a new type of discrete energy levels with $0^+$ 
to the levels of the Bogoliubov--de Gennes (BdG) modes with various $J^P$.
The combined spectrum of both levels reproduces the observed 
energy levels above the Hoyle state in $^{12}$C very well, by adjusting 
a single newly introduced parameter of our phenomenological model, 
the strength of the trapping potential.

As in Ref.~\cite{Nakamura2016}, we start with the phenomenological 
model in which the $\alpha$
particles are trapped inside the nuclei by the external isotropic harmonic
potential, 
\begin{align}
V_\ex(r)=\frac{1}{2} m \Omega^2 r^2\,, 
\label{eq:Vex}
\end{align}
and the $\alpha$--$\alpha$ interaction
is given by the Ali--Bodmer potential \cite{Ali1966}, 
\begin{align}
U (|\bm x -\bm x'|)& = V_r e^{-\mu_r^2 |\bm x -\bm x'|^2} 
- V_a e^{-\mu_a^2 |\bm x -\bm x'|^2}\,.
\label{eq:AliBodmer}
\end{align}
The repulsive Coulomb potential affects numerical results 
very little and is suppressed in this work.

Let $\psi(x)$ $(x=(\bx,t))$ be the field operator of the $\alpha$ cluster,
and the model Hamiltonian is 
\begin{align}
&\hat{H}=\intx \hpsi^\d(x) \left(-\frac{\nabla^2}{2m}+
V_\ex(\bx)- \mu \right) \hpsi(x) \notag\\
&\,\,+\frac12 \intxxd \hpsi^\d(x) 
\hpsi^\d(x') U(|\bx-\bx'|) \hpsi(x') \hpsi(x) \,,
\label{Hamiltonian}
\end{align}
where $m$ and $\mu$ denote the mass of the $\alpha$ cluster 
and the chemical potential, respectively. We set $\hbar=c=1$
throughout this paper. The total Hamiltonian $\hat{H}$ possesses
the global phase symmetry, namely that $\hat{H}$ is invariant under
$\hpsi\, \rightarrow\, e^{i \theta} \hpsi$ with a constant $\theta$. 
When the $\alpha$ clusters are condensed, the original 
field operator $\hpsi$ must be divided into a condensate c-number 
component $\xi$ and an excitation component $\hphi$,
$\hpsi=\xi+\hphi$, according to
that the criterion $\bra0 \hpsi \ket0=\xi$. 
It is vital for our formulation
that the function $\xi$, called the order parameter, 
is given by the vacuum expectation value of the field operator. 
Note that
the original gauge symmetry is spontaneously broken.
 The function $\xi$ is assumed to be stationary and isotropic in this paper, 
 and is 
normalized to the condensed particle number $N_0$ as
$\intx |\xi(\bx)|^2 = N_0$. 
Any constant phase of $\xi$ is allowed, refelcting the original global
phase symmetry, and as physical results are not affected by its choice, 
we take a real $\xi$ throughout this paper.
The Hamiltonian (\ref{Hamiltonian}) is 
classified according to power degree of $\hphi$:
\begin{align}
&\hat{H} = \hat{H}_2 + \hat{H}_{3,4}\notag\\
&\hat H_{2} = \frac12 \intxxd {\hat {\bar \Phi}}(x)
\Tc(\bx,\bx') {\hat \Phi}(x') \,, \label{eq:H2}\\
&\hat H_{3,4} =\frac 12 \intxxd U(|\bx-\bx'|) \notag\\
&\quad \times \left[\left\{
2 \xi(\bx')+\hphi^\d(x')\right\}\hphi^\d(x)\hphi(x)
\hphi(x') +{\rm h.c.} \right]
\,,
\end{align}
with $t=t'$, and 
\begin{align}
V_H(\bx) &= \intxd U(|\bx-\bx'|)\xi^2(\bx')\,, 
\label{eq:VH1} \\
{\hat \Phi} (x) &= \BM \hphi(x) \\
\hphi^\dagger (x) \EM\,, \quad {\hat {\bar \Phi}}
(x) ={\hat \Phi}^\dagger (x)\sigma_3\,,\\
\Tc(\bx,\bx') &=\BM \Lc(\bx,\bx') & \Mc(\bx,\bx') \\
-\Mc(\bx, \bx') & -\Lc(\bx,\bx') \EM\,,\\
\Mc(\bx,\bx') &= U(|\bx-\bx'|)\xi(\bx) \xi(\bx')\,,\\
\Lc(\bx,\bx') &= \delta(\bx-\bx') \bigl\{-{\nabla^2}/{2m}+V_\ex(\bx) \notag\\
&\hspace{1.5cm}-\mu + V_H(\bx) \bigl\} +\Mc(\bx,\bx')\,,
\end{align}
where $\sigma_i$ $(i=1,2,3)$ is the Pauli matrix.
We have the Gross--Pitaevskii (GP) equation \cite{GP}
\begin{equation}\label{eq:GP}
\left\{ -{\nabla^2}/{2m}+V_\ex(\bx) -\mu + V_H(\bx)
\right\} \xi(\bx) = 0 \,,
\end{equation}
because the $\hphi$-linear term in ${\hat H}$ must vanish, 
otherwise the ground state could not be stationary.

The field operator $\hphi$ is expanded by the complete set of 
the BdG eigenfunctions,
\begin{align} 
& \intxd \Tc(\bx,\bx') \by_\bn(\bx') = \omega_\bn
\by_\bn(\bx) \,, \label{eq:BdG}\\
& \qquad \by_\bn(\bx)= \BM u_\bn(\bx) \\ v_\bn(\bx) \EM
\label{eq:bmy}
\,.
\end{align}
The index $\bn = (n,\, \ell,\, m)$ is a 
 triad of the main, azimuthal, and magnetic quantum numbers
for isotropic $\xi$. 
Similarly as the BdG equation is introduced for fermionic systems,
Eq.~(\ref{eq:bmy}) is the BdG equation for bosonic systems. 
The bosonic eigenfunction is normalized as 
$\intx (|u_\bn|^2-|v_\bn|^2)=1$ (see Eq.~(\ref{eq:yzOrthoNormal1})) 
for the commutation relations, while the fermionic eigenfunction 
is normalized as $\intx (|u_\bn|^2+|v_\bn|^2)=1$ for the 
anti-commutation relations.

Another eigenfunction, denoted by $\bz_{\bn}$,
is introduced:
\begin{align}
& \intxd \Tc(\bx,\bx') \bz_\bn(\bx') = -\omega_\bn
\bz_\bn(\bx) \,, \\
& \qquad \bz_\bn(\bx)= \sigma_1 \by^\ast_\bn(\bx)
= \BM v_\bn^\ast(\bx) \\ u^\ast_\bn(\bx) \EM
\,.
\end{align}
The inner product is defined as
$((\ba,\bb))\equiv
\intx \ba^\dagger(\bx)\sigma_3 \bb(\bx)$\,,
and the orthonormal relations are
\begin{align}
&((\by_\bn,\by_{\bn'}))=-((\bz_\bn,\bz_{\bn'}))
=\delta_{\bn\bn'}\,, 
\label{eq:yzOrthoNormal1}
\\
& ((\by_\bn,\bz_{\bn'}))=0\,.
\label{eq:yzOrthoNormal2}
\end{align}
We also have the eigenfunction with zero eigenvalue,
\begin{align}\label{eq:BdGy0}
\intxd \Tc(\bx,\bx') \by_0(\bx') =0 \,, \quad 
\by_0(r)= \BM \xi(\bx) \\ -\xi(\bx) \EM\,,
\end{align}
which is orthogonal to all the eigenfunctions including itself,
\begin{align}
((\by_0,\by_0))=((\by_0,\by_{\bn}))=((\by_0,\bz_{\bn}))=0\,.
\end{align}
For the completeness of the set of BdG eigenfunctions,
the adjoint eigenfunction $\by_{-1}$ is necessary,
\begin{align}
&\intxd \Tc(\bx,\bx') \by_{-1}(\bx') =I\by_0(\bx) \,, \label{eq:BdGy-1} \\
&\by_{-1} (\bx) = \BM \eta(\bx) \\ \eta(\bx) \EM \,, \label{eq:BdGy-1-2}\\
&((\by_{-1},\by_{-1}))
=((\by_{-1},\by_{\bn}))=((\by_{-1},\bz_{\bn}))=0\,,
\end{align}
where the constant $I$ is determined by the condition,
\begin{align}
((\by_{-1},\by_{0}))=1 \,.
\end{align}
The function $\eta(\bx)$ and the constant $I$ can also be calculated as
\begin{align}
\eta(\bx)=\frac{\partial \xi(\bx)}{\partial N_0}\,,
\qquad I=\frac{\partial \mu}{\partial N_0}\,.
\label{eq:etaI}
\end{align}
The completeness relation reads as
\begin{align}
&\sigma_3 \delta(\bx-\bx')
=\by_0(\bx)\by^\dagger_{-1}(\bx')+\by_{-1}(\bx)\by^\dagger_{0}(\bx')
\notag \\
&\qquad +
\sum_{\bn} 
\left\{\by_\bn(\bx)\by^\dagger_\bn(\bx')-
\bz_\bn(\bx)\bz^\dagger_\bn(\bx')\right\}
\,,
\end{align}
and the field operators are expanded as
\begin{align}
{\hat \Phi}(x) &= 
-i{\hat Q}(t) \by_0(\bx)+{\hat P}(t)\by_{-1}(\bx) \notag \\
&\qquad +
\sum_{\bn} 
\left\{ {\hat a}_{\bn}(t)\by_\bn(\bx)
+{\hat a}^\dagger_{\bn}(t)\bz_\bn(\bx)\right\} \,,
\label{eq:PhiExp}
\end{align}
where the commutation relations,
\begin{align}
[{\hat Q}\,,\,{\hat P}]=i \,,\quad
[{\hat a}_{\bn}\,,\,{\hat a}^\dagger_{\bn'}]=
\delta_{\bn \bn'}\,,
\label{eq:CCRQPa}
\end{align}
are derived from the canonical commutation relations of $\hphi(x)$.

The inevitable appearance of the eigenfunction 
with zero eigenvalue $\by_0(\bx)$ and the pair of canonical 
operators ${\hat Q}$ and ${\hat P}$ is concluded from a general
argument using the NG theorem. For this, we derive the 
Ward-Takahashi relation, adding a
symmetry breaking term 
$-\epsilon \int\!d^3x\, \left\{{\hat \psi}(x)+
{\hat \psi}^\dagger(x) \right\}$ with an infinitesimal parameter $\epsilon$
to the total Hamiltonian ${\hat H}$ in Eq.~(\ref{Hamiltonian}) \cite{UMT},
\begin{align}
&-2\xi(\bx)=  -i \epsilon 
\int\!d^4y\,\notag \\
& \quad  \bra{0}{\rm T}\left[\left\{{\hat \varphi}_H(y)+
{\hat \varphi}^\dagger_H(y)\right\} \left\{ {\hat \varphi}_H(x)-
{\hat \varphi}^\dagger_H(x)\right\}\right]\ket{0}\,,
\label{eq:WTcommon}
\end{align}
where the suffix $H$ represents the Heisenberg operator.
When $\xi$ is a constant, Eq.~(\ref{eq:WTcommon}), with the spectral form
of the full propagator, leads 
to the existence of a gapless mode in zero momentum limit 
in the Fourier space, which is well-known as the NG 
theorem for a homogeneous system. Equation~(\ref{eq:WTcommon}) 
for the present inhomogeneous system indicates 
that there should exist an eigenfunction, whose components are 
proportional to $\xi(\bx)$. This implies the existence of $\by_0(\bx)$,
having zero eigenvalue, and subsequently the existence of $\by_{-1}(\bx)$,
as above. Then we are forced to introduce ${\hat Q}$ and ${\hat P}$
in the expansion of the field operator. This is an implication of
the NG theorem for an inhomogeneous system.

This way the pair of canonical operators ${\hat Q}$ and ${\hat P}$
originate from the SSB of the global phase, and we call them 
the zero mode operators or NG operators.
As was explained in Ref.~\cite{NTY2014,Nakamura2016}, the substitution of 
Eq.~(\ref{eq:PhiExp}) into Eq.~(\ref{eq:H2}) gives 
$\hat{H}_2 = {I\hat{P}^2}/{2} 
+ \sum_\bn \omega_\bn \hat a_\bn^\d \hat a_\bn \,,$
whose NG operator part causes serious defects.
To avoid the defects, we replace the term ${I\hat{P}^2}/{2}$ 
in the unperturbed Hamiltonian
with 
\begin{align} 
&\hat H_u^{QP} = - \left(\delta\mu + 2C_{2002} + 2C_{1111} \right) \hat P+
\frac{I-4C_{1102}}{2}\hat P^2 \notag \\
&\,\, + 2C_{2011}\hat Q\hat P\hat Q 
+ 2C_{1102}\hat P^3 + \frac{1}{2}C_{2020}\hat Q^4 -2C_{2011}
\hat Q^2 \notag \\
&\,\,+ C_{2002}\hat Q\hat P^2\hat Q + \frac{1}{2} C_{0202}\hat P^4 \,,
\label{eq:HuQP}
\end{align}
where 
\begin{align}
&C_{iji'j'}= 
\int\! d^3x d^3x'\, U(|\bx-\bx'|)\notag \\
&\qquad \times \{\xi(\bx)\}^i
\{\eta(\bx)\}^j
\{\xi(\bx')\}^{i'}\{\eta(\bx')\}^{j'}
\,,
\label{eq:Cijij}
\end{align}
and $\delta \mu $ is a counter term that the 
criterion $\bra0\hpsi\ket0=\xi\,$ determines. The Hamiltonian $H_u^{QP}$ 
is obtained from gathering all the terms consisting only 
of ${\hat Q}$ and ${\hat P}$ in ${\hat H}_2$ and ${\hat H}_{3,4}$.
We set up the eigenequation of $H_u^{QP}$\,,
\begin{align} \label{eq:HuQPeigen}
\hat H_u^{QP} \ket{\Psi_\nu} = E_\nu \ket{\Psi_\nu}\qquad 
(\nu=0,1,\cdots)\,.
\end{align}
The eigenstates $\{\ket{\Psi_\nu}\}$ are collective states of 
${\hat Q}$ and ${\hat P}$, simply called {\it zero mode states}, and  
span their own state subspace, referred to as the NG subspace. The 
state of the total system  $\ket{\mathcal S}$
is expressed as a direct product, 
\begin{align}
\ket{\mathcal S}= \ket{\Psi_\nu} 
\ket{\cdot}_\ex \,,
\end{align}
where $\ket{\cdot}_\ex$ is a Fock state associated 
with the BdG mode operator $\hat a_\bn$. The discrete spectrum of the 
zero mode states in Eq.(\ref{eq:HuQPeigen}) is 
our original consequence
that is to be compared with the observed energy levels.
We may point out a resemblance between the breakings
 of the rotational symmetry in the deformed nuclear ground state \cite{Ring1980}
and the global phase symmetry in our model. Because both are  SSBs 
in finite systems, the quantum degrees of freedom to restore 
the broken symmetries appear as the zero mode operators, which give rise to
a rotational band of the states in case of the broken rotational symmetry and
the collective states $\ket{\Psi_\nu}$ with $E_\nu$ in our present case.

The vacuum state $\ket{\Psi_0}\ket{0}_\ex$  is 
identified as the Hoyle state just above the three $\alpha$ threshold  in the case  of 
$^{12}$C.
The states $\ket{\Psi_\nu}\ket{0}_\ex
\,\, (\nu=1,2,\cdots)\,$, are NG (or zero mode) excited states
with the excitation energy from the vacuum $E_\nu-E_0$\,. 
The excitation in the NG subspace 
changes neither 
the value of the angular momentum $J$ nor the sign of the parity $P$.
The state $\ket{\Psi_0}({\hat a_\bn^\dagger}\ket{0}_\ex)$, called 
the BdG state, has the excitation energy $\omega_{\bn}$, 
measured from the vacuum state.

\section{BOGOLIUBOV--DE GENNES EQUATION}\label{sec-BdG}

We give the BdG equations here in some details which were not presented 
explicitly in Ref.~\cite{Nakamura2016}.
We put the following separable form of the BdG eigenfunction
in Eq.~(\ref{eq:bmy}),
\begin{align}
\by_\bn(\bx)= \BM {\mathcal U}_{n\ell}(r) \\ 
{\mathcal V}_{n\ell}(r) \EM
\, Y_{\ell m}(\theta,\varphi)\,.
\end{align}
For convenience the function ${\tilde U}_{\ell m}(r,r',\theta,\varphi)$
is introduced as
\begin{align}
&{\tilde U}_{\ell m}(r,r',\theta,\varphi)/ r^{\prime 2} \equiv
\notag \\
& \quad \int\! d\Omega'\,
U(\sqrt{r^2+r^{\prime 2} -2rr'\cos(\theta'-\theta)})
Y_{\ell m}(\theta',\varphi')
\,. \label{eq:defUlm}
\end{align}
For performing the surface integral, the 
direction of the $z'$-axis is taken along the vector $\bx$
and then $\cos(\theta'-\theta)$ becomes $\cos \theta'$. We define
the eigenfunctions of ${\hat L}_n={\hat {\bm L}}
\cdot {\bx}/r$, denoted by $Y^n_{\ell m}(\theta,\varphi)$, as
${\hat L}_n Y^n_{\ell m}(\theta,\varphi)=m Y^n_{\ell m}(\theta,\varphi)$.
The Wigner $D$-matrix gives the following relations,
\begin{align}
Y^n_{\ell m}(\theta',\varphi')=
\sum_{m'} D^{\ell,\ast}_{m'm} (\varphi,\theta,0)
Y_{\ell m'}(\theta',\varphi')\,, \\
Y_{\ell m}(\theta',\varphi')=
\sum_{m'} D^{\ell}_{mm'} (\varphi,\theta,0)
Y^n_{\ell m'}(\theta',\varphi')\,.
\end{align}
The last relation is substituted into Eq.~(\ref{eq:defUlm}), and 
we integrate it over the variable $\varphi'$ to have
\begin{align}
&{\tilde U}_{\ell m}(r,r',\theta,\varphi)/ r^{\prime 2}
= 2\pi Y_{\ell m}(\theta,\varphi) \notag \\
&\quad \times 
\int_{-1}^1 dw\,
U(\sqrt{r^2+r^{\prime 2} -2rr'w})\,P_{\ell}(w)
\,,\label{eq:Ulm}
\end{align}
where the relation $D_{m0}^\ell (\varphi,\theta,0)
= \sqrt{\frac{4\pi}{2\ell+1}} Y_{\ell m}(\theta,\varphi)$ has been used.
For the isotropic $\xi(\bx)= \xi(r)$, $V_H$ in Eq.~(\ref{eq:VH1}) is also isotropic, 
given by
\begin{align}
V_H(r)=\frac{N_0}{\sqrt{4\pi}} \int\!dr'\, {\tilde U}_{00}(r,r')
{\tilde \xi}^2(r') 
\,, 
\end{align}
where
\begin{align}
& \xi(r) =\sqrt{\frac{N_0}{4\pi}}{\tilde \xi}(r)\,, \\
& {\tilde U}_{00}(r,r') =
\frac{\sqrt{\pi}r'}{r}\Bigl\{ \frac{V_r}{2\mu^2_r}\qty
(e^{-\mu^2_r(r-r')^2} -e^{-\mu^2_r(r+r')^2} ) \notag \\
&\qquad - \frac{V_a}{2\mu^2_a}\qty(e^{-\mu^2_a(r-r')^2} - e^{-\mu^2_a(r+r')^2})
\Bigr\}\,.
\end{align}
The Fock term in Eq.~(\ref{eq:BdG}) is manipulated as
\begin{align}
&\intxd \Mc(\bx,\bx')u_{\bn}(\bx') \notag \\
&\quad = \frac{N_0}{4\pi} {\tilde \xi}(r)
\int\! dr'
\, {\tilde \xi}(r') {\cal U}_{n\ell}(r') 
{\tilde U}_{\ell m}(r,r',\theta,\varphi) \notag \\
&\quad =F_\ell[r:{\cal U}_{n\ell}]
{\tilde \xi}(r)Y_{\ell m}(\theta,\varphi)
\end{align}
with a linear functional of $f$,
\begin{align}
&F_\ell[r: f]=\frac{N_0}{2} 
\int\! dr'
\, r^{\prime 2} {\tilde \xi}(r') f(r')\notag \\
&\quad \times \int_{-1}^1 dw\,
U(\sqrt{r^2+r^{\prime 2} -2rr' w})\, P_{\ell}(w)
\,,
\end{align}
and similarly for $v_\bn$\,. The BdG equation~(\ref{eq:BdG}) is 
reduced to
\begin{align}
&h_\ell\,{\mathcal U}_{n\ell}(r)+
F_\ell\left[r: {\mathcal U}_{n\ell}
+ {\mathcal V}_{n\ell}\right] {\tilde \xi}(r)
=\omega_\bn {\mathcal U}_{n\ell}(r) \,,
\label{eq:BdG_u}\\
&h_\ell\,{\mathcal V}_{n\ell}(r)+
F_\ell\left[r: {\mathcal U}_{n\ell}+ 
{\mathcal V}_{n\ell}\right] {\tilde \xi}(r)
=-\omega_\bn {\mathcal V}_{n\ell}(r)
\label{eq:BdG_v} \,,
\end{align}
with 
\begin{align}
&h_\ell = -\frac{1}{2m}\left(\frac{d^2}{dr^2} 
+\frac{2}{r}\frac{d}{dr}-\frac{\ell(\ell+1)}{r^2}\right)\notag \\
& \qquad \quad +
V_\ex(r)- \mu+V_H(r) \,.
\end{align}

\section{ELECTRIC TRANSITION PROBABILITIES}
\label{sec-ElecTranProb}

The formulation presented in Secs.~\ref{sec-ModelFormulation}
and \ref{sec-BdG} enables us to calculate the $\gamma$-decay 
transitions among the states with $\alpha$ cluster condensate.
\par
The decay rate of an electric transition with a photon angular
momentum $J$ for an unpolarized initial state and 
summing all final polarization states \cite{Bohr1969B}
is generally given by
\begin{align}
{\bar \Gamma}_{fi}({\rm E}:k,J)
=\frac{8\pi(J+1)}{J((2J+1)!!)^2}k^{2J+1}B({\rm E}J:J_i\, \rightarrow\, J_f)
\,,
\end{align}
where $k$ is the photon energy $k=E_i-E_f$, calculated from the initial
and final state energies of the nucleus $E_i$ and $E_f$\,. 
The symbol $B$ stands for the reduced transition probability,
\begin{align}
&B({\rm E}J:J_i\, \rightarrow\, J_f) = \notag \\
&\quad \frac{1}{2J_i+1}
\left|\bra{f(J_f)} |
{\hat {\cal M}}({\rm E}:kJ)| \ket{i(J_i)}\right|^2 \,,
\label{eq:BE2}
\end{align}
where ${\hat {\cal M}}$ is the multipole moment, and 
$\ket{i}$, $\ket{j}$, $J_i$, and $J_f$ are 
the initial and final nuclear states and spins, respectively.

The transitions, 
$\ket{\Psi_0} ({\hat a}^\dagger_{12m} \ket{0}_\ex)$ 
$\rightarrow$ 
$\ket{\Psi_0} \ket{0}_\ex$ and $\ket{\Psi_1} \ket{0}_\ex$,
correspond in our approach to the transitions, 
2$_2^+$$\rightarrow$ 0$_2^+$ and 0$_3^+$ in $^{12}$C, respectively, 
which will be discussed in the next sections. 
The reduced transition probabilities for these
processes are calculated as
\begin{align}
&B({\rm E}2:2\, \rightarrow\, 0)= \notag \\
&\left| \bra{f(J_f=0,M_f=0)} 
{\hat {\cal M}}({\rm E}:k 20) \ket{i(J_i=2,M_i=0)}\right|^2 \,,
\label{eq:E2-1}
\end{align}
and we have for 2$_2^+$$\rightarrow$0$_2^+$
\begin{align}
&\bra{f(0,0)} 
{\hat {\cal M}}({\rm E}:k 20) \ket{i(2,0)}= \notag \\
& \qquad \frac{60e\sqrt{N_0}}{mk^3} \int\!dr\, r \left[
{\tilde \xi} (r) \left\{ \frac{d}{dr} \mathcal{U}_{12}(r)j_2(kr) \right. \right.\notag \\
&\qquad \qquad\quad \left. + \mathcal{U}_{12}(r)
\left( \frac{j_2(kr)}{r} + kj^\prime_2 (kr) \right) \right\} \notag \\ 
&\qquad\quad\qquad \left. + \frac{d}{d r}{\tilde \xi}(r)\mathcal{V}_{12}(r) j_2(kr) \right]\,,
\label{eq:E2-2}
\end{align}
and for 2$_2^+$$\rightarrow$0$_3^+$
\begin{align}
&\bra{f(0,0)} 
{\hat {\cal M}}({\rm E}:k 20) \ket{i(2,0)}= \notag \\
&\frac{60e\sqrt{N_0}}{mk^3} \int\!dr\, r \left[
\left\{i \bra{\Psi_1} \hat{Q}\ket{\Psi_0} {\tilde \xi}(r) + \bra{\Psi_1}\hat{P}\ket{\Psi_0}
{\tilde \eta}(r) \right\} \right. \notag \\
& \times 
\left\{\frac{d}{dr} \mathcal{U}_{12}(r)j_2(kr) + \mathcal{U}_{12}(r)
\left( \frac{j_2(kr)}{r} + kj^\prime_2 (kr) \right) \right\}
\notag \\
& + \left\{ i \bra{\Psi_1} \hat{Q}\ket{\Psi_0} \frac{d}{dr}{\tilde \xi}(r)
+ \bra{\Psi_1}\hat{P}\ket{\Psi_0}
\frac{d}{dr}{\tilde \eta}(r) \right\} \notag \\
& \qquad\left. \times \mathcal{V}_{12}(r) j_2(kr) \right]\,,
\end{align}
where $j_\ell$ is the spherical Bessel function and 
$j_\ell'(z)=\frac{dj_\ell}{dz}$, and
$\eta(\bx)=\eta(r) = \sqrt{\frac{N_0}{4\pi}}\, {\tilde \eta}(r)$.

The monopole E0 transition probabilities for 
the processes 0$^+_2$$\rightarrow$0$^+_3$ and 0$^+_4$ are given by
\begin{align}
& M({\rm E}0:0^+_2\, \rightarrow\, 0^+_{\nu+2})=
\left|\bra{\Psi_\nu} 
2e \intx \hpsi^\d(\bx)\hpsi(\bx)  \ket{\Psi_0} \right|^2\notag \\
&= 4e^2 \left|I_{QQ} \bra{\Psi_\nu}{\hat Q}^2 \ket{\Psi_0}
+I_{PP} \bra{\Psi_\nu}{\hat P}^2 \ket{\Psi_0}\right. \notag \\
&\qquad \left. +I_{P} \bra{\Psi_\nu}{\hat P} \ket{\Psi_0}
  \right|^2 \,, 
\label{eq:E0}
\end{align}
where $\nu=1,2$ and
\begin{align}
&I_{QQ}=\intx r^2 \xi^2(\bx)\,, \quad 
I_{PP}=\intx r^2 \eta^2(\bx)\,, \notag \\
&I_{P}=2 \intx r^2 \xi(\bx) \eta(\bx)\,.
\end{align}

\section{BOSE--EINSTEIN CONDENSATION OF  ALPHA CLUSTERS IN $^{12}$C:\\ REALISTIC 70\% CONDENSATION CASE}
\label{sec-BEC12C70}
\label{sec-12C} 

First we study the BEC of $\alpha$ clusters in the Hoyle state 
at 7.654 MeV excitation energy  of $^{12}$C and the excited states above it. 
Using a three $\alpha$ cluster orthogonality condition model,
Yamada and Schuck \cite{Yamada2005} confirmed the result of 
Matsumura and Suzuki \cite{Matsumura2004} that about 70\% of the 
three $\alpha$ particles in the Hoyle state are sitting in the 0s state.   
 Here it is to be noted that this large probability 70\% itself, compared 
to a superfluid liquid HeII, does not necessarily mean the  
realization of the BEC of three $\alpha$ clusters.  Because these 
traditional cluster models do not have an order parameter to characterize 
the phase transition in its theory, it is impossible to judge in principle
whether  the system is in the NG phase with its global phase 
being locked or not. We attempt this 70\%, i.e., $N_0=0.7N$ 
in our calculations, investigate
 whether BEC is realized for the $N=3$ system. 

\subsection{Alpha-alpha and trapping potentials}

The original nuclear force is supposed to fill the two separate roles 
in our phenomenological approach: One is to trap the $\alpha$ clusters 
inside the nucleus, represented by the external harmonic 
potential $V_{\mathrm ex}(r)$ in Eq.~(\ref{eq:Vex}) with the
parameter $\Omega$. The other is a residual $\alpha$--$\alpha$ interaction,
for which we take the Ali--Bodmer potential $U(r)$ in Eq.~(\ref{eq:AliBodmer}) 
for the $s$-wave, which is specified by the four parameters, the strengths 
of the repulsive and attractive parts, $V_r$ and $V_a$, and 
their respective inverse ranges, $\mu_r$ and $\mu_a$ \cite{Ali1966}. 
These four parameters have been determined to fit the phase shifts of 
$\alpha$--$\alpha$ scattering \cite{Ali1966}.

In the present calculations we adjust $V_r$, which is the most sensitive 
to our analysis among the four parameters, while the remaining parameters
are kept fixed. 
The used potential parameter set $d_0$ of Ref.~\cite{Ali1966} 
are $V_a=130$\,MeV, $\mu_a=0.475$\,fm$^{-1}$, and $\mu_r=0.7$\,fm$^{-1}$. 
The two parameters $\Omega$ and $V_r$ 
play the role to balance between concentration by $V_{\mathrm ex}(r)$ and 
repulsion by $U(r)$, which are crucial for a stable condensate. 

\begin{table}[t] 
\caption{The fitted parameters of $\Omega$ and $V_r$ for three rms radius 
$ {\bar r}$ of $^{12}$C with $N_0=0.7N$.}
\begin{tabular}{c|c|c|c}
\hline
\hline
${\bar r}$\, [fm] &$\Omega$\, [MeV]& $V_r$\,[MeV]& common parameters \\ \hline
3.8 & 2.42 & 415& $V_a= 130$ MeV \\
\cline{1-3} 
3.5 & 2.84 & 398& $\mu_a=0.475$\,fm$^{-1}$ \\
\cline{1-3} 
3.2 & 3.38 & 380& $\mu_r=0.7$\,fm$^{-1}$ \\
\hline
\hline
\end{tabular}\label{table:70parameter12C}
\end{table}

\begin{figure}[tbh!]
\begin{center}
\includegraphics[width=8.0cm]{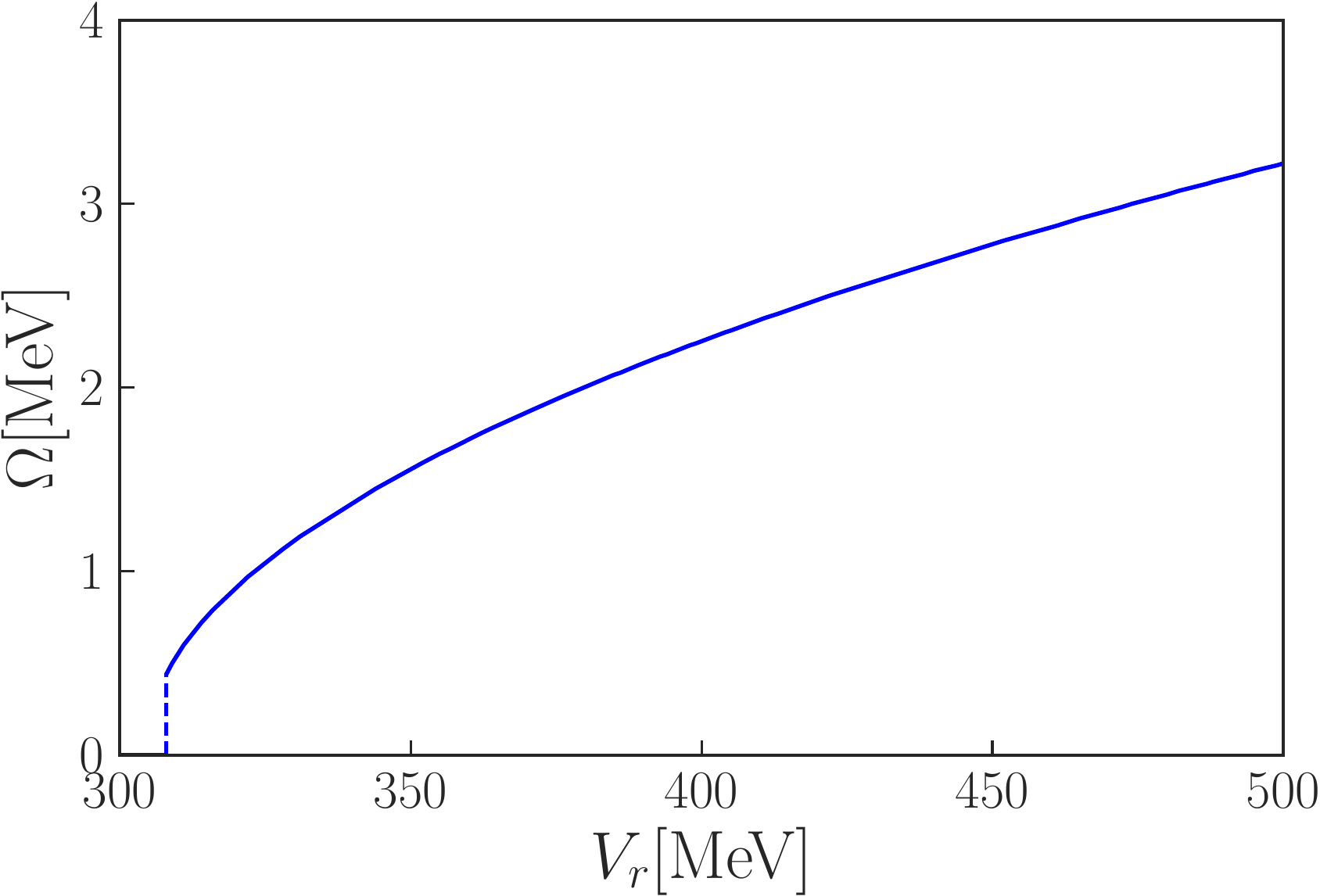} 
\caption{ (Color online) The relation between the confining potential parameter 
$\Omega$ and the repulsive potential $V_r$ for $^{12}$C is 
plotted for the case of ${\bar r}=3.8$ fm with $N_0=0.7N$. As  
shown by the vertical  dashed lines, the BEC system collapses for 
$V_r$ smaller than the critical value around 310 MeV.}
\label{fig:70C12_rmsfit_N70p_38}
\end{center} 
\end{figure}

The vacuum state $\ket{\Psi_0}\ket{0}_{\rm ex}$ is identified 
as the Hoyle state in the calculations. 
The rms radius of the Hoyle state, denoted by ${\bar r}=\sqrt{\langle r^2 \rangle}$,
 is calculated from $\xi(r)$. As the $\alpha$ cluster model calculations 
\cite{Uegaki1977,Fukushima1978,Tohsaki2001,Matsumura2004,Kanada2007}
report the range of ${\bar r}= 3.2$--$3.8\, \mathrm{fm}$ typically,
we consider the three cases of ${\bar r}=3.8\,,\,3.5$ and 3.2 $\mathrm{fm}$. 
The two parameters $\Omega$ and $V_r$ are determined to reproduce 
the experimental $0_3^+$ state, i.e., the first $0^+$ state above the Hoyle state,
which is considered to be the first NG excited state 
$\ket{\Psi_1}\ket{0}_{\rm ex}$.
The parameters $\Omega$ and $V_r$ are given in Table~\ref{table:70parameter12C} 
when $N_0=0.7N$ is assumed.

\begin{figure}[tbh!] 
\begin{center}
\includegraphics[width=8cm]{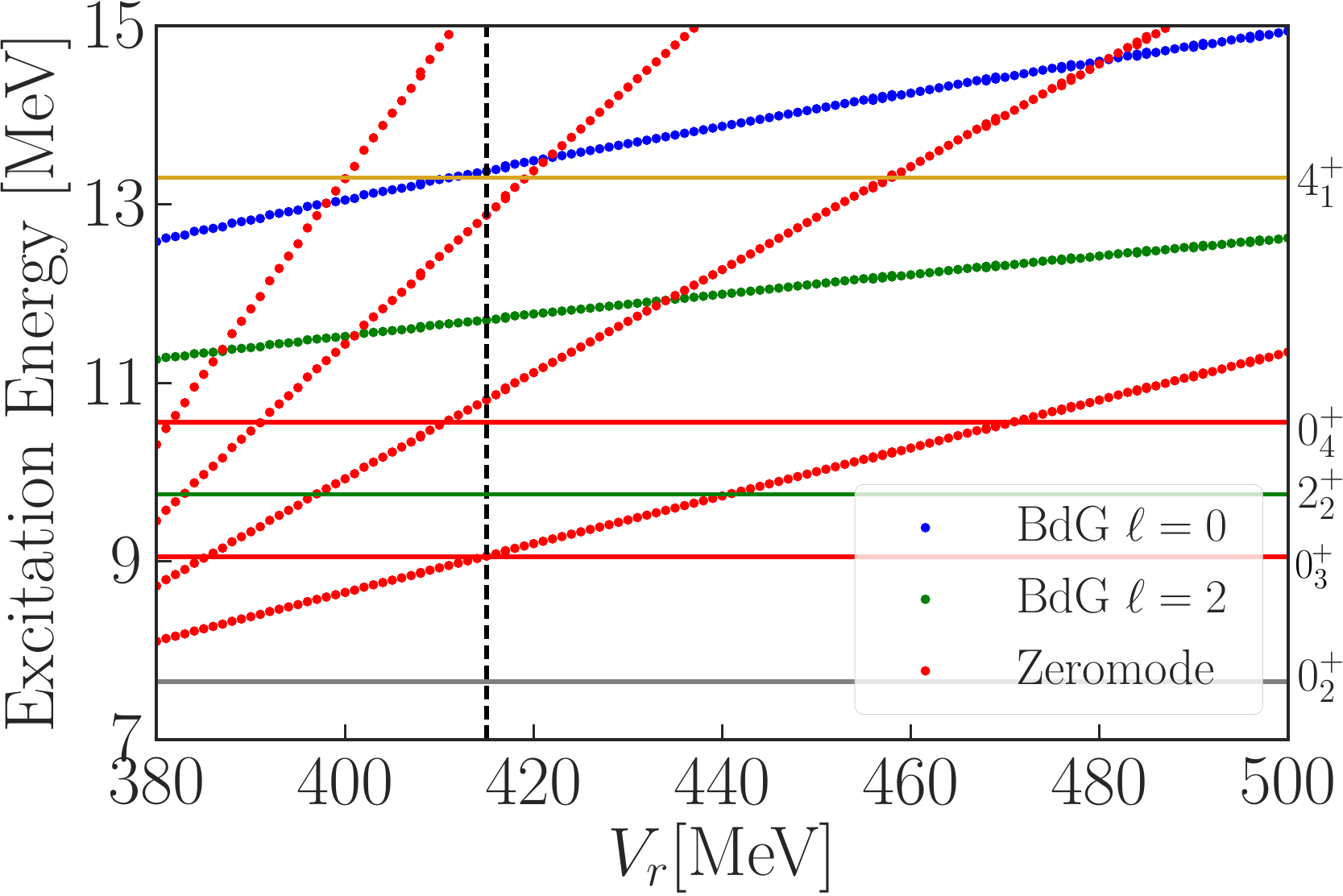}
\caption{(Color online) The energy levels of $^{12}$C calculated 
with $N_0=0.7N$ as a function of $V_r$ for ${\bar r}=3.8$ fm are 
compared with the observed energy levels (horizontal lines) taken from 
Refs.~\cite{Itoh2011,Freer2009,Zimmerman2011,Zimmerman2013,Itoh2013,Freer2011}.
}
\label{fig:70Vr_Omega_E}
\end{center}
\end{figure}

Figure~\ref{fig:70C12_rmsfit_N70p_38} 
shows how $\Omega$ and $V_r$ are constrained when ${\bar r}$ is fixed to be 3.8 fm. 
An increase in $\Omega$ requires an increase in $V_r$ to keep a 
constant ${\bar r}$\,, and they are related roughly as $\Omega^2\, \propto\, V_r$\,.
As is noted in Ref.~\cite{Nakamura2016} where 100\% condensation was assumed, 
we see in Fig.~1 that the BEC for the $N_0=0.7N$ case also becomes unstable 
when $V_r$ is smaller than some critical value.
As displayed in Fig.~\ref{fig:70Vr_Omega_E}, the parameters are determined 
to reproduce the experimental excitation energy of $0^+_3$ from the vacuum, 
$E_1-E_0$. 
The parameters in Table I for ${\bar r}=3.5$ and 3.2 fm are determined similarly. 

\subsection{The order parameter, BdG eigenfunctions and wavefunctions of the zero mode states}

\begin{figure*}[tbh!]
\begin{center}
\begin{tabular}{c}
\begin{minipage}{0.325\hsize}
\begin{center}
\includegraphics[clip, width=5.7cm]{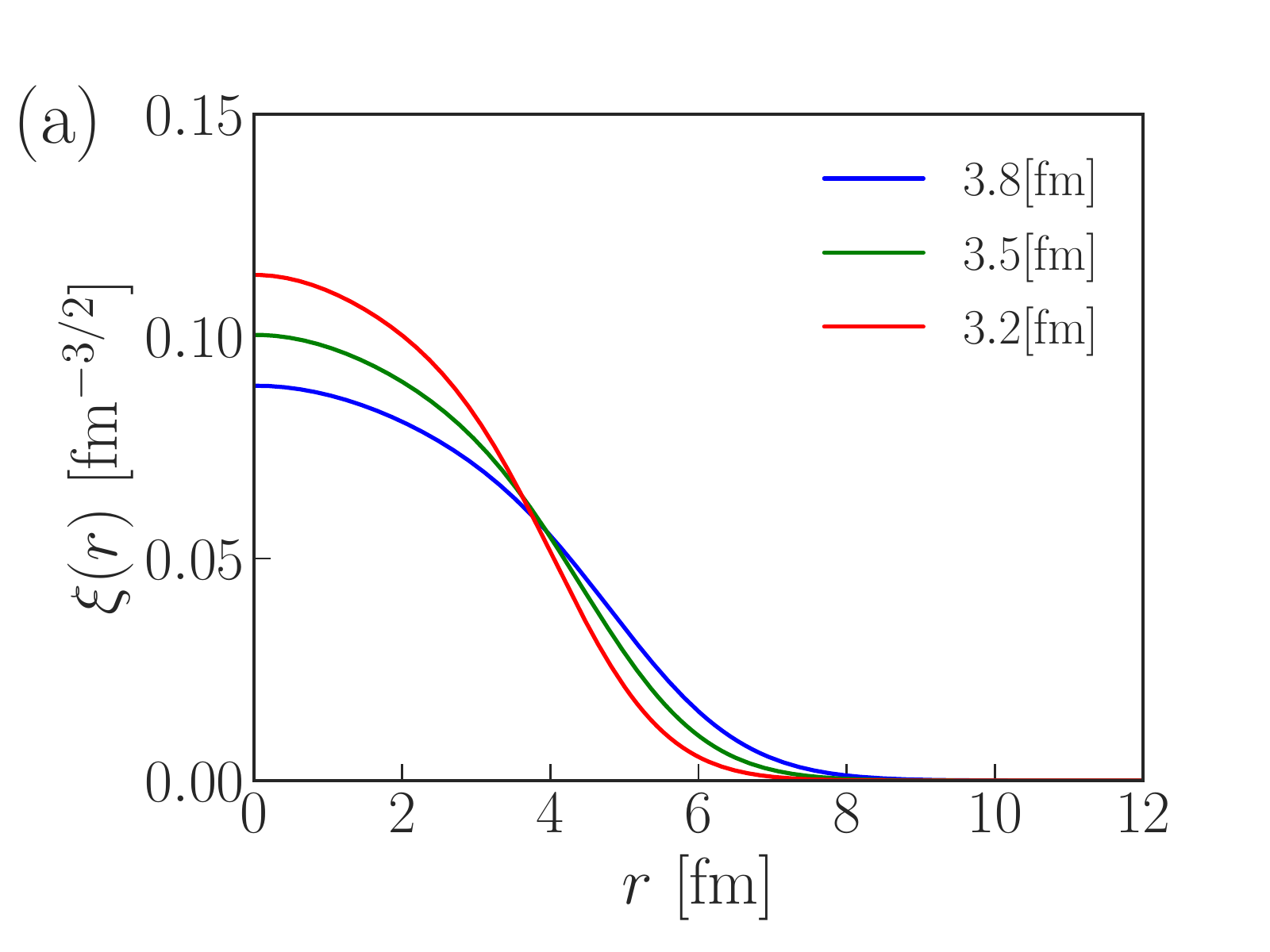}
\hspace{1cm} 
\end{center}
\end{minipage}
\begin{minipage}{0.325\hsize}
\begin{center}
\includegraphics[clip, width=5.7cm]{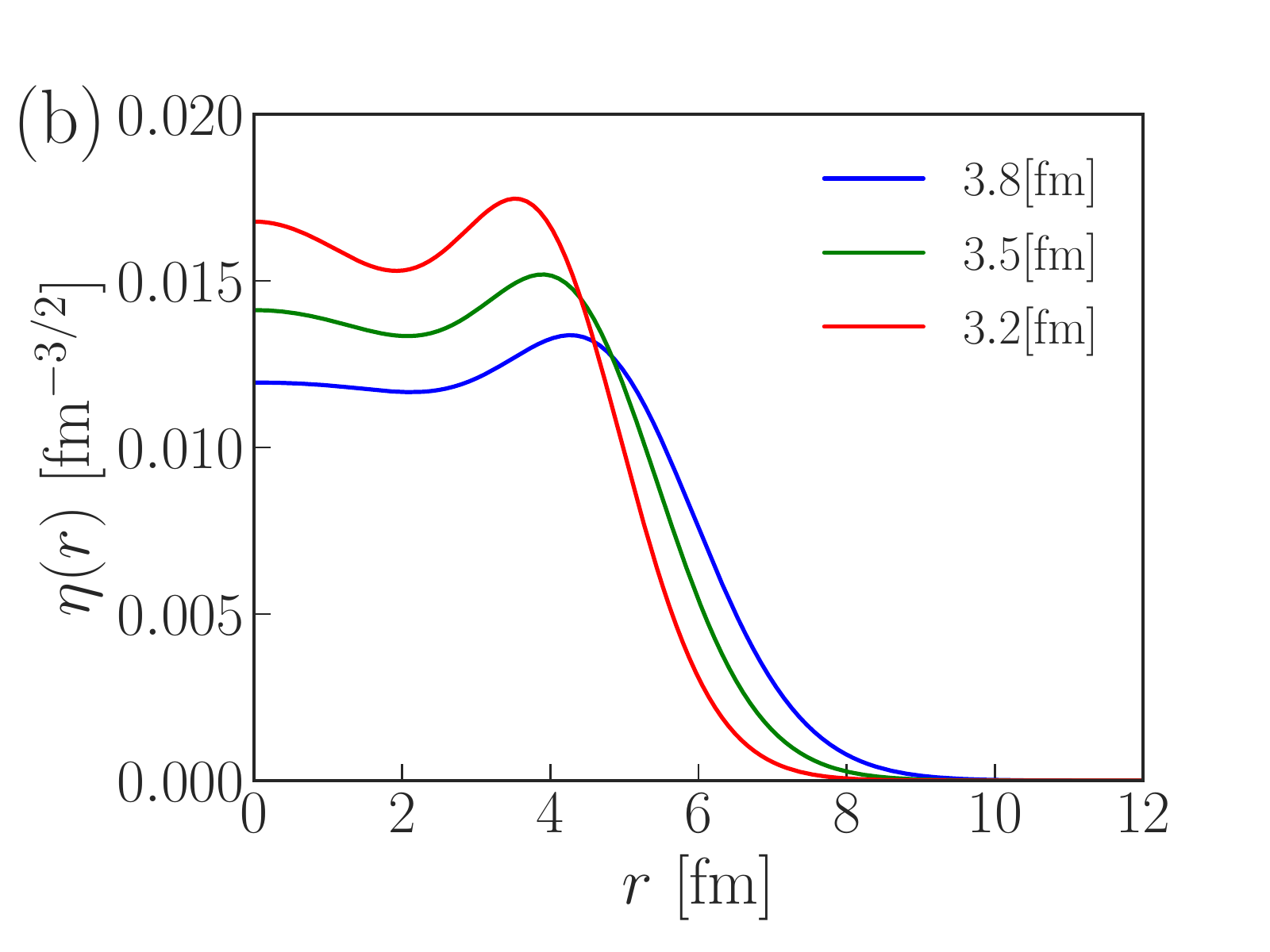}
\hspace{1cm} 
\end{center}
\end{minipage}
\begin{minipage}{0.325\hsize}
\begin{center}
\includegraphics[clip, width=5.7cm]{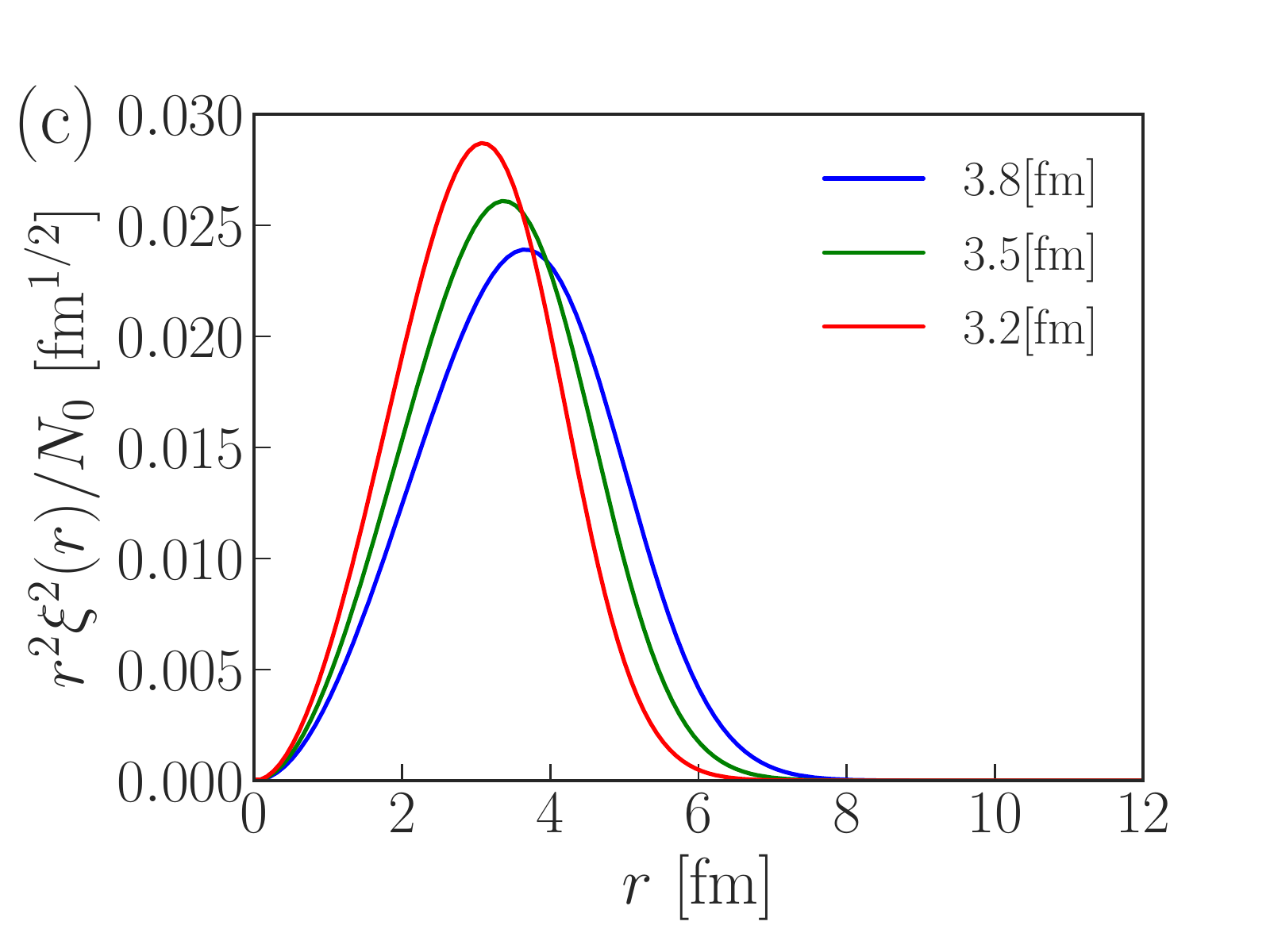}
\hspace{1cm} 
\end{center}
\end{minipage}
\end{tabular}
\end{center}
\caption{(Color online) The calculated (a) eigenfunction with zero eigenvalue
$\xi(r)$,
 (b) its adjoint eigenfunction $\eta(r)$ and (c) radial density distribution 
of the condensate $r^2 | \xi(r)|^2/N_0$
for ${\bar r}=3.8\,,\,3.5$  and 3.2  $\mathrm{fm}$ with $N_0=0.7N$ in $^{12}$C.}
\label{fig:70xieta_32_38}
\end{figure*}
\begin{figure*}[!tbh] 
\begin{center}
\begin{tabular}{c}
\begin{minipage}{0.325\hsize}
\begin{center}
\includegraphics[clip, width=5.7cm]{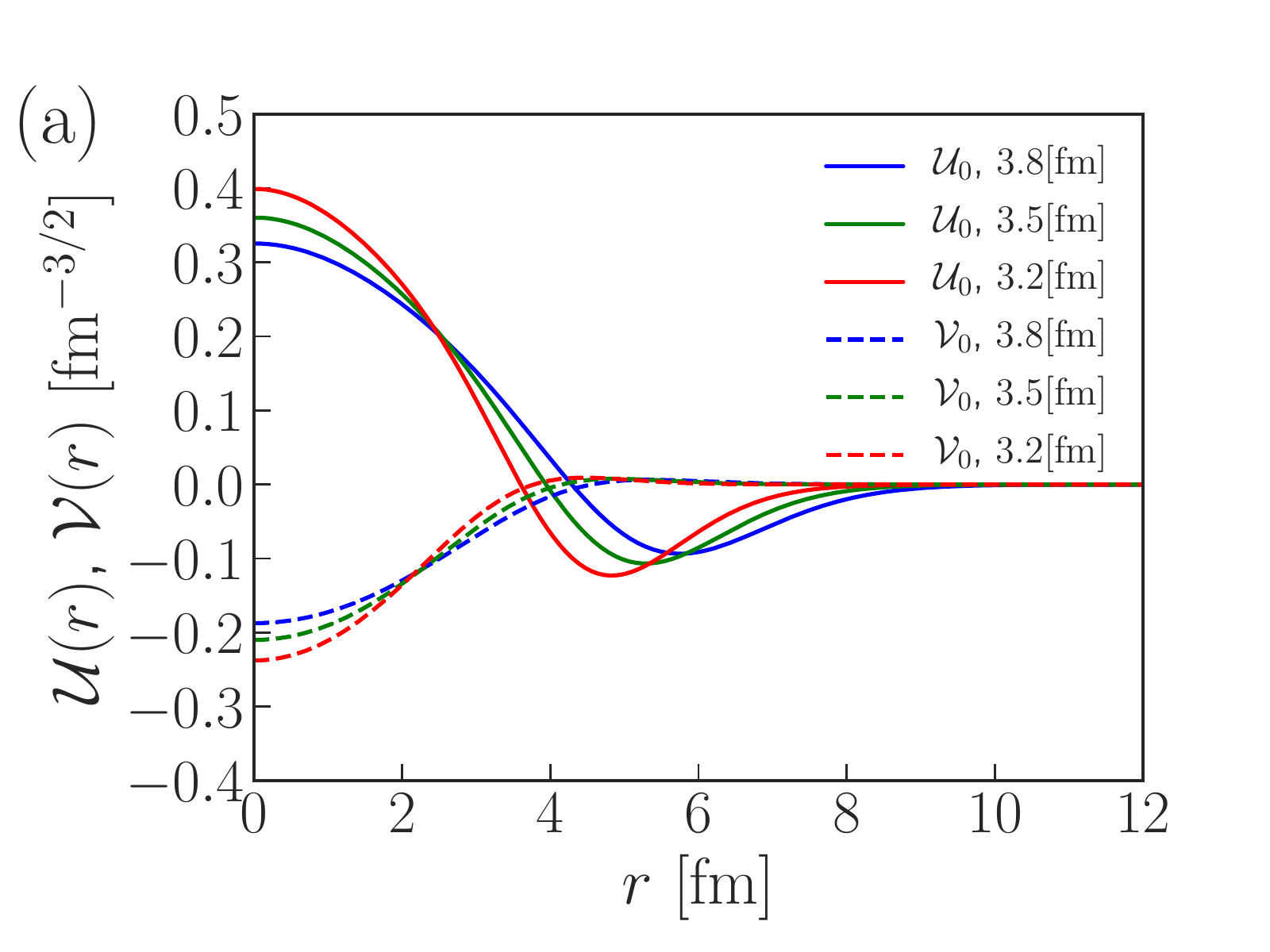}
\hspace{1cm} 
\end{center}
\end{minipage}
\begin{minipage}{0.325\hsize}
\begin{center}
\includegraphics[clip, width=5.7cm]{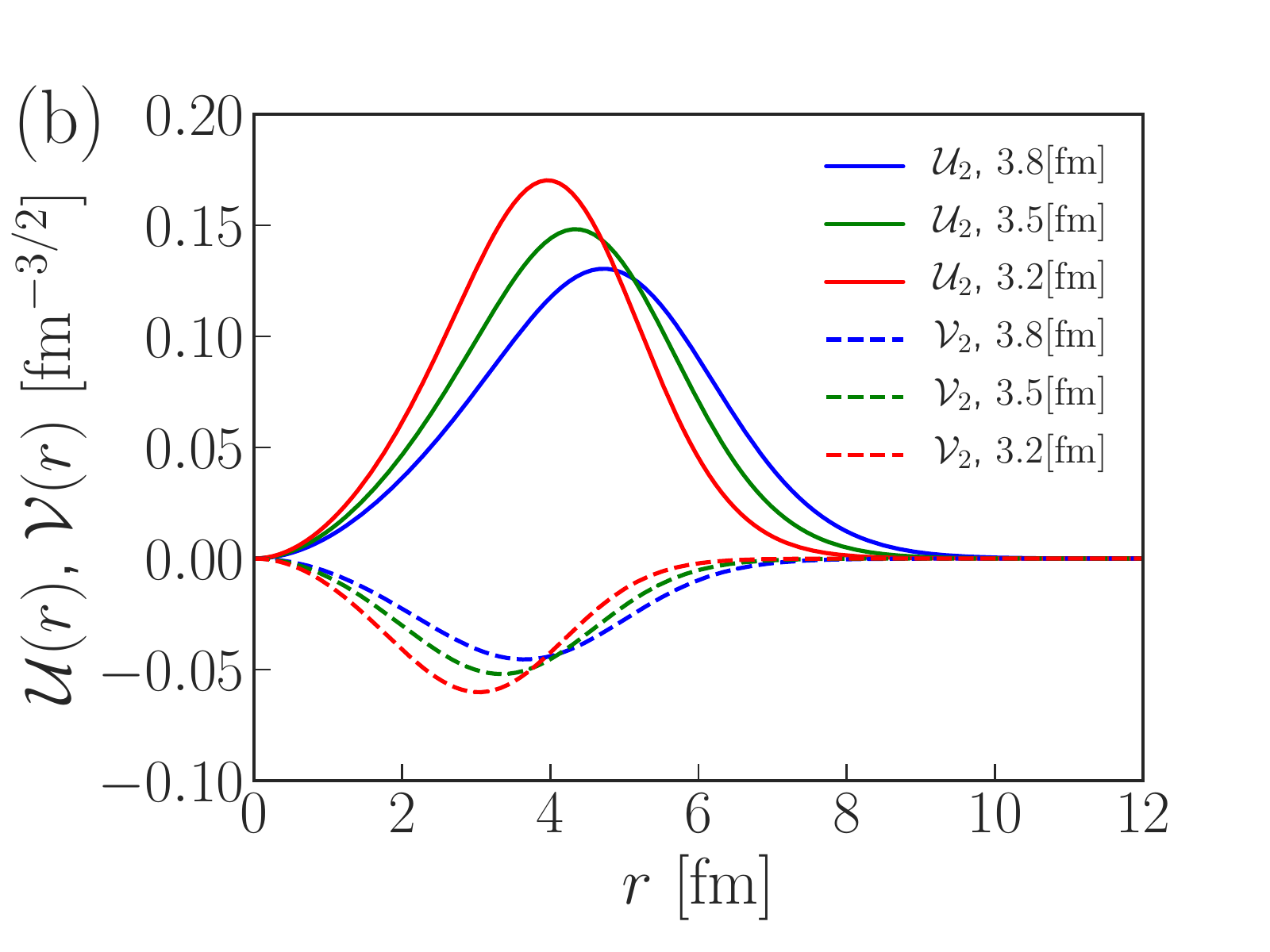}
\hspace{1cm} 
\end{center}
\end{minipage}
\begin{minipage}{0.325\hsize}
\begin{center}
\includegraphics[clip, width=5.7cm]{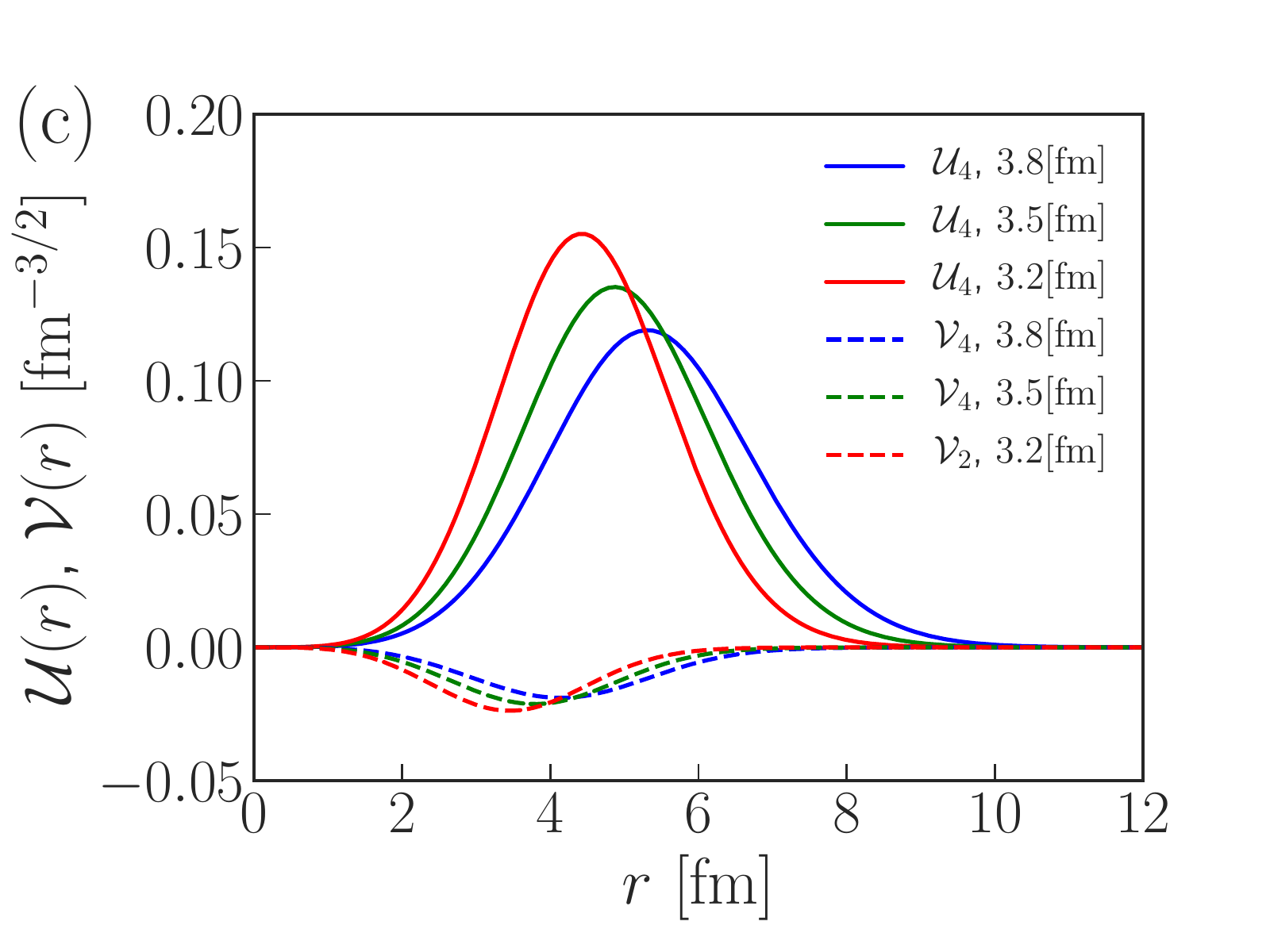}
\hspace{1cm} 
\end{center}
\end{minipage}
\end{tabular}
\end{center}
\caption{(Color online) Numerically calculated BdG wavefunctions, 
${\mathcal U}_{1\ell}(r)$ (solid lines) and ${\mathcal V}_{1\ell}(r)$ 
(dashed lines) for (a) $\ell=0$, (b) $\ell=2$, and (c) $\ell=4$ 
for ${\bar r}=3.8\,,\,3.5$  and 3.2 $\mathrm{fm}$ with $N_0=0.7N$ in $^{12}$C.
In the legends of the figures, the suffix $n=1$ is omitted and only 
$\ell$ is given for ${\mathcal U}_{1\ell}$ and ${\mathcal V}_{1\ell}$.}
\label{fig:70BdG_32_38}
\end{figure*}

\begin{figure*} [!tbh] 
\begin{center}
\begin{tabular}{c}
\begin{minipage}{0.325\hsize}
\begin{center}
\includegraphics[clip, width=5.7cm]{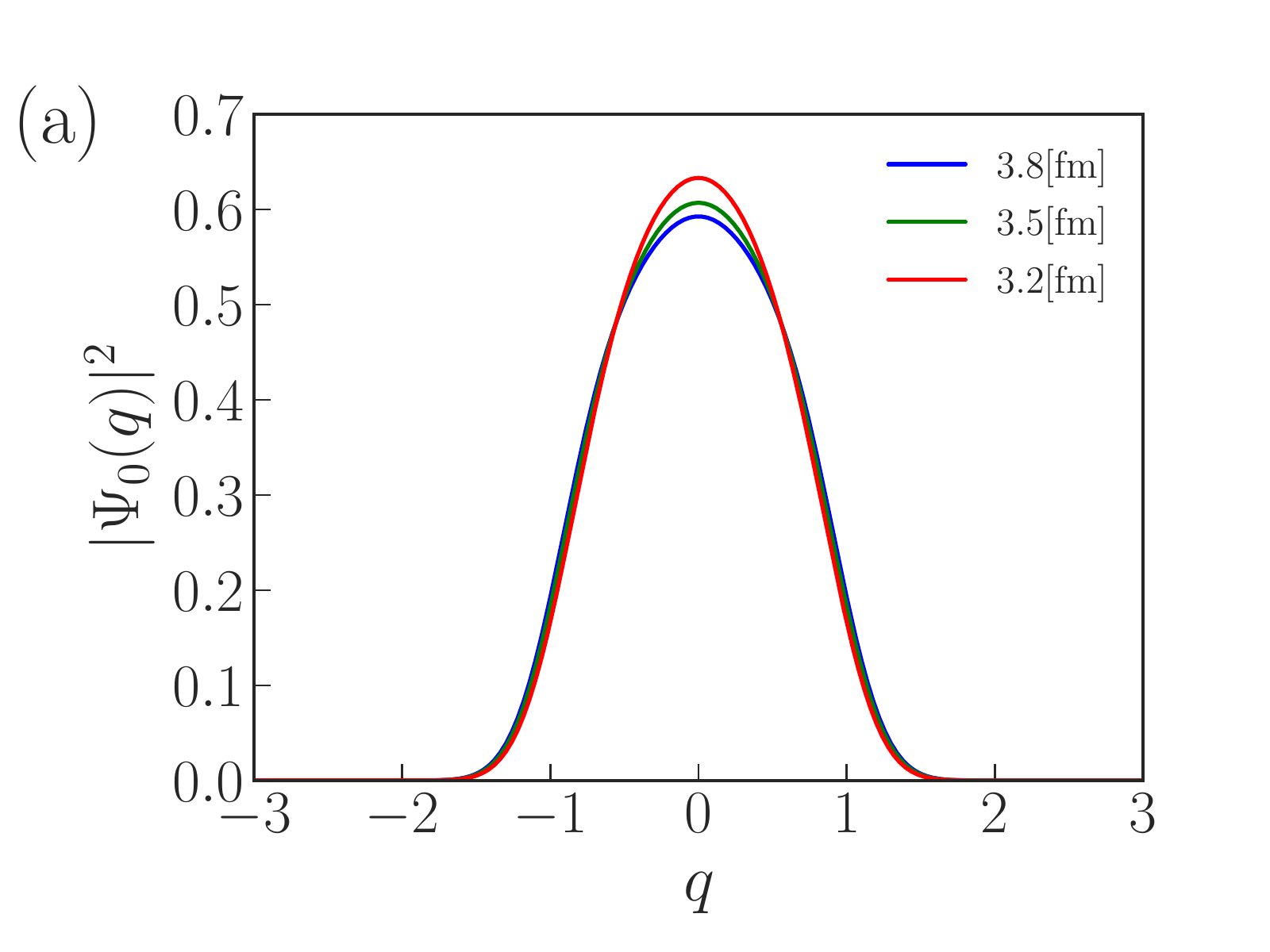}
\hspace{1cm} 
\end{center}
\end{minipage}
\begin{minipage}{0.325\hsize}
\begin{center}
\includegraphics[clip, width=5.7cm]{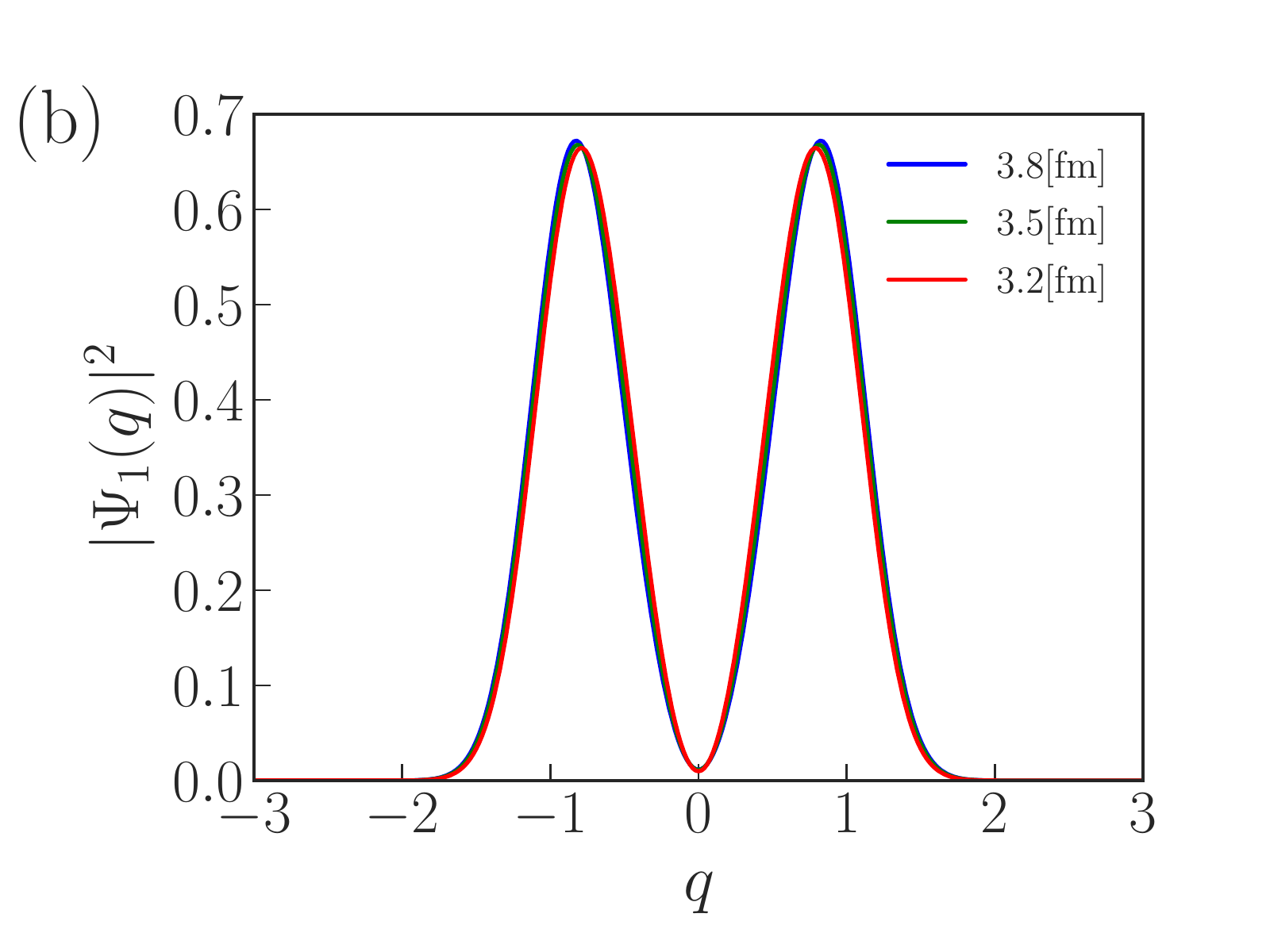}
\hspace{1cm} 
\end{center}
\end{minipage}
\begin{minipage}{0.325\hsize}
\begin{center}
\includegraphics[clip, width=5.4cm]{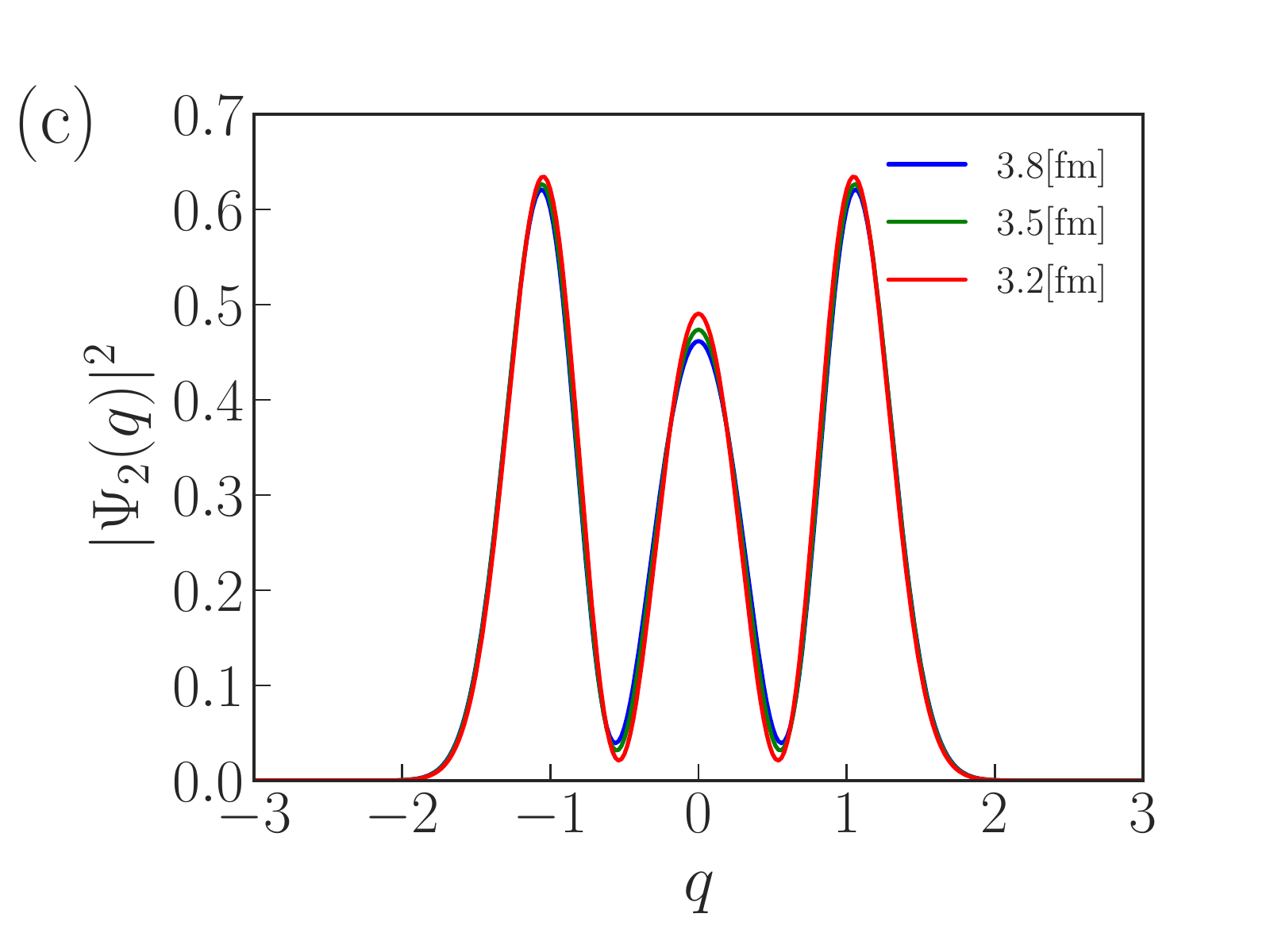}
\hspace{1.2cm} 
\end{center}
\end{minipage}
\end{tabular}
\end{center}
\caption{(Color online) The squares of numerically calculated wavefunctions
of the zero mode states, 
(a) $|\Psi_0(q)|^2$ , (b) $|\Psi_1(q)|^2$, and (c) $|\Psi_2(q)|^2$, 
for ${\bar r}=3.8\,,\,3.5$  and 3.2 $\mathrm{fm}$ with $N_0=0.7N$ in $^{12}$C.} 
\label{fig:70ZeroMode_32_38}
\end{figure*}

\par
The calculated wave functions are shown 
in Figs.~\ref{fig:70xieta_32_38}--\ref{fig:70ZeroMode_32_38} in sequence.

The eigenfunction with zero eigenvalue 
in Eq.~(\ref{eq:BdGy0}) or the order parameter 
in Eq.~(\ref{eq:GP}), $\xi(r)$, is shown in Fig.~\ref{fig:70xieta_32_38} (a).
In Fig.~\ref{fig:70xieta_32_38} (b) the adjoint eigenfunction $\eta(r)$ 
in Eq.~(\ref{eq:BdGy-1-2}), which is calculated as a derivative of $\xi$ 
with respect to the number of the $\alpha$ clusters $N_0$ in Eq.~(24), 
represents the fluctuation of the number of $\alpha$ clusters \cite{Kobayashi2009}.
The $|\xi(r)|^2$ represents the condensate fraction density, and 
the superfluid density \cite{Josephson1966,Ceperley1995,Mulkerin2017} 
is given by $|\xi(r)|^2$/$N_0$.  In Fig.~\ref{fig:70xieta_32_38} (c) 
the radial density distribution of the condensate, 
defined by $ r^2 \xi^2(r)/N_0$ is displayed. We see from 
Fig.~\ref{fig:70xieta_32_38} (a) that the superfluid density distribution 
of the condensate $\alpha$ clusters in the vacuum Hoyle state extends 
to about 8 fm, which is consistent with the picture of the diffused 
gas-like structure. As the size of the condensate becomes larger 
from ${\bar r}=3.2$ to 3.8 fm, the density in the central region 
is depressed while that in the surface is enhanced. 
In Fig.~~\ref{fig:70xieta_32_38} (b) we note that the number 
fluctuation around the average $N_0$ is roughly flat up to 4 fm 
with about 0.05, 0.045 and 0.04 for ${\bar r} = 3.8$, 3.5 and 3.2 fm, 
respectively, and with a small depression at around 2--2.5 fm, 
and extends to about 8fm. The magnitude of $\eta$ is considerably 
large even at 6 fm compared with $\xi$, which means that the number 
fluctuation occurs significantly not only in the central region 
but also in the surface region beyond the rms radius up to about 6 fm. 
As seen in Fig.~\ref{fig:70xieta_32_38} (c), 
the radial distribution of the superfluid density diffuses 
more widely for larger size of the condensate.

\par
In Fig.~\ref{fig:70BdG_32_38} the wave functions of the first BdG 
excitation modes ($n=1$) with $\ell=0\,,\, 2$ and $4$ are shown. 
The radial behavior of ${\mathcal U}_{1\ell}(r)$, which represents 
the radial extension of the state, is rather similar to that of 
$\xi$ in Fig.~\ref{fig:70xieta_32_38} (a). The peak position of 
${\mathcal U}_{1\ell}(r)$ moves outward as $\ell$ increases.
This is reasonably understood by considering that the states 
with larger $\ell$ correspond to the higher excitation energy,
as seen in Fig.~\ref{fig:70Vr_Omega_E}. 
On the other hand, in Fig.~\ref{fig:70BdG_32_38} 
we see that ${\mathcal V}_{1\ell}(r)$ is drastically 
small compared to  ${\mathcal U}_{1\ell}(r)$ as $\ell$ increases 
and  is strongly damped beyond $r=4$ fm in the surface region 
for all the  cases of ${\bar r}$=3.2 -- 3.8 fm. 
The size of the condensate is dominantly determined by 
the behavior of the wave function ${\mathcal U}_{1\ell}(r)$. 
In fact, as the size of the condensate becomes larger 
from ${\bar r}=$3.2 to 3.8 fm, the peak of  ${\mathcal U}_{1\ell}(r)$ 
moves outward  and the amplitude at the surface is increased, 
while that in the internal region decreases. 
The magnitude of ${\mathcal V}_{1\ell}(r)$, which represents the 
quantum fluctuations of the $\alpha$ clusters of 
the condensate, decreases for larger $\ell$. 
For ${\ell}=0$ the magnitude of ${\mathcal V}_{10}(r)$ in the 
internal region is not small compared to ${\mathcal U}_{10}(r)$, 
which means that the quantum fluctuation is significant 
for ${\ell}=0$. The node of the ${\mathcal U}_{1\ell}(r)$ for $\ell=0$ 
is due to the orthogonality to the nodeless
$\xi$ and $\eta$. 

As in Table~\ref{table:70parameter12C}, $V_r$ changes only slightly
in our fitting,  and the wave functions are determined 
mainly by the value of $\Omega$, not by the value of $V_r$\,.
It is natural, as seen in Figs.~\ref{fig:70xieta_32_38} and \ref{fig:70BdG_32_38}, that the peaks at the center are enhanced or the peak positions are shifted closer to the center with the higher peaks, as ${\bar r}$ is smaller. We note the relation ${\bar r}\,\propto\,
1/\sqrt{\Omega}$ there, which is identical with the relation for the ground state wave function of a simple harmonic oscillation.

 We introduce the eigenstate of ${\hat Q}$, denoted by
$\ket{q}$, as  
${\hat Q}\ket{q}=q\ket{q}$. In order to solve Eq.~(\ref{eq:HuQPeigen}), 
we move to the $q$-diagonal representation, in which the state is represented by 
the wavefunction $\Psi_\nu(q)=\braket{q}{\Psi_\nu}$, and the operators
${\hat Q}$ and ${\hat P}$ are represented by $q$ and 
$\frac{1}{i}\frac{\partial}{\partial q}$, respectively, 
consistently with Eq.~(\ref{eq:CCRQPa}).
Figure~\ref{fig:70ZeroMode_32_38} represents $|\Psi_{\nu}(q)|^2$
($\nu=0\,,\,1\,,\,2$) numerically calculated. 
We see that the excitation of the NG mode is caused
by the nodal excitation of $\Psi_{\nu}(q)$ with respect to $q$ in the NG subspace. 
It is important to note that this nodal excitation is anharmonic as seen 
in $\hat H_u^{QP}$ in Eq.~(\ref{eq:HuQP}), which brings the excitation 
energy of the $\nu=1$ state  lower and  closer to the vacuum, and the $\nu=2$ state 
closer to the $\nu=1$ state in Fig.~\ref{fig:70Vr_Omega_E}  
(and Fig.~\ref{fig:ex_energy_C_N70p_32_38} later). 
In fact, in Fig.~\ref{fig:70Vr_Omega_E} (Fig.~\ref{fig:ex_energy_C_N70p_32_38}) 
the energy intervals between the $\nu=0$ and $\nu=1$ states, and between the $\nu=1$ 
and $\nu=2$ states are smaller than those for other higher NG mode states for $\nu>2$. 
It is worth noting that the NG wave functions and 
the NG excitation energies depend very little on ${\bar r}$ and $\Omega$\,.
This is because the coefficients of $\hat H_u^{QP}$ in Eq.~(\ref{eq:HuQP})
include the $\Omega$-dependent $\xi(r)$ and $\eta(r)$ only in the integrands
and the integration values are insensitive to $\Omega$\,.
On the other hand, the coefficients of $\hat H_u^{QP}$ have a factor $V_r$,
and the NG energy levels rise with an increasing $V_r$\,. 

\subsection{Electric transition probabilities}

Using the obtained wave functions, the reduced transition probabilities 
 in Eq.~(\ref{eq:BE2}) with Eqs.~(\ref{eq:E2-1}) and (\ref{eq:E2-2})
and the monopole transition probabilities in Eq.~(\ref{eq:E0})
are calculated numerically, The results are 
summarized in Table~\ref{table:70B} in comparison with other 
theoretical calculations in 
Refs.~\cite{Kanada2007} and \cite{Funaki2015}. 
There the parameters are $\Omega=2.42$ MeV and $V_r=415$ MeV in 
the case of ${\bar r}=3.8$ fm in Table~\ref{table:70parameter12C}.
We note the difference of our results between the E2 transitions 
$2_2^+\, \rightarrow \, 0_2^+$ and $2_2^+\, \rightarrow \, 0_3^+$: 
Whereas the former is the transition to change only the BdG state, 
the latter involves transition in both of the NG and BdG states. 
The difference between the two monopole transitions 
$0_2^+\,\rightarrow\, 0_3^+$ and $0_2^+\,\rightarrow\, 0_4^+$
is explained as follows: The most dominant 
term $I_p\bra{\Psi_\nu} {\hat P} \ket{\Psi_0}$ interferes with
the other two terms constructively in the former, but destructively
in the latter.

\begin{table}[tbh]
\caption{Calculated reduced transition probabilities $B({\rm E}2:2\, \rightarrow\, 0)$
 and monopole transition probabilities 
$M({\rm E}0:0\, \rightarrow\, 0)$
in $^{12}$C with 70\%  and  100\% 
(see Subsec.~\ref{subsec-100condensation})
condensation in units of $e^2\,{\rm fm}^4$ and ${\rm fm}^2$, respectively,
are displayed in comparison with
 Ref.~\cite{Kanada2007} and Ref.~\cite{Funaki2015}.}
\label{table:70B}
\begin{ruledtabular}
\begin{tabular}{ccccc}
Transition &70\%&100\% &Ref.~\cite{Kanada2007} & Ref.~\cite{Funaki2015} 
\\\hline
$B({\rm E}2:2_2^+\, \rightarrow \, 0_2^+)$ &121&158 &100 & 295-340\\
$B({\rm E}2:2_2^+\, \rightarrow \, 0_3^+)$ & 76&62 &310 & 88-220\\
$M({\rm E}0:0_2^+\, \rightarrow \, 0_3^+)$ & 1.59&2.34 &34.5& 2.0 \\
$M({\rm E}0:0_2^+\, \rightarrow \, 0_4^+)$ & 0.072&0.145 &0.57& {\rm ---}\\
\end{tabular}
\end{ruledtabular}
\end{table}

\section{ROBUSTNESS OF BOSE--EINSTEIN CONDENSATE STRUCTURE OF $^{12}$C VS CONDENSATION RATES}
\label{sec-12CN0}

So far it has been assumed that the $\alpha$ clusters inside $^{12}$C 
are condensed with $N_0=0.7N$ based on the preceding cluster model calculations 
 \cite{Matsumura2004,Yamada2005} that about 70\% of three $\alpha$ 
clusters inside $^{12}$C are in the $0$s state. However, experimentally
 the condensation rate  has not been directly measured. Therefore  
it seems important to investigate whether the structure obtained 
under 70\% condensation  is robust  for the different  condensation 
rates. Here we study and compare the three cases of condensation, 
$N_0= 3.0$ (condensation rate 100\%),  2.5 (83\%) and  2.0 (67\%) in $^{12}$C. 

\subsection{Three cases of condensation rate: $N_0=$3, 2.5 and 2}

Similarly as in the previous section, we determine the confining and 
$\alpha$--$\alpha$ potentials by  adjusting the parameters 
of $\Omega$ and $V_r$, for different  $N_0$ with  ${\bar r}=3.8$ [fm] and $E_1$. 
The obtained potential parameters are  summarized in Table~\ref{table:parameter12CN0}.
The  values of $\Omega$ and $V_r$ depend only slightly on $N_0$.

\begin{table}[tbh!]
\caption{The fitted parameters of $\Omega$ and $V_r$ used in the calculations
 with different  three condensation rates of $N_0= 3.0\,,\, 2.5\,,\, 2.0$ 
in $^{12}$C with fixed ${\bar r}=3.8$ fm.}
\begin{tabular}{c|c|c|c}
\hline
\hline
$N_0$ &$\Omega$\, [MeV]& $V_r$\,[MeV]& common parameters \\ \hline
3.0 & 2.62 & 403& $V_a= 130$ MeV \\
\cline{1-3} 
2.5 & 2.53 & 410& $\mu_a=0.475$\,fm$^{-1}$ \\
\cline{1-3} 
2.0 & 2.40 & 417& $\mu_r=0.7$\,fm$^{-1}$ \\
\hline
\hline
\end{tabular}\label{table:parameter12CN0}
\end{table}

\begin{figure}[tbh!]
\begin{center}
\includegraphics[width=8cm]{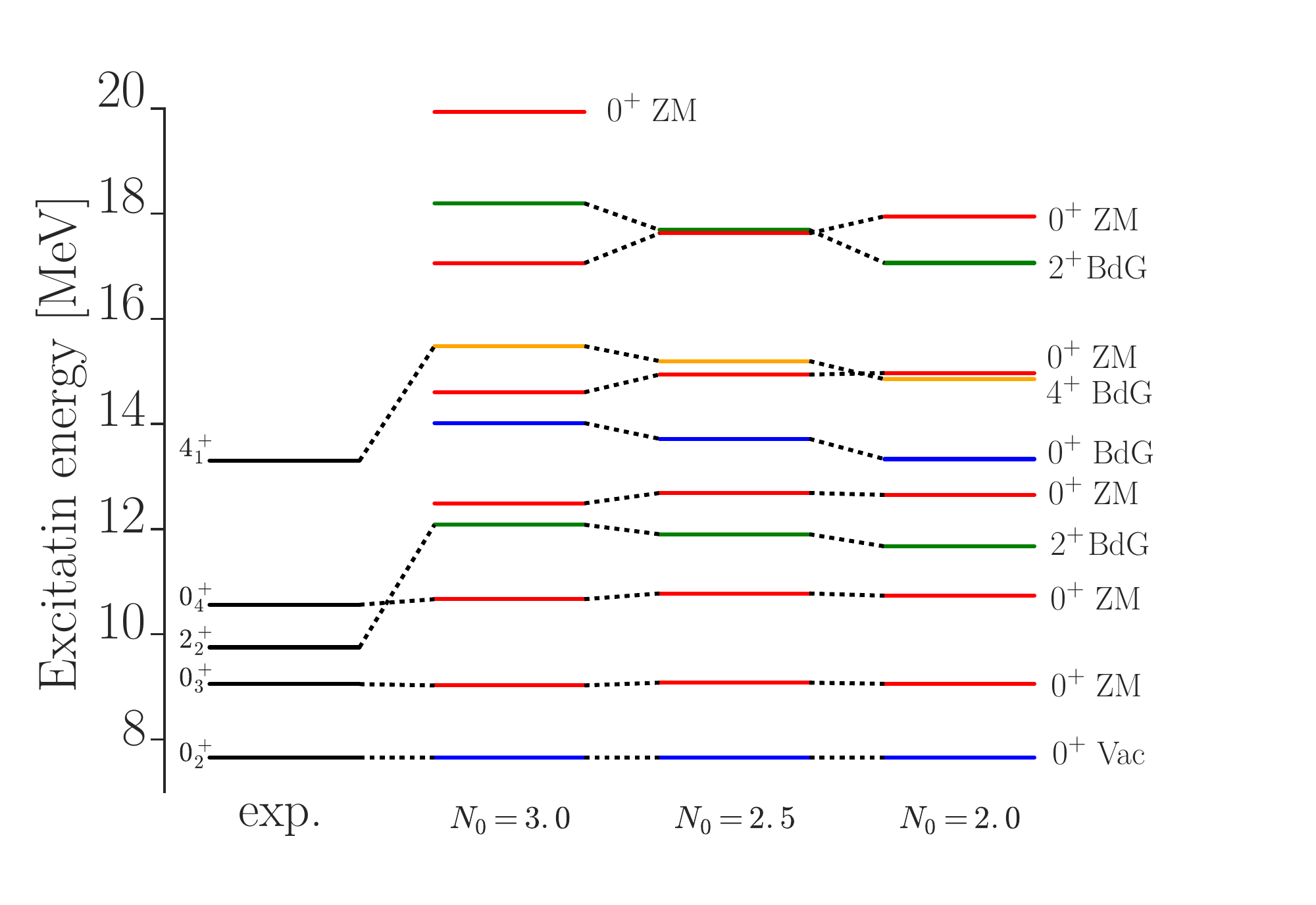}
\caption{(Color online) The energy levels of $^{12}$C calculated with the 
parameters in Table~\ref{table:parameter12CN0} for the 
three condensation rates, $N_0=3.0$ (100\%), 2.5 (83\%) and 2.0 (67\%), 
with fixed ${\bar r}=3.8\, \mathrm{fm}$.}
\label{fig:energyN0}
\end{center}
\end{figure}

In Fig.~\ref{fig:energyN0} the calculated energy levels of $^{12}$C
 for different three condensation rates  are displayed. The  
 zero mode excitation levels  remain constant or rise only slightly 
as $N_0$ becomes smaller. This is  because ${\bar r}$ and $V_r$ are 
fixed and the fitted $\Omega$ is therefore almost constant for varying $N_0$\,. 
This   shows that the nature of the low-lying zero mode states
 just above the Hoyle state, the experimental $0_3^+$ and the $0_4^+$ states,
 is robust with respect to the changes of the condensation rates. 
On the other hand,  the BdG excitation energies decrease only slightly 
as $N_0$ becomes smaller. This is reasonable  because the Hartree 
and Fock terms of the self-interaction in BdG equations (\ref{eq:BdG_u}) 
and (\ref{eq:BdG_v}), which push up energy levels due to a dominant 
contribution of the repulsive force, are proportional to $N_0$\,. 
This is also understood as follows. Because $\Omega$ 
becomes smaller as $N_0$ decreases as in Table~\ref{table:parameter12CN0}, 
the BdG vibrational energy
 becomes smaller. As a result, the choice of $N_0=2.0$, about 70\% 
condensation, gives the most favorable BdG energy levels in our calculations.

\subsection{Detailed study of 100\% condensation case}
\label{subsec-100condensation}

To see how the energy level structure, wave functions 
and electric transitions,
 calculated  with $N_0=0.7N$ 
in the previous section,  are affected for the different 
three condensation rates, we will focus on comparing them here 
with those in the 
100\% condensation case.  A brief result of 100\% condensation has 
been given in Ref.\cite{Nakamura2016}. We will present here 
the results rather in detail because it was found that there were 
small errors in numerical calculations involving the BdG modes 
with $\ell=2$ and 4, although the results for the zero mode states
 and the BdG modes with $\ell=0$ are not altered.

\begin{figure}[tbh!]
\begin{center}
\includegraphics[width=8.0cm]{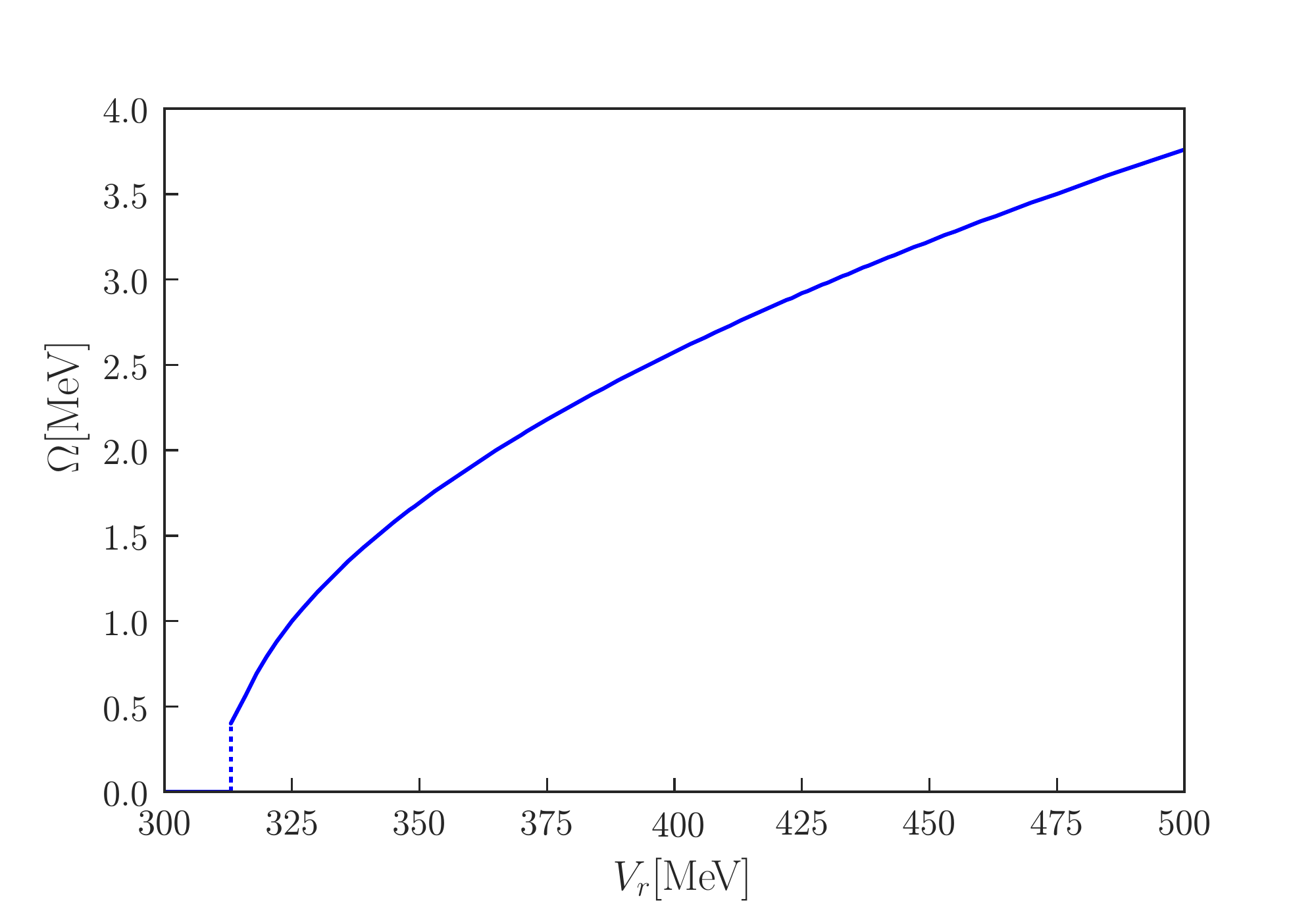}
\caption{The relation between the confining potential parameter 
$\Omega$ and the repulsion potential $V_r$ for $^{12}$C is plotted 
for the case of ${\bar r}=3.8$ fm with 100\% condensation ($N_0=N$).
}
\label{fig:rmsfit}
\end{center}
\end{figure}

\begin{figure}[tbh!]
\begin{center}
\includegraphics[width=8cm]{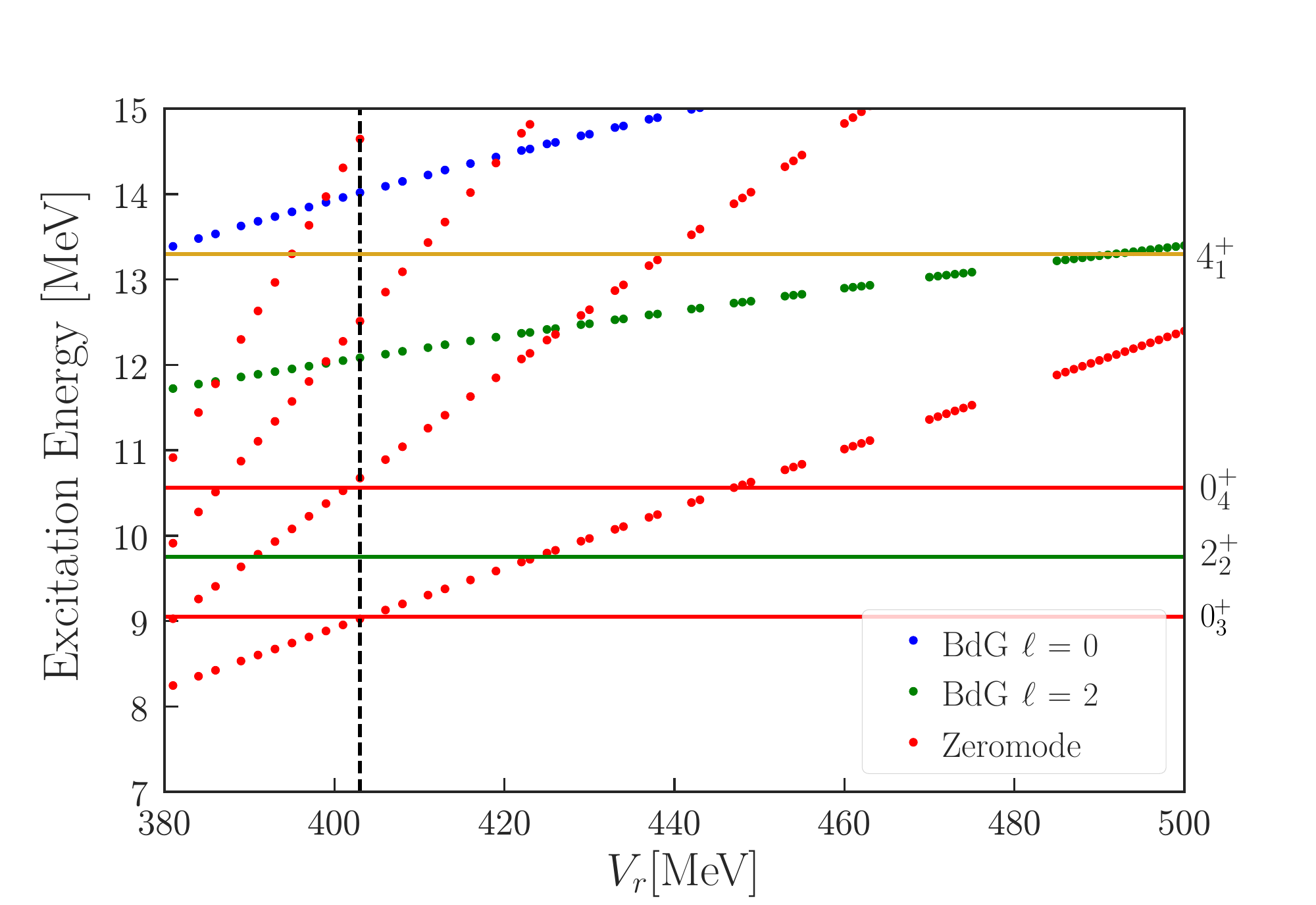}
\caption{ (Color online) The energy levels of $^{12}$C calculated with 
100\% condensation ($N_0=N$) as a function of $V_r$ for ${\bar r}=3.8$ 
fm are compared with the observed energy levels (horizontal lines) 
taken from 
Refs.~\cite{Itoh2011,Freer2009,Zimmerman2011,Zimmerman2013,Itoh2013,Freer2011}.
}
\label{fig:Vr_Omega_E}
\end{center}
\end{figure}

As shown in Fig.~\ref{fig:rmsfit}, the two parameters $\Omega$ and $V_r$ are 
constrained for $N_0=N$ when ${\bar r}$ is fixed to be $3.8\, \mathrm{fm}$, 
similarly to the $N_0=0.7N$ case in Fig.~\ref{fig:70C12_rmsfit_N70p_38}. 
The parameters are determined 
as in Fig.~\ref{fig:Vr_Omega_E}, similarly to the 
70\% case in Fig.~\ref{fig:70Vr_Omega_E},  and 
the obtained  parameters are given in Table~\ref{table:100parameter12C}. 

\subsubsection{Energy levels}
\begin{figure}[tbh!]
\begin{center}
\includegraphics[width=8cm]{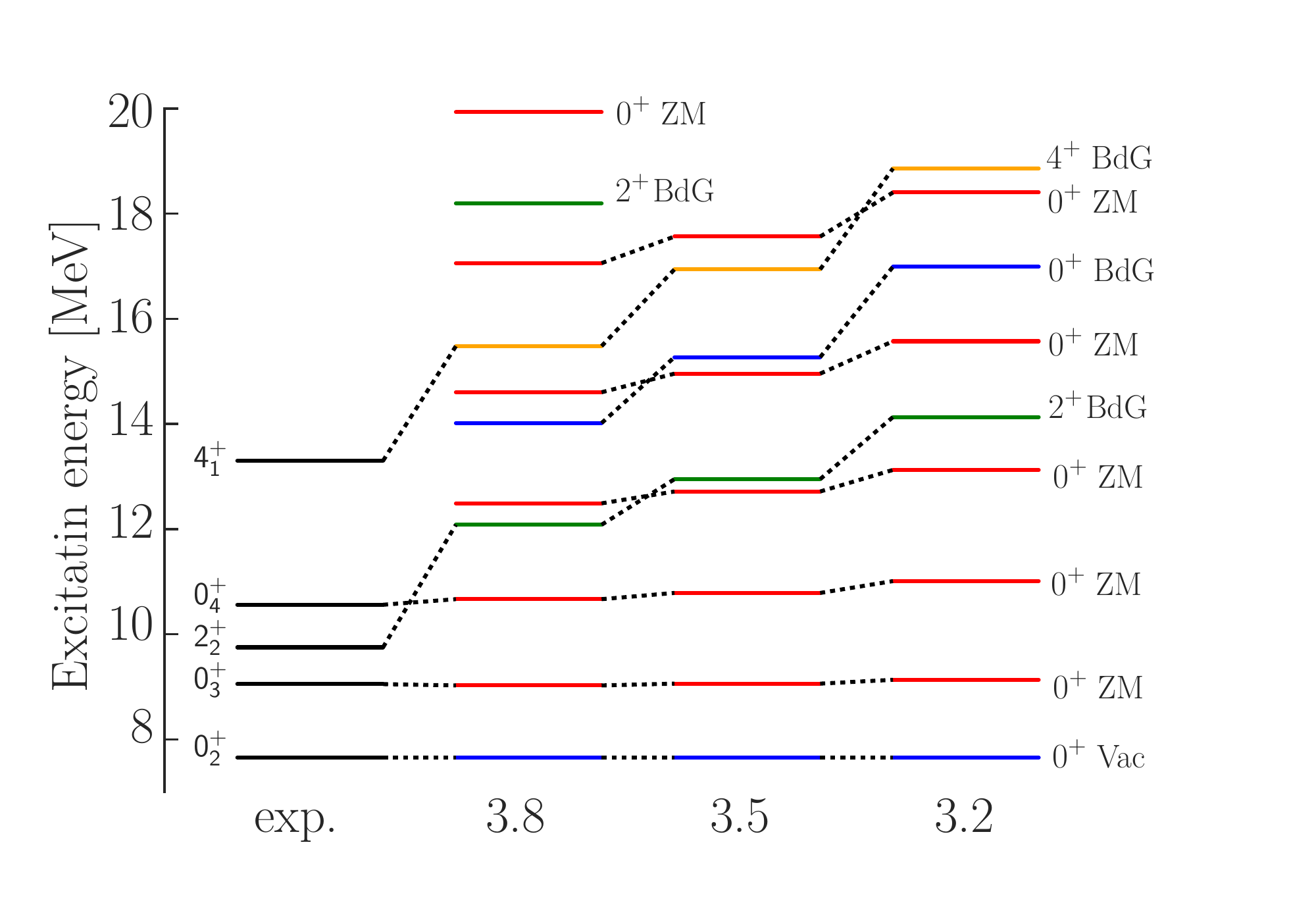}
\caption{ (Color online) The energy levels of $^{12}$C, 
calculated from the parameters in Table~\ref{table:100parameter12C} with 
fixed 100\% condensation ($N_0=N$)  for ${\bar r}=3.8$, 3.5 and 3.2 fm, 
are compared with the experimental energy levels from 
Refs.\cite{Itoh2011,Freer2009,Zimmerman2011,Zimmerman2013,Itoh2013,Freer2011}.
Vac, ZM and BdG mean the vacuum Hoyle state, zero mode states and 
BdG excited states, respectively.
}
\label{fig:energy}
\end{center}
\end{figure}

The energy levels, calculated from the parameters
 in Table \ref{table:100parameter12C}, are displayed for each of 
${\bar r}=3.8\,,\, 3.5
\,,\,3.2$ fm in Fig.~\ref{fig:energy}.
The $0^+_4$ state is reproduced well as the second NG excited state 
$\ket{\Psi_2}\ket{0}_{\rm ex}$, which is  
similar to the $N_0=0.7N$ case for the whole calculated range of the ${\bar r}$. 
The calculated levels of the $2^+_2$ and $4^+_1$ 
states are  identified as the lowest excitations of the BdG modes 
with $\ell=2$ and $\ell=4$, respectively. 
The BdG energy levels 
show that the levels increase
for smaller ${\bar r}$. The increase in $\Omega$ affects more dominantly on
the levels than the decrease in $V_r$\,, similarly to the 70\% condensation case.

\begin{table}[tbh!]
\caption{ The fitted parameters of $\Omega$ and $V_r$ for three rms radius 
${\bar r}=3.8$, 3.5 and 3.2 fm of $^{12}$C with 100\% condensation ($N_0=N$).
}
\begin{tabular}{c|c|c|c}
\hline
\hline
${\bar r}$\, [fm] &$\Omega$\, [MeV]& $V_r$\,[MeV]& common parameters \\ \hline
3.8 & 2.62 & 403& $V_a= 130$ MeV \\
\cline{1-3} 
3.5 & 3.10 & 389& $\mu_a=0.475$\,fm$^{-1}$ \\
\cline{1-3} 
3.2 & 3.77 & 375& $\mu_r=0.7$\,fm$^{-1}$ \\
\hline
\hline
\end{tabular}\label{table:100parameter12C}
\end{table}

\begin{figure*}[tbh!]
\begin{center}
\begin{tabular}{c}
\begin{minipage}{0.325\hsize}
\begin{center}
\includegraphics[clip, width=5.7cm]{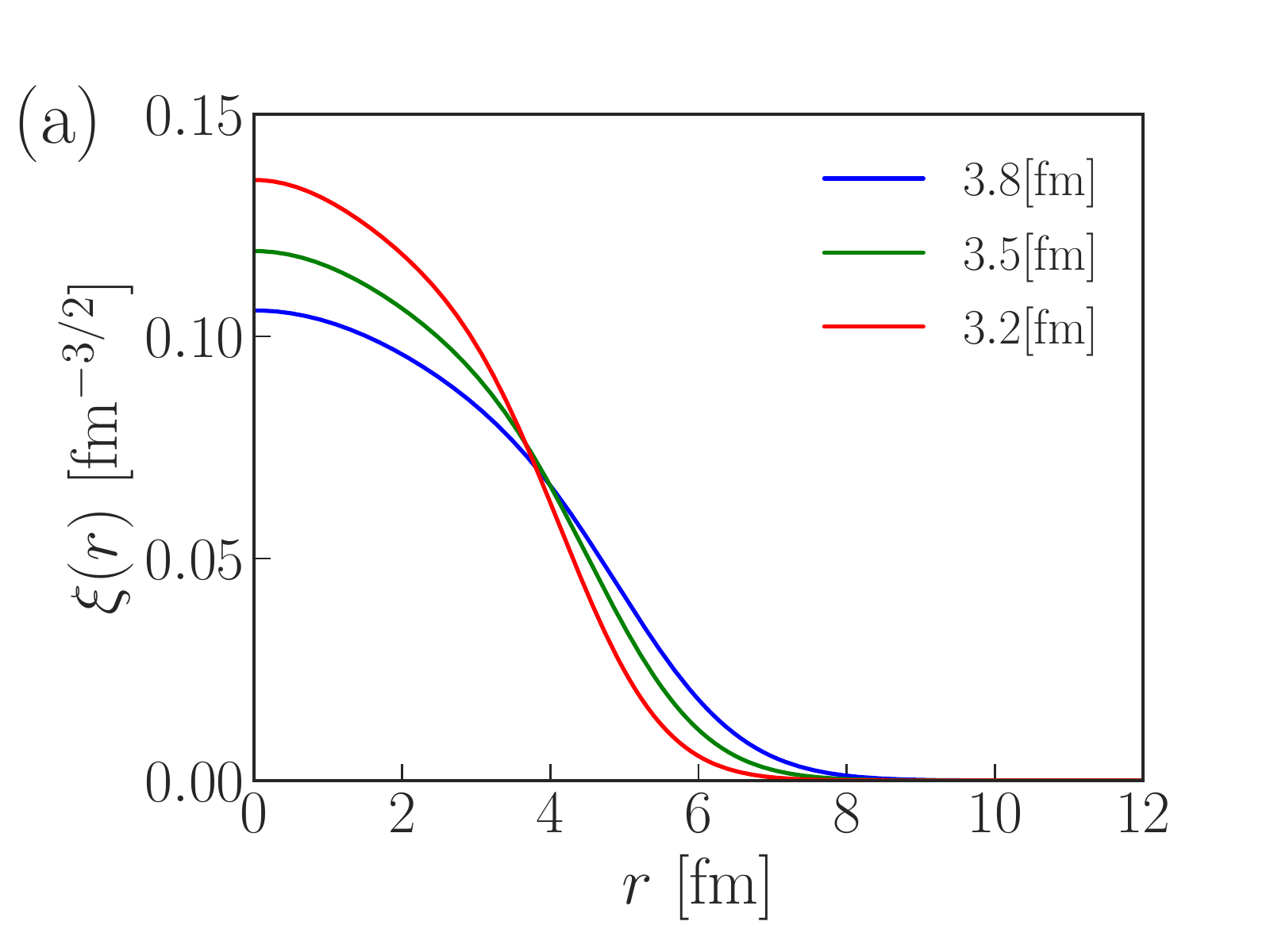}
\hspace{1cm} 
\end{center}
\end{minipage}
\begin{minipage}{0.325\hsize}
\begin{center}
\includegraphics[clip, width=5.7cm]{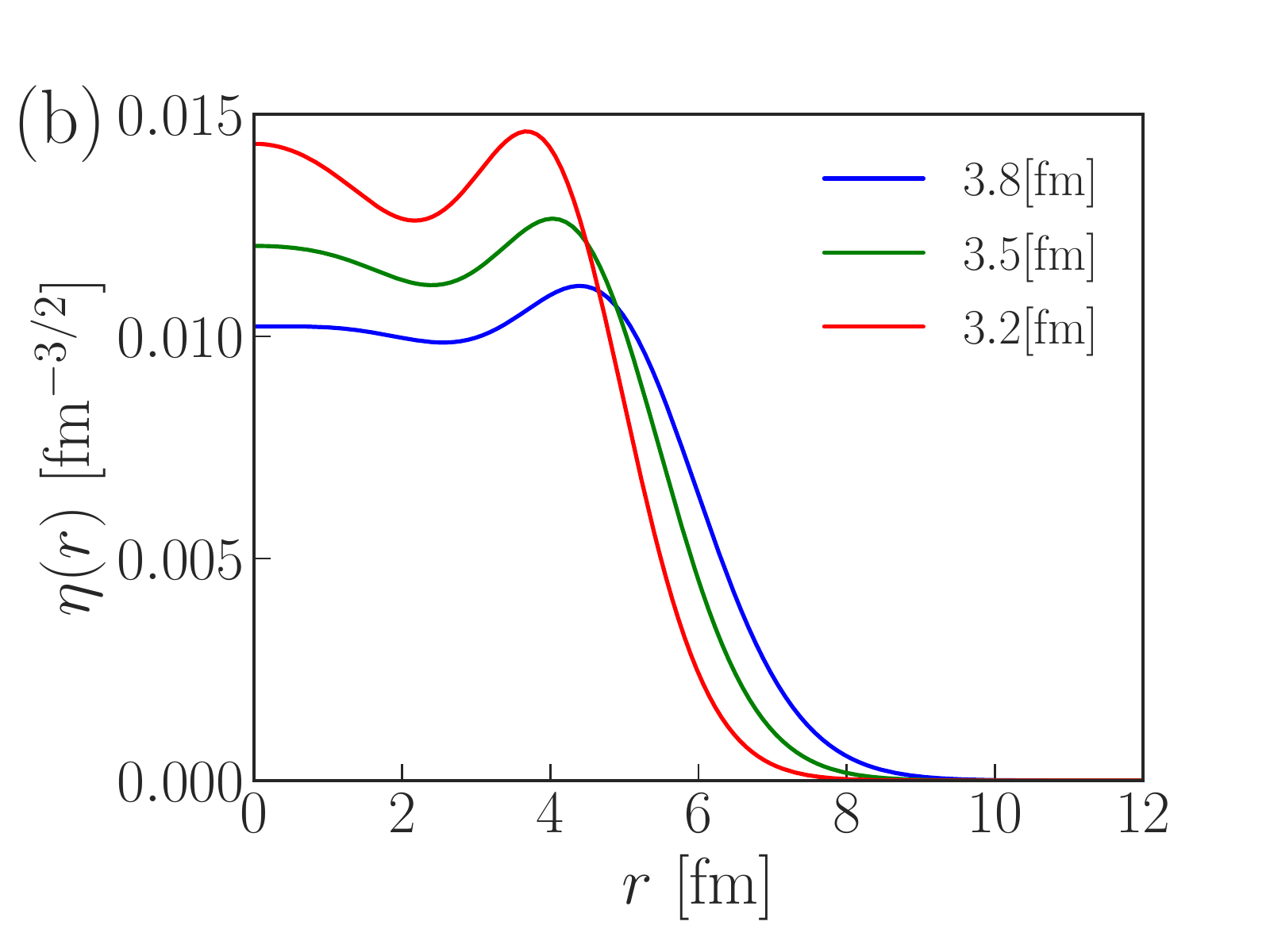}
\hspace{1cm} 
\end{center}
\end{minipage}
\begin{minipage}{0.325\hsize}
\begin{center}
\includegraphics[clip, width=5.7cm]{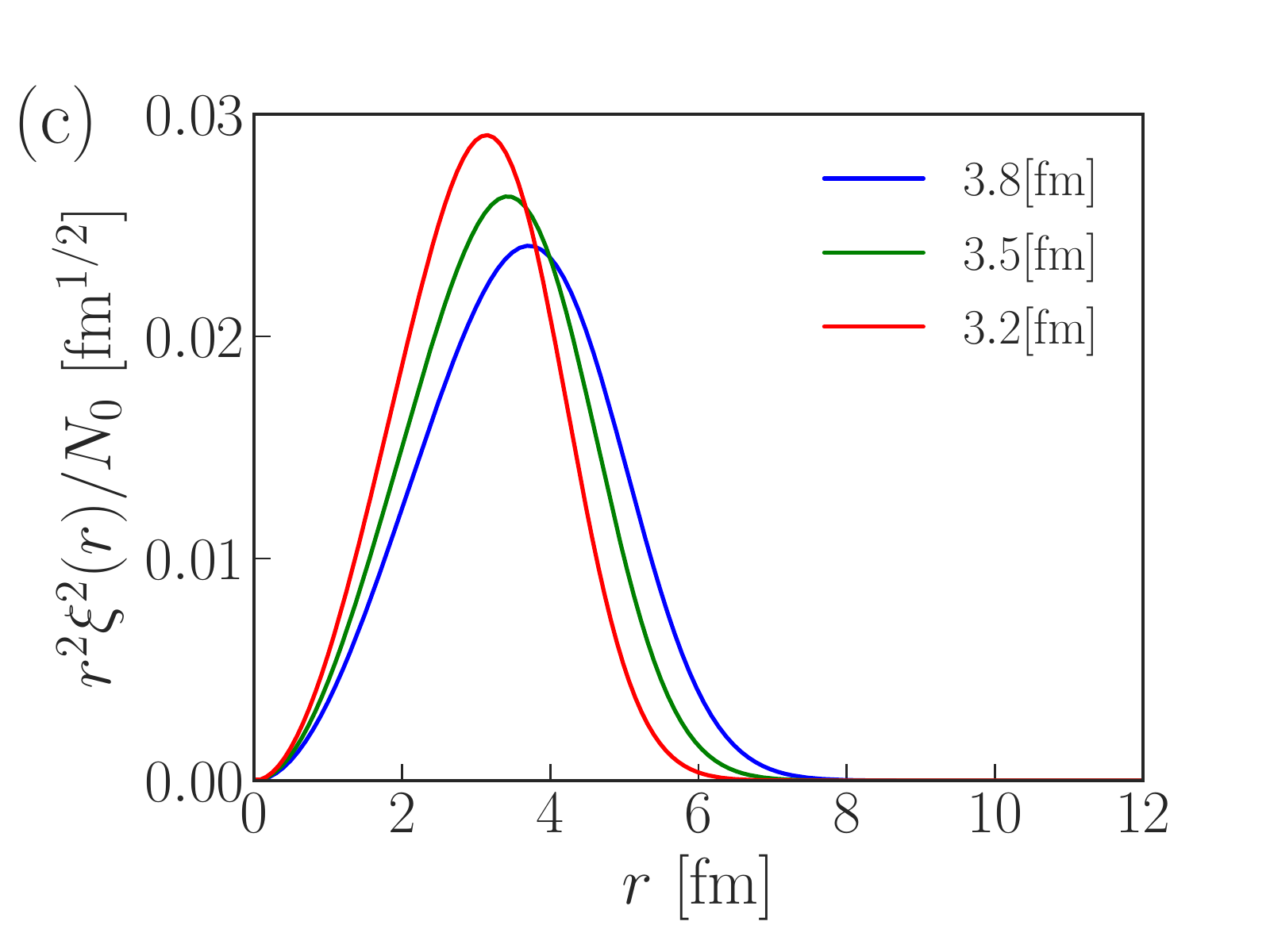}
\hspace{1cm} 
\end{center}
\end{minipage}
\end{tabular}
\end{center}
\caption{(Color online) Numerically calculated (a) $\xi(r)$, (b) $\eta(r)$ and (c) 
the radial density distribution of the condensate $r^2 | \xi(r)|^2/N_0$
with ${\bar r}=3.8\,,\,3.5\,,\,3.2, \mathrm{fm}$ for100\% condensation ($N_0=N$).}
\label{fig:xieta_32_38}
\end{figure*}

\begin{figure*}[tbh!]
\begin{center}
\begin{tabular}{c}
\begin{minipage}{0.325\hsize}
\begin{center}
\includegraphics[clip, width=5.7cm]{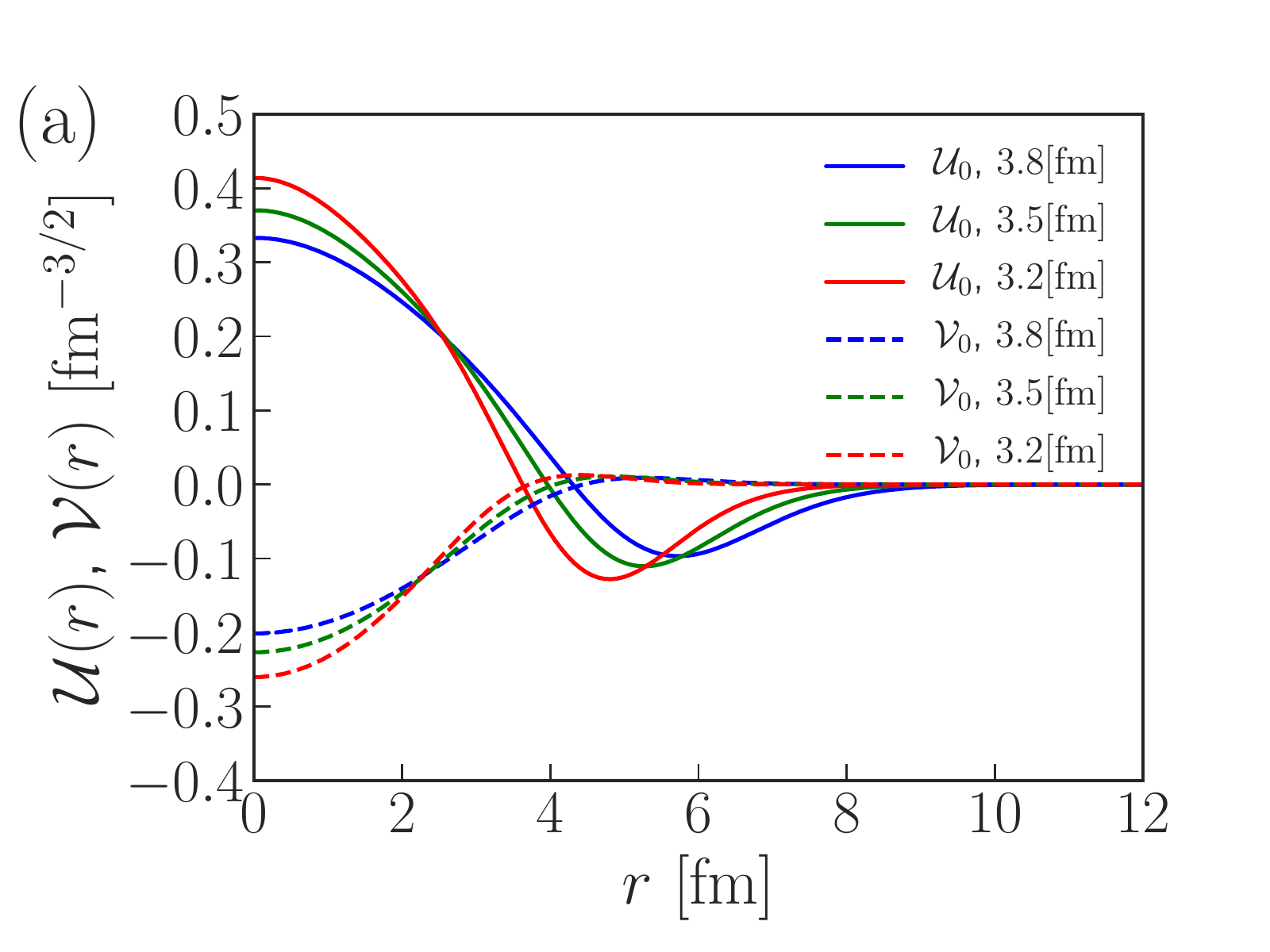}
\hspace{1cm} 
\end{center}
\end{minipage}
\begin{minipage}{0.325\hsize}
\begin{center}
\includegraphics[clip, width=5.7cm]{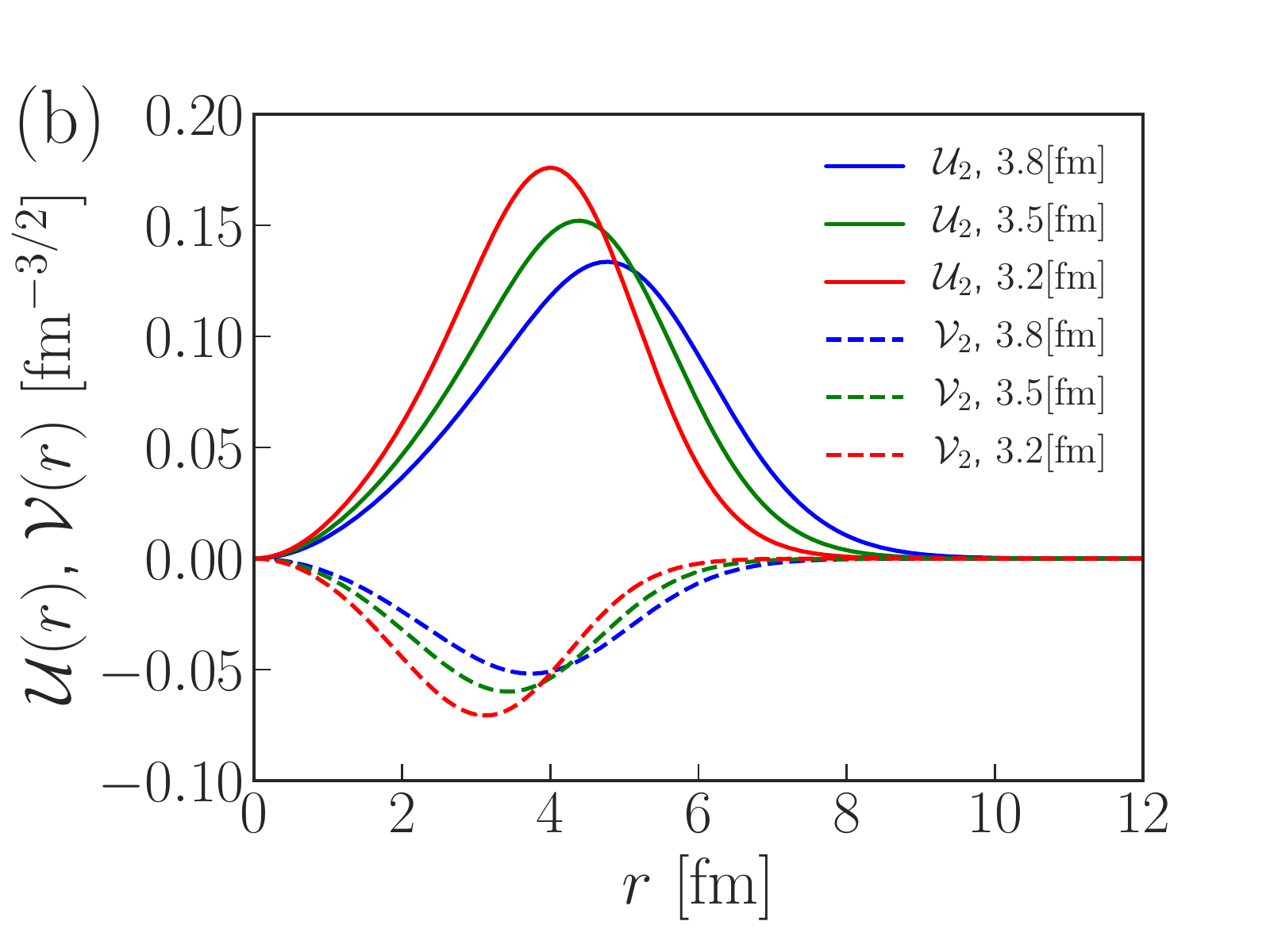}
\hspace{1cm} 
\end{center}
\end{minipage}
\begin{minipage}{0.325\hsize}
\begin{center}
\includegraphics[clip, width=5.7cm]{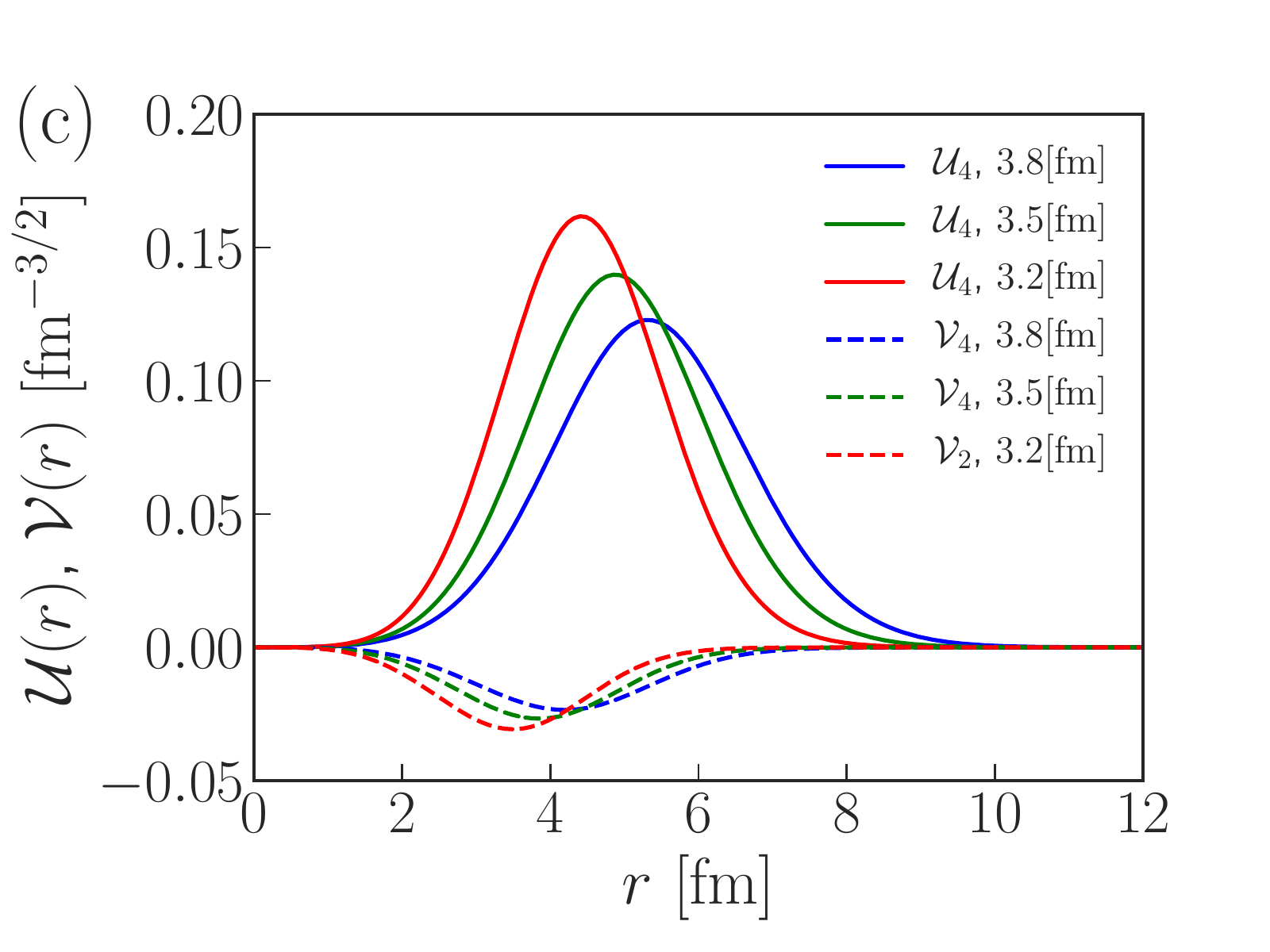}
\hspace{1cm} 
\end{center}
\end{minipage}
\end{tabular}
\end{center}
\caption{(Color online) Numerically calculated BdG wavefunctions, 
${\mathcal U}_{1\ell}(r)$ (solid lines) and ${\mathcal V}_{1\ell}(r)$ 
 (dashed lines) for (a) $\ell=0$, (b) $\ell=2$, and (c) $\ell=4$, 
with ${\bar r}=3.8\,,\,3.5\,,\,3.2, \mathrm{fm}$ for100\% condensation ($N_0=N$).}
\label{fig:BdG_32_38}
\end{figure*}

\begin{figure*}[tbh!]
\begin{center}
\begin{tabular}{c}
\begin{minipage}{0.325\hsize}
\begin{center}
\includegraphics[clip, width=5.7cm]{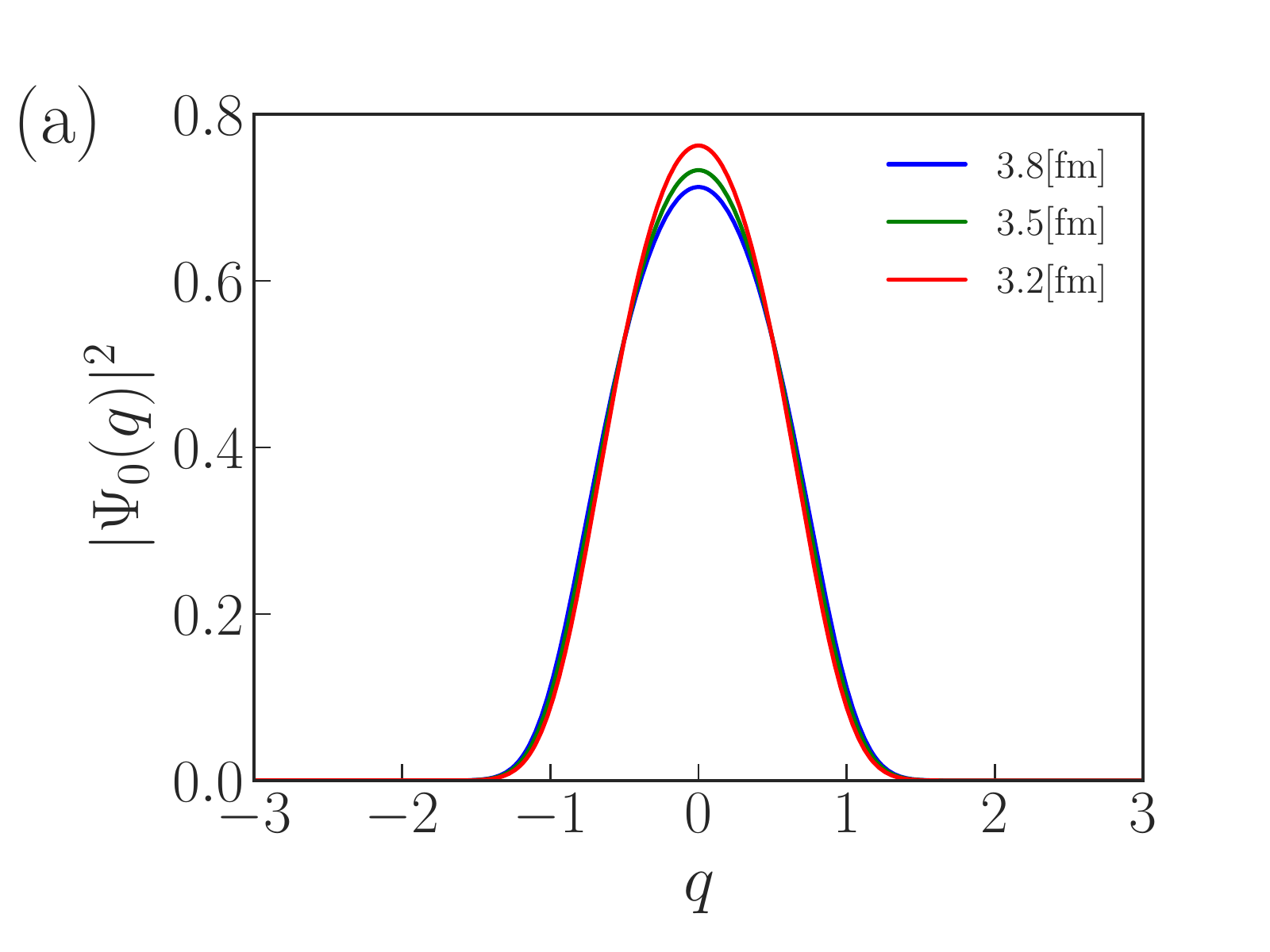}
\hspace{1cm} 
\end{center}
\end{minipage}
\begin{minipage}{0.325\hsize}
\begin{center}
\includegraphics[clip, width=5.7cm]{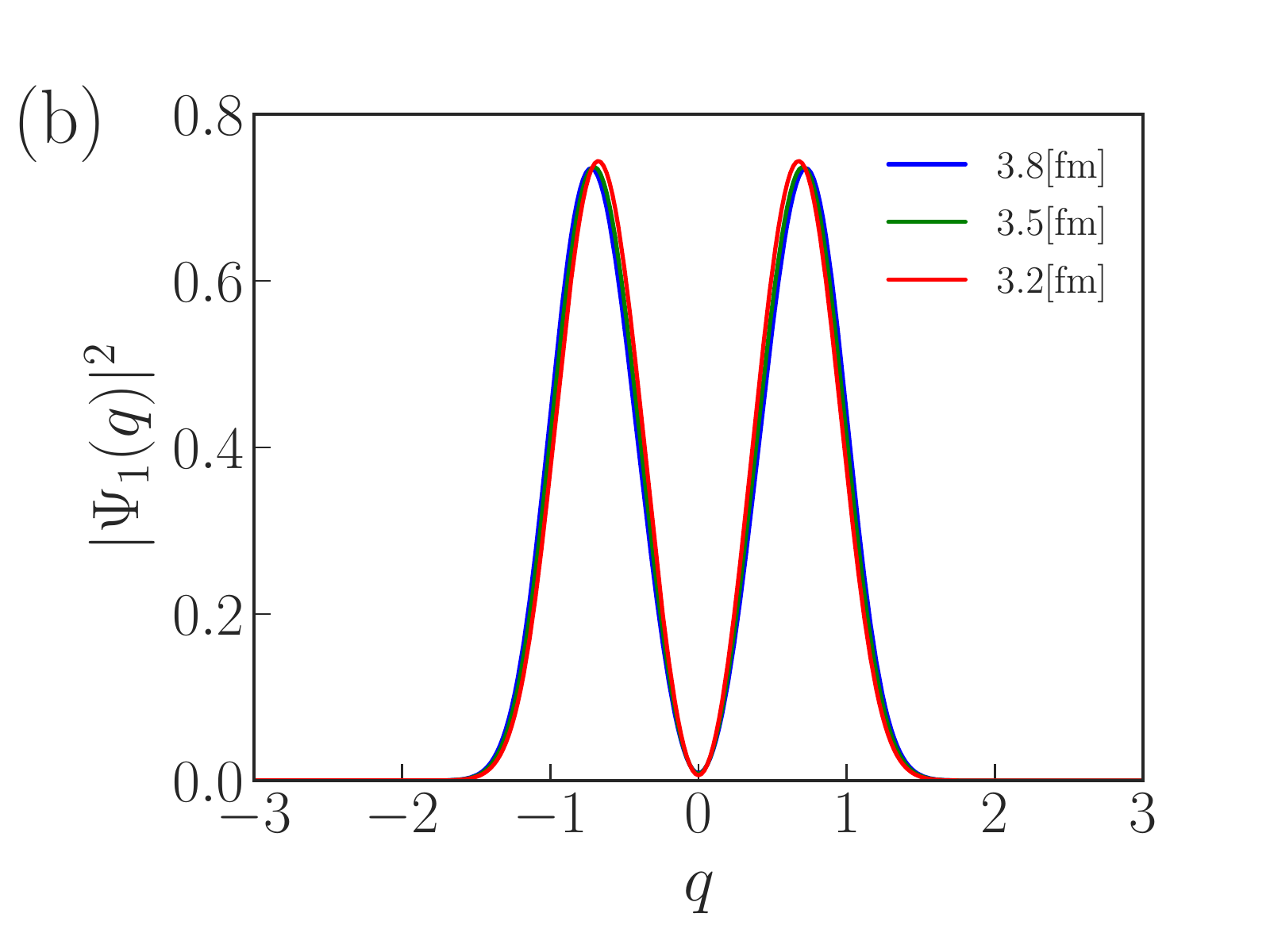}
\hspace{1cm} 
\end{center}
\end{minipage}
\begin{minipage}{0.325\hsize}
\begin{center}
\includegraphics[clip, width=5.7cm]{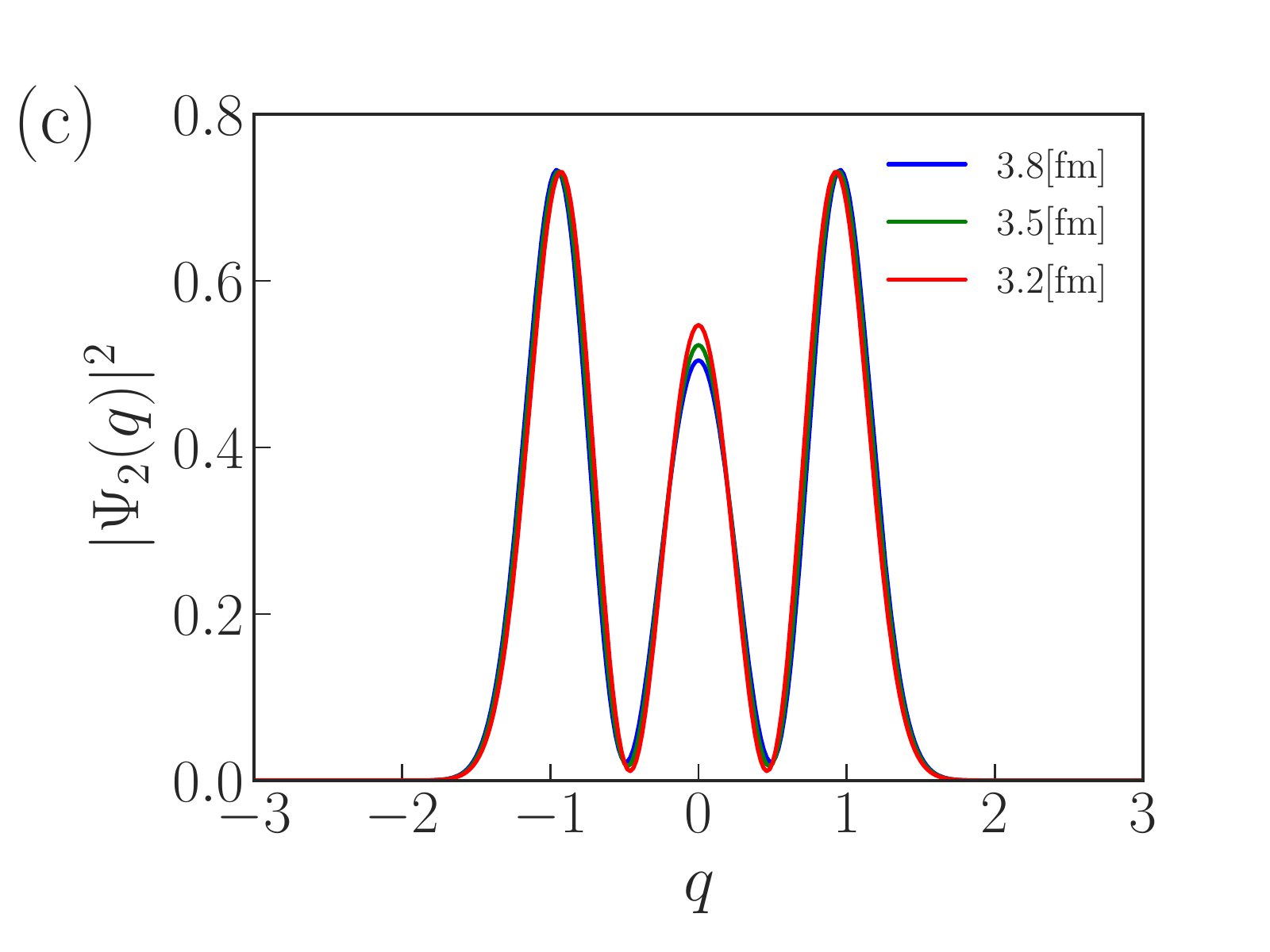}
\hspace{1cm} 
\end{center}
\end{minipage}
\end{tabular}
\end{center}
\caption{(Color online) The squares of numerically calculated wavefunctions
of the zero mode states, 
(a) $|\Psi_0(q)|^2$, (b) $|\Psi_1(q)|^2$, and (c) $|\Psi_2(q)|^2$, 
with ${\bar r}=3.8\,,\,3.5\,,\,3.2, \mathrm{fm}$ for100\% condensation ($N_0=N$).} 
\label{fig:ZeroMode_32_38}
\end{figure*}

\subsubsection{The order parameter, BdG eigenfunctions and wavefunctions of the zero mode states}

The results of the calculated wave functions with 100\% condensation are shown 
in sequence. The eigenfunction with zero eigenvalue in Eq.~(\ref{eq:BdGy0}) 
or the order parameter in Eq.~(\ref{eq:GP}),  $\xi(r)$,  
and its adjoint eigenfunction $\eta(r)$ in Eq.~(\ref{eq:BdGy-1-2})
are depicted in Fig.~\ref{fig:xieta_32_38} (a) and (b), respectively.
 The radial density distribution of the condensate 
is also shown in  Fig.~\ref{fig:xieta_32_38} (c). 
By comparing  Figs.~\ref{fig:70xieta_32_38} and \ref{fig:xieta_32_38},
 we see that the essential features of the behavior of $\xi(r)$ and $\eta(r)$ 
with ${\bar r}=3.8$, 3.5 and 3.2 fm  change  little, irrespective of
the condensation rates 70\% and 100\%.
In Fig.~\ref{fig:BdG_32_38} the wave functions of the first BdG 
excitation modes ($n=1$) with $\ell=0\,, 2$ and $4$ are displayed. We see 
that the  features of the BdG wave functions there for the three 
different sizes of the condensate 
with ${\bar r}=3.8$, 3.5 and 3.2 fm are similar to those 
in Fig.~\ref{fig:70BdG_32_38} with  condensation rate 70\%.
Figure~\ref{fig:ZeroMode_32_38} represents $|\Psi_{\nu}(q)|^2$
($\nu=0\,,\,1\,,\,2$), numerically calculated from solving 
Eq.~(\ref{eq:HuQPeigen}). 
 The probability densities
  $|\Psi_{\nu}(q)|^2$ ($\nu=0\,,\,1\,,\,2$) depend very little on ${\bar r}$, 
i.e.,   $\Omega$. Furthermore we see that Fig.~\ref{fig:ZeroMode_32_38} 
  resembles  closely Fig.~\ref{fig:70ZeroMode_32_38} for  all the three cases 
with ${\bar r}=3.8$, 3.5  and 3.2 fm. This means that 
the wavefunctions of the zero mode states depend little on the size 
of the condensate and condensation rate. This is parallel to the result
that the excitation energies of the low-lying zero mode states are almost independent 
of the condensate size and the condensation rates.  This is understood 
when we consider that the origin of the zero mode states 
is due to the SSB of the global phase of the condensate, which is 
independent of both the sizes and the condensation rates of the system.

\subsubsection{Electric transition probabilities}
\label{subsubsec-ET100}

Calculated $B({\rm E}2)$  and $M({\rm E}0)$
 values are displayed in the column of Table~\ref{table:70B}.

We note there that as the condensation 
rate increases from 70\% to 100\%, the E2 
transition of $2_2^+$$\rightarrow$ 
$0_2^+$ increases from 121 to 158, while the transition 
of $2_2^+$$\rightarrow$ $0_3^+$ decreases from 76 to 62.
Also the ratio B(E2: $2_2^+$$\rightarrow$ $0_2^+$)/
 B(E2: $2_2^+$$\rightarrow$ $0_3^+$) changes from 1.59 to 2.5. 
With increasing condensation rate the transition of 
$2_2^+$$\rightarrow$ $0_2^+$
 is enhanced more than the $2_2^+$$\rightarrow$ $0_3^+$ transition. 

As for the monopole transitions, the strength of 
the monopole transition probabilities increases
with increasing condensation rate, while
the condensation rate does not affect the ratio $M({\rm E}0:
0^+_2 \, \rightarrow \, 0^+_3)/M({\rm E}0:0^+_2 \, \rightarrow \, 0^+_3)$ much.

Thus, in our approach, the relative ratio of the E2 transitions and 
the strength of the E0 transitions sensitively reflects the 
condensation probabilities.  
The sensitivity of these quantities of the  transitions 
may serve to the experimental 
determination of condensation rate of the Hoyle state.
This sensitivity of the relative ratio of the E2 transitions
is related to the vibrational  nature of the BdG $2^+$ state.

\subsection{Energy levels of $^{12}$C}

\begin{figure}[tbh!]
\begin{center}
\includegraphics[width=8cm]{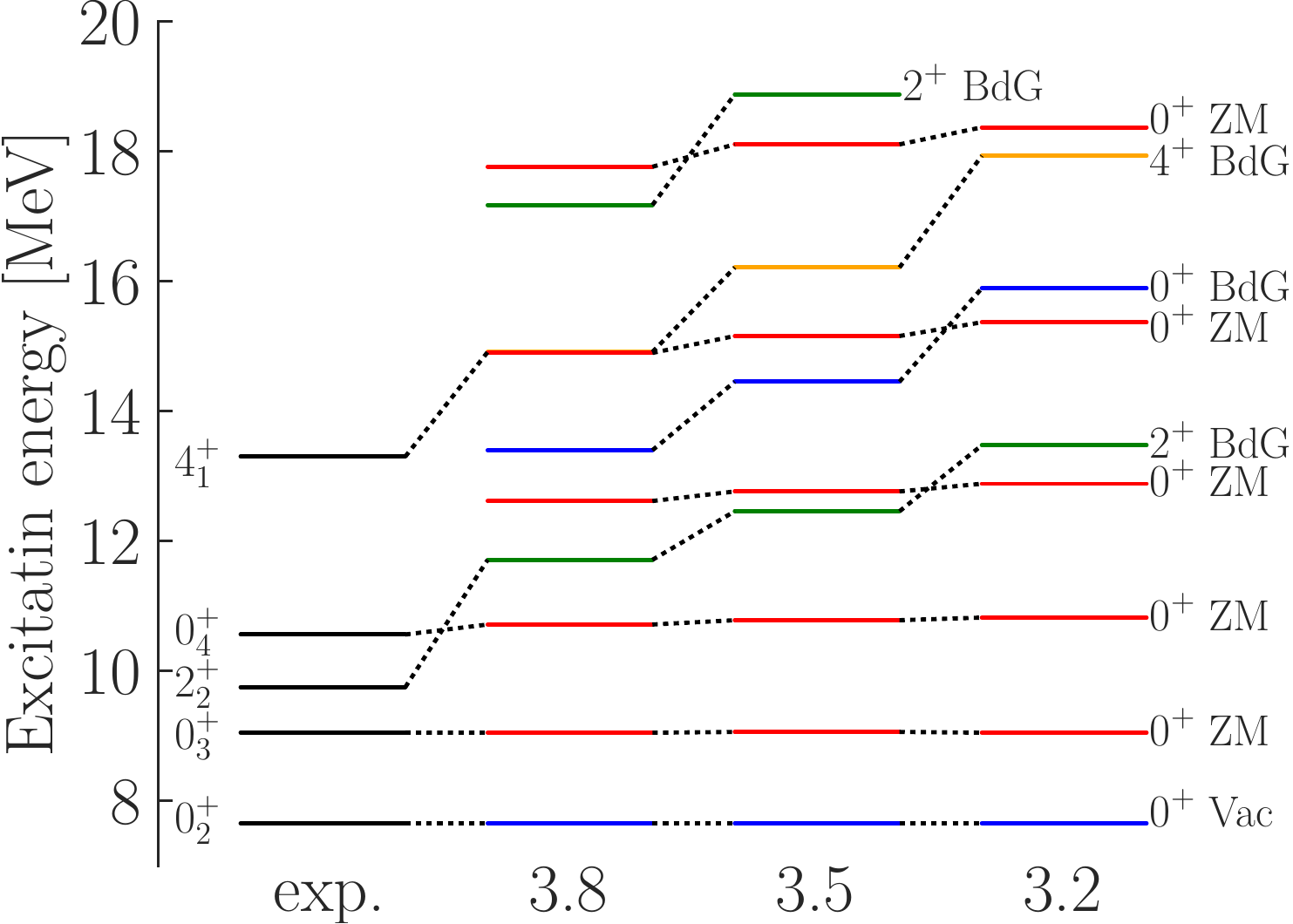} 
\caption{ (Color online) The energy levels of $^{12}$C, 
calculated from the parameters in Table~\ref{table:70parameter12C} with fixed 
$N_0=0.7N$ for ${\bar r}=3.8$, 3.5 and 3.2 fm are compared with 
the experimental energy levels from 
Refs.~\cite{Itoh2011,Freer2009,Zimmerman2011,Zimmerman2013,Itoh2013,Freer2011}.
Vac, ZM and BdG mean the vacuum Hoyle state, zero mode states and 
BdG excited states, respectively.}
\label{fig:ex_energy_C_N70p_32_38}
\end{center}
\end{figure}

In Fig.~\ref{fig:ex_energy_C_N70p_32_38} the energy 
levels calculated with the parameters in Table~\ref{table:70parameter12C} 
for each of ${\bar r}=3.8\,,\,$ 3.5 and 3.2 fm are shown in 
comparison with the experimental energy levels. 
The $0^+_4$ state is identified as the second NG excited state 
$\ket{\Psi_2}\ket{0}_{\rm ex}$,
and its predicted energy level agrees well with the observed value for 
the whole calculated range of the ${\bar r}$. The small excitation energy 
of $0_3^+$ state, less than 2 MeV, from the Hoyle, and the similar small 
energy between the $0_4^+$ and $0_3^+$ states are consistent with 
the picture that these two $0^+$ states are due to the collective 
states of the NG operators. The almost same small excitation energy of the two $0^+$ 
states despite of the change of the size of the vacuum Hoyle state also 
supports that they are the zero mode states due to the SSB of the global 
phase of the vacuum. On the other hand, the calculated energy levels 
$2^+$ and $4^+$ of the lowest excitations of the BdG modes with $\ell=2$ 
and $\ell=4$ correspond to the observed $2^+_2$ and $4^+_1$ states, respectively. 
The agreement between the calculations and experiment is better for 
${\bar r}=3.8$ fm and is deteriorated for ${\bar r}=3.5$ and 3.2 fm 
because the excitation energy increases with a larger $\Omega$ value in 
Table~\ref{table:70parameter12C} in accordance with the vibrational 
nature of these states. Thus the observed $2^+_2$ and $4^+_1$ states are 
both interpreted as vibrational states on the BEC Hoyle state.
The BdG energy levels in Fig.~\ref{fig:ex_energy_C_N70p_32_38} show 
that the levels increase
for smaller ${\bar r}$. The increase in $\Omega$ affects more dominantly on
the levels than the decrease in $V_r$\,.

\subsection{Robustness of the zero mode states for various condensation rates}

We found that the energy levels in the realistic 70\% condensation case 
barely change from those in the 100\% condensation case. 
 This means that  in much  small condensation rates such 
as 50\%, 30\% and 20\% , the similar zero mode states could  appear.  
Although  the condensation rate in $^{12}$C is reported to be as large 
as more than 60\% in the theoretical cluster model calculations, it is 
intriguing to calculate the energy level structure under such hypothetical 
condensation rates, because in heavier mass region such as $^{52}$Fe, 
in which we are interested from the viewpoint of universality of BEC 
of $\alpha$ cluster,  the condensation rate could be much smaller than $^{12}$C. 
\begin{figure}[tbh!]
\begin{center}
\includegraphics[width=8cm]{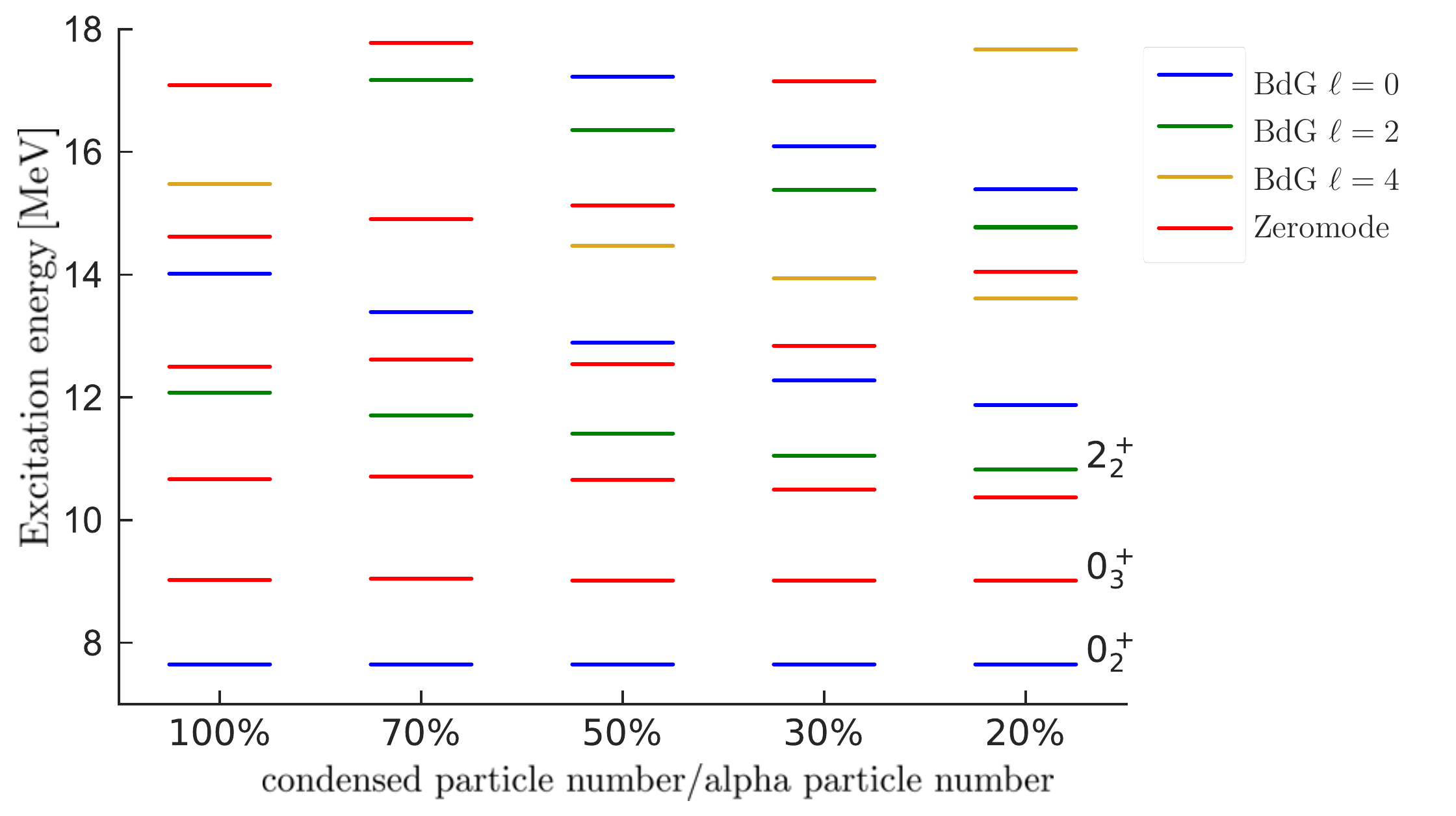}
\caption{(Color online) The energy levels of $^{12}$C calculated 
with different condensation rates with ${\bar r}=3.8$ fm.
}
\label{fig:energy_12C_p}
\end{center}
\end{figure}

 In Fig.~\ref{fig:energy_12C_p} the energy levels calculated under 
condensation rates of 20\%, 30\% and 50\% in comparison
 with those of  70\% and 100\% are displayed. We see that  
the structure of the energy levels
of the first two low-lying zero mode states  does not depend
 on the condensation rate,
 even for the smaller condensation rates.
This is naturally understood because the zero mode states emerge once 
the global symmetry is spontaneously broken. 

\begin{figure}[tbh!]
\begin{center}
\includegraphics[width=8cm]{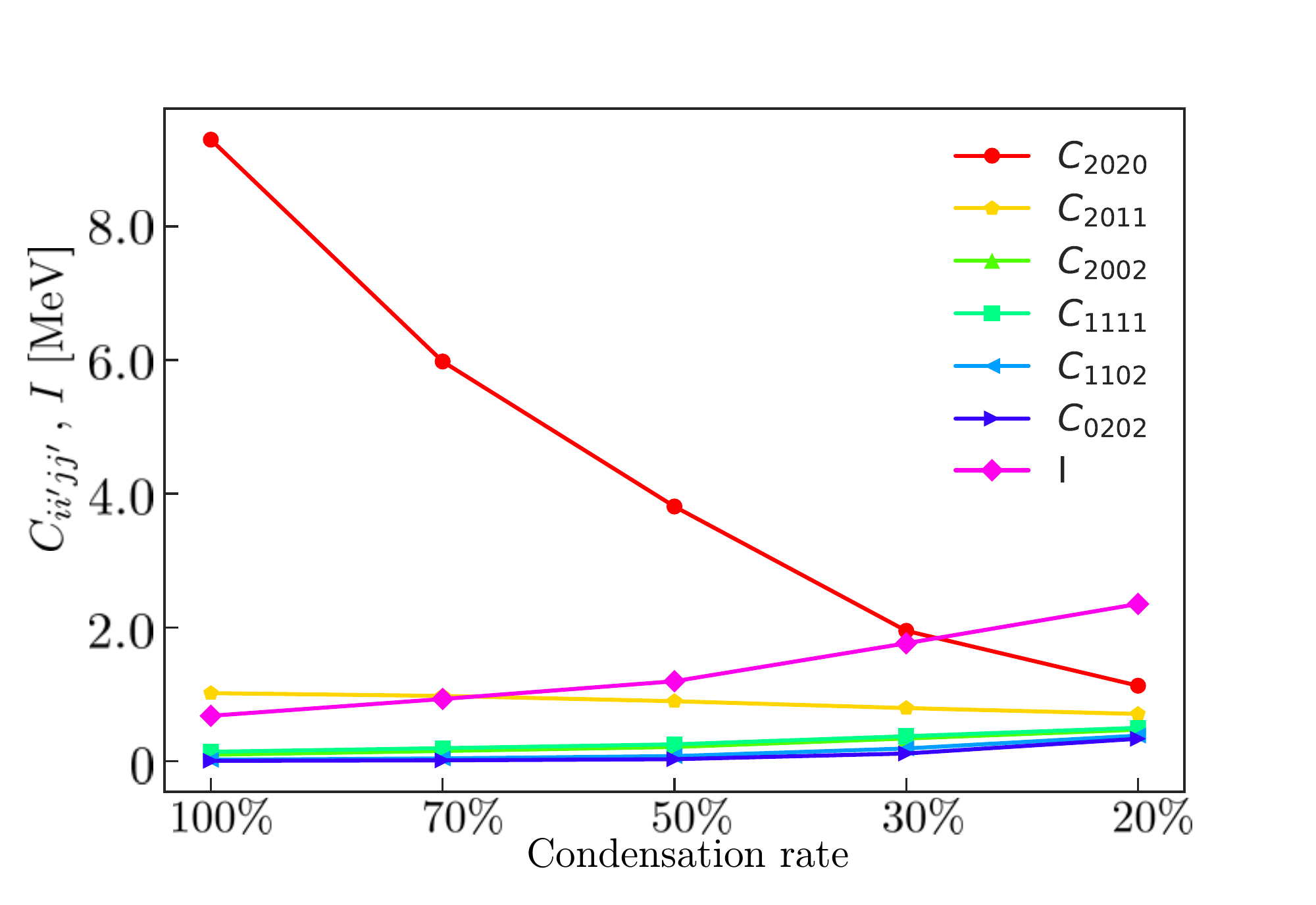}
\caption{(Color online) How the coefficient $C_{iji^{\prime}j^{\prime}}$ 
of the zero mode Hamiltonian in Eq.~(\ref{eq:HuQP}) change  
in the calculations of  Fig.~\ref{fig:energy_12C_p}  with  different 
condensation rates of the three $\alpha$ clusters in $^{12}$C is  displayed.
}
\label{fig:Zeromode_coefficient}
\end{center}
\end{figure}

As for the BdG modes the excitation energy decreases as the condensation 
rate decreases. However, the electric transitions to the BdG states 
are rather sensitive to the condensation rate  
of the three $\alpha$ clusters in $^{12}$C.
The transition probability changes,
 depending on the coefficient $C_{2020}$ of the $Q^4$ term in Eq.(\ref{eq:HuQP}). 
As seen  in  Fig.\ref{fig:Zeromode_coefficient}, this term  is 
overwhelmingly large than other terms. 

\par
It is worth noting  that Bose-Einstein condensation  of $\alpha$ clusters occurs even when the condensation rate is not  100\%.  In such a case   $\alpha$ clusters that are  not involved in condensation are sitting at the excited higher energy levels other than the lowest $0s$ state of the trapping potential.
This   partial involvement of $\alpha$ clusters in causing  the Bose-Einstein condensation of $\alpha$ clusters reminds us the superfluidity of liquid HeII, in which the observation of the momentum distribution by neutron inelastic  scattering  \cite{Sears1982} showed that only  about 13\%  of the atoms are in the lowest state at 1K and others are staying at the excited states. 
 Also in the superfluid nuclei in heavy mass region where the  paring interaction is strong,  Cooper pairs created only near the Fermi surface are responsible for causing superfluidity of nuclei.  Many other nucleons are not needed to be paired as a Cooper pair.  In fact, even several  valence  nucleons in the open shells outside the inert core in heavy nuclei cause superfluidity \cite{Ring1980,Broglia1973,Brink2005}.  
The present theoretical finding that Bose-Einstein condensation of $\alpha$ clusters is  caused  by partial condensation of $\alpha$ cluster as small as  20-30\% in nuclei seems to be similar to Bose-Einstein condensation in other systems in nature. This is   interesting and important,
suggesting that 
 $\alpha$ cluster condensation may be realized in a wide range of mass region in nuclei  and also could be  in nuclear matter  and $\alpha$ cluster matter \cite{Clark1966,Tohsaki1998,Carstoiu2009} at low densities.

\section{BOSE--EINSTEIN CONDENSATION OF MANY ALPHA CLUSTERS IN $^{16}$O--$^{52}$F\lowercase{e}}
\label{sec-Nalpha}

There is no reason that the BEC of $\alpha$ clusters is limited to  
the Hoyle and the related states of three $\alpha$ clusters, $N=3$. 
It  may occur for multi-$\alpha$ cluster systems with $N>3$. 
Experimental and theoretical efforts have been devoted to search 
for such condensate states in heavier nuclei. 
Experimentally, however, 
it is not easy to detect multi-$\alpha$ clusters emitted with very 
low kinetic energies near the threshold in coincidence. Also it is 
not easy   technically for the microscopic cluster models such as 
the RGM and GCM to solve the multi-$\alpha$ cluster problems up to 
$N=13$ systematically. On the other hand, the present field theoretical 
superfluid   $\alpha$ cluster model, in which the order parameter 
is embedded, has no difficulty in application to nuclei 
with a large number of $\alpha$ clusters like  $N=13$.

For $^{16}$O with $N=4$, experimentally much attention
has been paid to  four $\alpha$ cluster states 
with a linear chain structure   for more 
than a decade since the observation by Chavallier {\it et al.} \cite{Chevallier1967}. 
In Ref.~\cite{Ohkubo2010} it was shown in the unified description of 
nuclear rainbows in $\alpha$+$^{12}$C scattering and $\alpha$ cluster 
structures in the bound and quasi-bound low energy region of $^{16}$O  
that the four $\alpha$ cluster states that have been considered to be a 
linear chain structure can be interpreted to have the $\alpha$+$^{12}$C 
($0_2^+$) structure  in  $\alpha$ cluster condensation. The observed $0^+$ 
state at 15.1 MeV just above the four $\alpha$ threshold was suggested to be 
a Hoyle analogue four $\alpha$ condensate, similarly
 to the four $\alpha$ cluster model OCM calculations in Ref.~\cite{Funaki2008C}.
As for the four $\alpha$ linear chain structure in $^{16}$O,
 recent calculations suggest that their excitation energy is much higher 
than ever considered and as high as above 30 MeV \cite{Ichikawa2011,Horiuchi2017}. 
Itoh {\it et al.} \cite{Itoh2014} reported 386 MeV inelastic $\alpha$ 
scattering to search for 
a four $\alpha$ condensate in $^{16}$O and  observed two $0^+$ states at 
16.7 MeV with the $\alpha$+$^{12}$C($0_2^+$) structure and 18.8 MeV with the
 $^8$Be+$^8$Be structure just above the four $\alpha$ threshold.

Beyond $^{16}$O Freer {\it et al.} \cite{Freer2005} searched for an $\alpha$ 
condensate state in $^{20}$Ne by detecting five $\alpha$ clusters in 
the $^{12}$C($^{12}$C,$^8$Be+$^{12}$C$(0_2^+))^4$He reactions and 
observed two $^{20}$Ne resonances at 35.2 and 36.5 MeV.  
A theoretical study  of four and five 
$\alpha$ condensates in $^{16}$O and $^{20}$Ne, using a 
quasi-Schuck wave function, was made by Itagaki 
{\it et al.} \cite{Itagaki2008}.
 Kawabata {\it et al.} \cite{Kawabata2013} searched for  six $\alpha$ 
condensates in $^{24}$Mg in 400 MeV  inelastic $\alpha$ scattering. 
Two $\alpha$ cluster condensation in $^{24}$Mg   and  a 
three $\alpha$ cluster condensation in $^{28}$Si were theoretically 
discussed using a quasi-Schuck wave function in 
Refs.~\cite{Itagaki2007,Ichikawa2012}. 
 Akimune {\it et al.} \cite{Akimune2014} reported nine $\alpha$ 
clusters search in $^{36}$Ar.
Also von Oertzen {\it et al.} 
\cite{Kokalova2003,Kokalova2006,Itagaki2010,Oertzen2011,Itagaki2011} 
discussed the observed experimental signature of $\alpha$ condensation 
in $^{52}$Fe and studied the three $\alpha$ cluster structure around 
the $^{40}$Ca core. Akimune {\it et al.} \cite{Akimune2013} 
made inelastic $\alpha$ scattering from $^{56}$Ni in inverse kinematic 
to observe $\alpha$ condensate state in $^{56}$Ni and strongly suggested 
the existence of an $\alpha$ gas state at high excitation energies in $^{56}$Ni.
Von Oertzen \cite{Oertzen2006} discussed the  conditions for a phase change 
with the formation of an $\alpha$ cluster condensate in excited compound nuclei
  up to $^{164}$Pb based on the systematics for binding energies per $\alpha$ 
cluster in $N=Z$ nuclei.  These investigations are encouraging and suggestive, 
however,   no clear  experimental  evidence of BEC of $\alpha$ clusters 
such as $\alpha$-cluster superfluidity or $\alpha$ cluster Josephson effect 
has been reported yet.
It is therefore important and intriguing to explore, based on a firm 
theoretical frame of superfluid cluster model with order parameter, 
the BEC of $\alpha$ clusters systematically in heavier mass region beyond  $N=3$.

\subsection{100\% condensation case}

We  calculate the energy levels of $N\alpha$ clusters of $^{16}$O 
in the 0p-shell, $^{20}$Ne--$^{40}$Ca ($N=4$--$10$) in the sd-shell 
of the light mass region, and $^{44}$Ti--$^{52}$Fe ($N=11$--$13$) 
in the  fp-shell of the medium-heavy mass region. 

\begin{figure}[tbh!]
\begin{center}
\includegraphics[width=8cm]{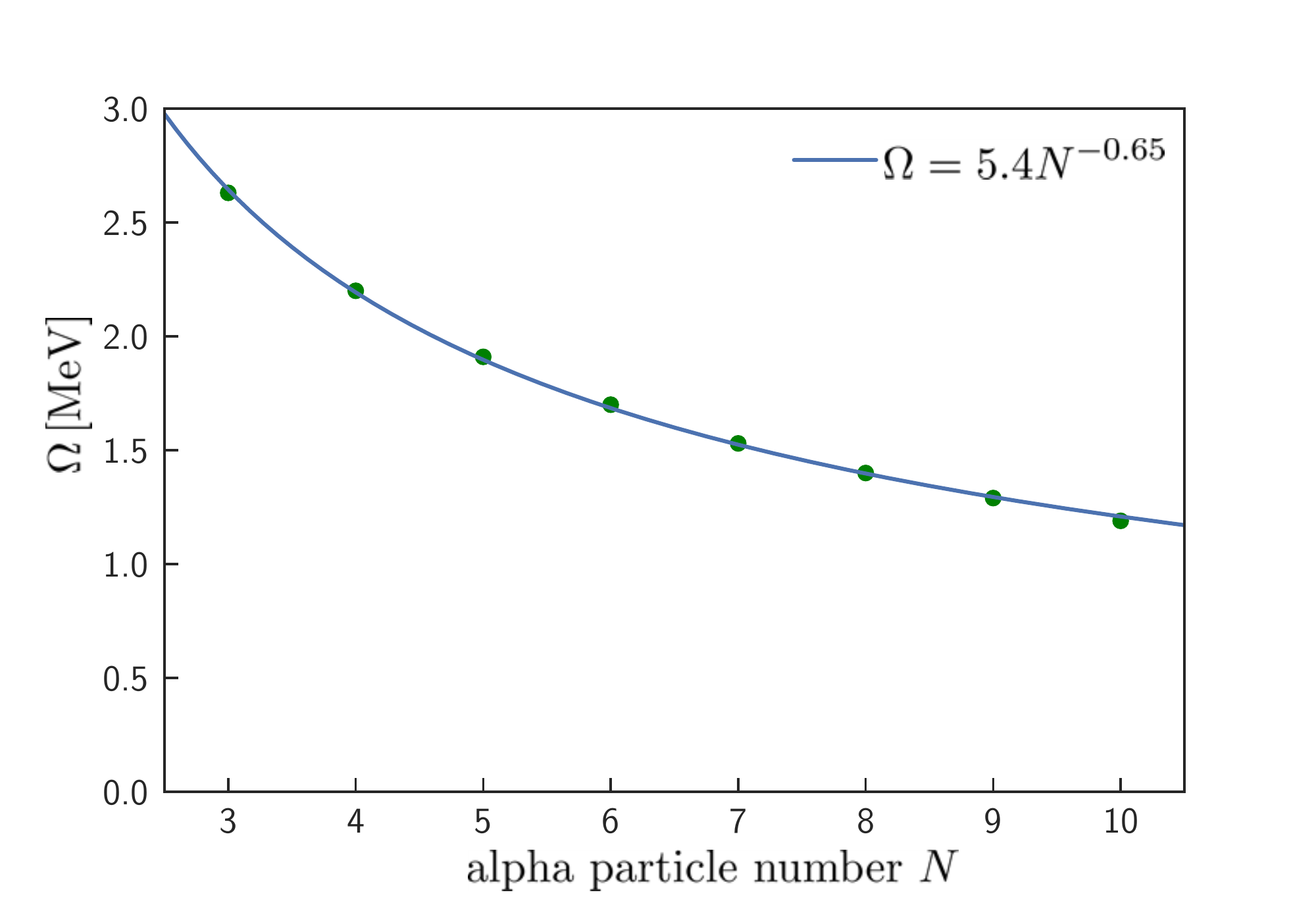}
\caption{The $N$-dependence of $\Omega$ of the confining potential, 
which is used in the energy level calculations of $N=3$--13 
 ($^{12}$C--$^{52}$Fe) in Fig.~\ref{fig:Nalpha_fit:Omega},  
is displayed by the crosses. This is obtained under a constraint 
${\bar r} \, \propto\, N^{1/3}$ and with fixed $V_r=400$ MeV for 
100\% condensation. The solid curve  $5.4N^{-0.65}$ is to guide the eye.
}
\label{fig:Omega}
\end{center}
\end{figure}

It is an unfounded claim that the parameters $\Omega$ and $V_r$ are 
independent of $N$ in our phenomenological model, but we have no general
argument to determine their $N$-dependences. Here we presume that ${\bar r}$
behaves as $N^{1/3}$, similarly as ${\bar r}$ of the ordinary nuclei
is proportional to $A^{1/3}$ with $A$ being the mass number. 
Fixing $V_r= 400$ MeV, we adjust $\Omega$ for each $N$
in such a manner that ${\bar r}$ behaves as $N^{1/3}$. Figure~\ref{fig:Omega}
shows the $N$-dependence of $\Omega$ thus obtained, which is represented 
by the curve $\Omega = 5.4\, N^{-0.65}$\,. 

\par
\begin{figure}[tbh!]
\begin{center}
\includegraphics[width=8cm]{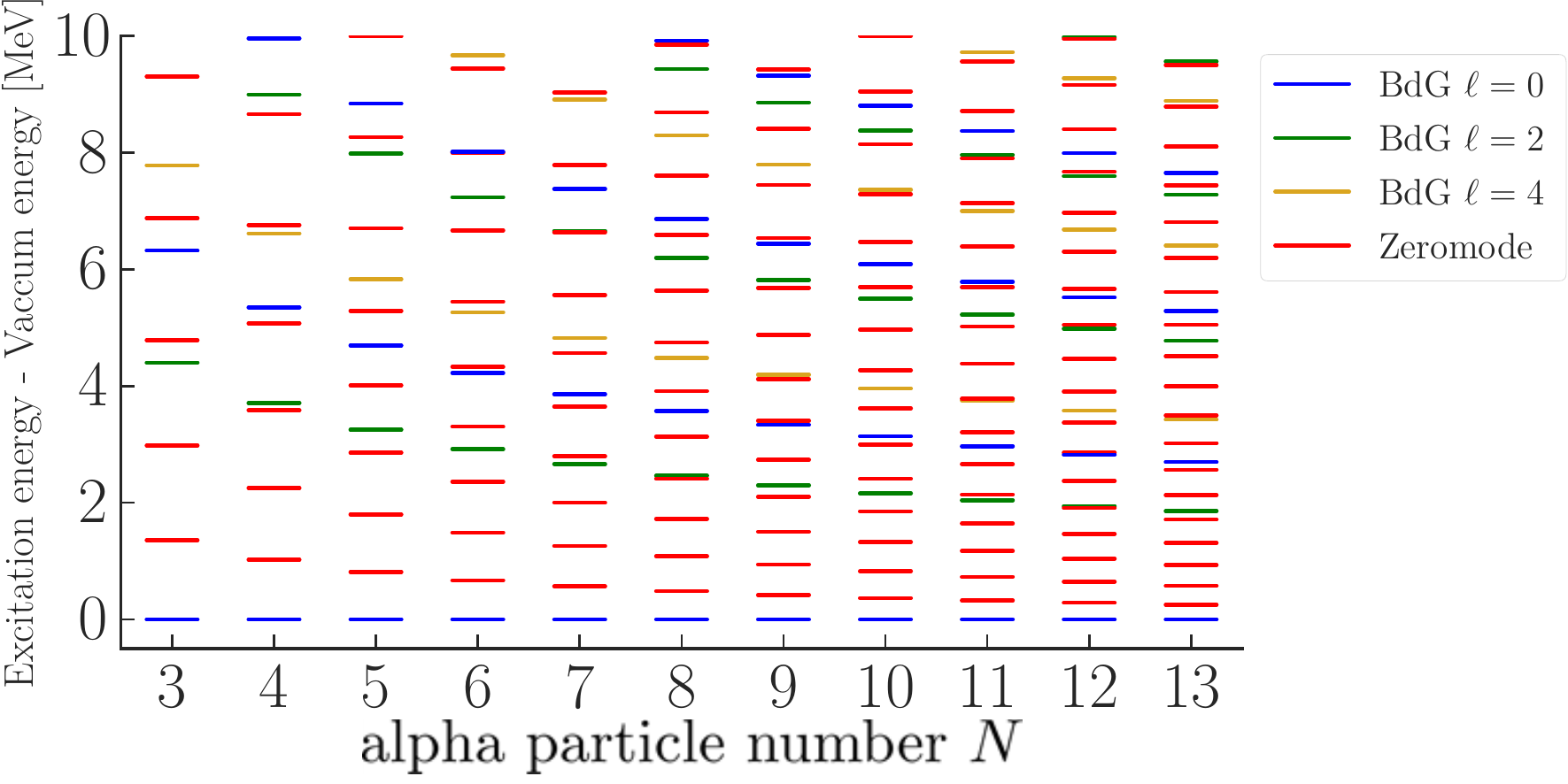}
\caption{(Color online) The energy levels calculated for $N=3$--13
 ($^{12}$C--$^{52}$Fe) 
with 100\% condensation using 
 $N$-dependent $\Omega$ given in Fig.~\ref{fig:Omega}.
Excitation energy is measured from the Hoyle-analog vacuum, i.e.,
the $N$-alpha condensate state near the $N$-alpha threshold.
}
\label{fig:Nalpha_fit:Omega}
\end{center}
\end{figure}

In Fig.~\ref{fig:Nalpha_fit:Omega} the energy levels, given by calculations 
using this $N$-dependent $\Omega$,
are displayed.
Figure~\ref{fig:Nalpha_fit:Omega} predicts the first zero mode state 
for each $N$ below 2 MeV near the threshold energy, whose energy
decreases as N becomes larger and which is expected to be  observed in experiment. 
We note that for $N>7$ the spectrum of the low-lying zero mode states
 becomes very similar. We therefore expect
that similar zero mode states would appear  in heavier  nucleus beyond  $^{52}$Fe.

\subsection{70\% condensation case}

We have assumed  100\% $\alpha$ cluster condensation $N_0=N$ above, 
but in reality  the condensation rate may not be necessarily 100\% and 
 may change from nucleus to nucleus. 
Similarly to $^{12}$C in the previous sections, we here consider
a case where $N\alpha$ clusters are condensed only partly, i.e.,
a typical case of $N_0=0.7 N$. 

\begin{figure}[tbh!]
\begin{center}
\includegraphics[width=8cm]{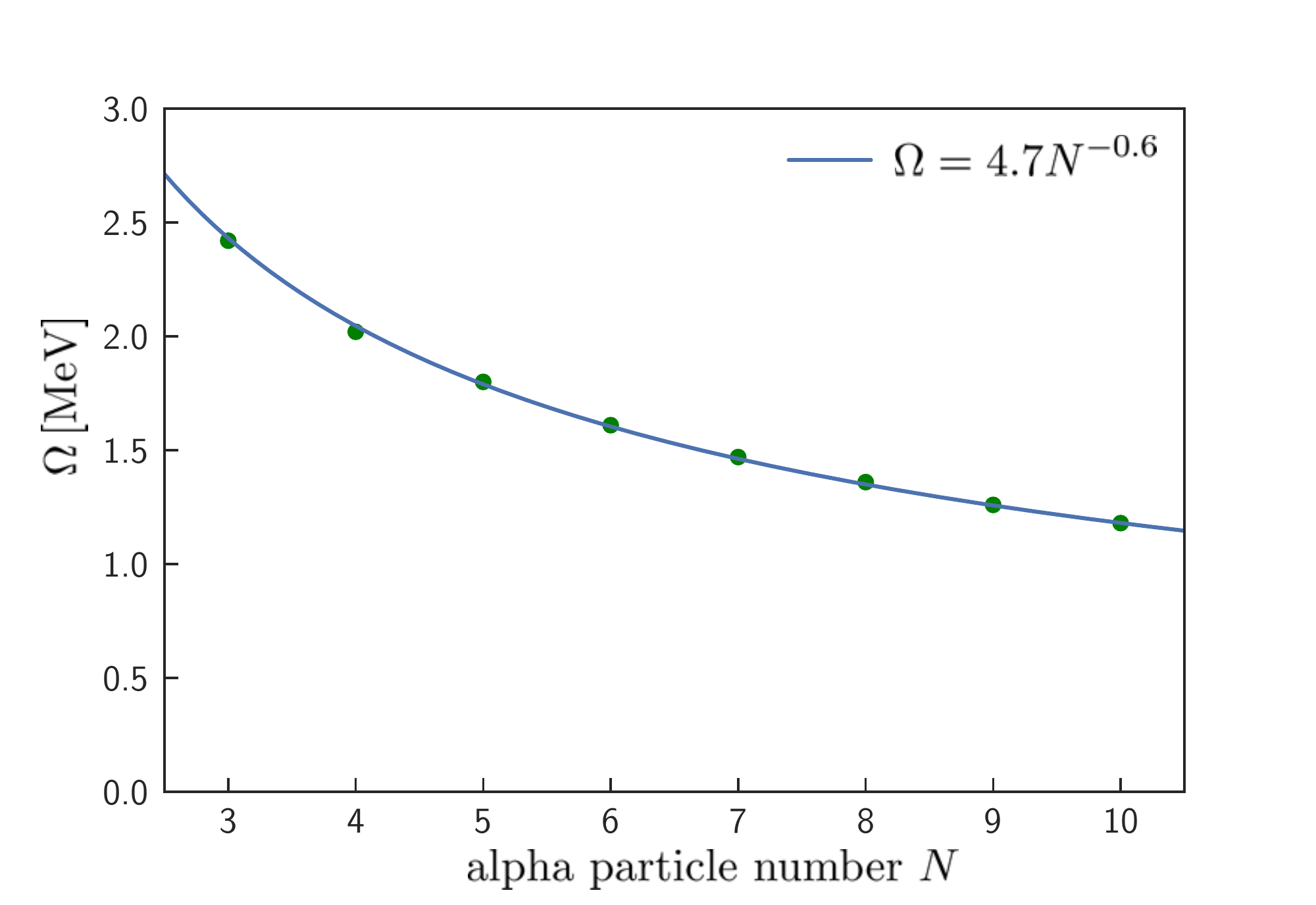}
\caption{
The $N$-dependence of $\Omega$ of the confining potential in case of $N_0=0.7 N$, 
which is used in the energy level calculations of $N=3$--13  ($^{12}$C--$^{52}$Fe)
 in Fig.~\ref{energy_nalpha_Omega_70p}, 
  is displayed by the crosses. This is obtained under a constraint 
${\bar r} \, \propto\, N^{1/3}$ and with fixed $V_r=400$ MeV. 
The solid curve  $4.70N^{-0.60}$ is to guide the eye. }
\label{fig:Omegafit_70p}
\end{center}
\end{figure}

\begin{figure}[tbh!]
\begin{center}
\includegraphics[width=8cm]{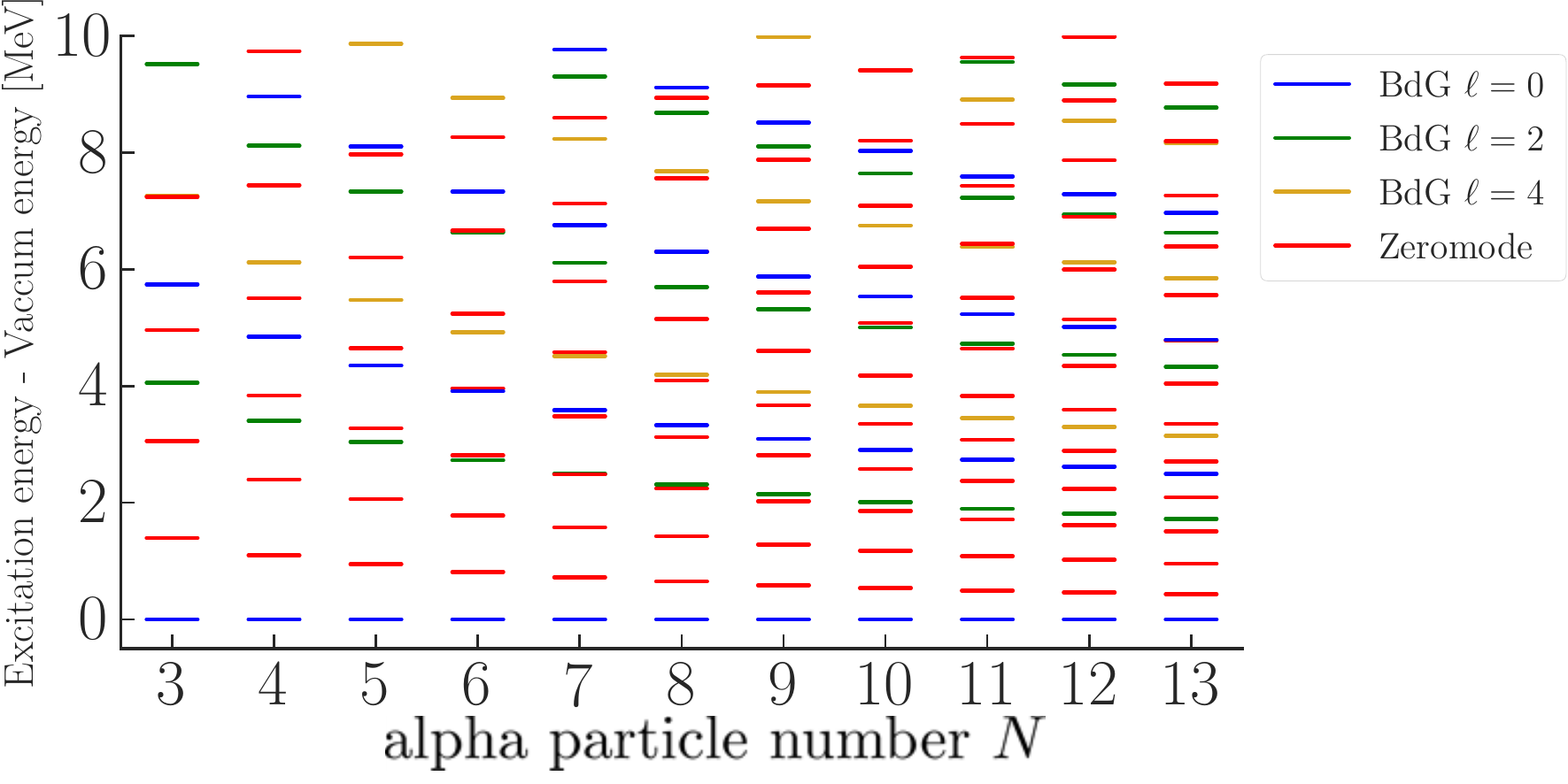}
\caption{(Color online) The energy levels calculated for $N=$3--13 ($^{12}$C--$^{52}$Fe) 
in case of 70\% condensation ($N_0=0.7 N$) using the interaction 
with $N$-dependent $\Omega$ in Fig.~\ref{fig:Omegafit_70p}.
Excitation energy is  measured from the Hoyle-analog vacuum, i.e.,
the $N$-alpha condensate state near the $N$-alpha threshold.
}
\label{energy_nalpha_Omega_70p}
\end{center}
\end{figure}

We attempt to determine the parameter $\Omega$ for each $N$ in such a manner
that ${\bar r}$ is proportional to $N^{1/3}$, setting $V_r=400$ MeV and $N_0=0.7 N$. 
The $N$-dependence of $\Omega$ is shown in Fig.~\ref{fig:Omegafit_70p}, 
and is traced by the curve  $\Omega=4.70\, N^{-0.60}$. 
In Fig.~\ref{energy_nalpha_Omega_70p} the energy levels calculated 
in the case of 70\% condensation with the $N$-dependent $\Omega$ is displayed.
From Figs.~\ref{fig:Nalpha_fit:Omega} and
\ref{energy_nalpha_Omega_70p}, we see that the energy level structure 
is affected 
little by the condensation rate.

\subsection{New  soft mode of Bose--Einstein condensation of alpha clusters}
\label{sec-NalphaN0} 

The zero mode operator
 due to the BEC of $\alpha$ clusters emerge universally for any $N$,
if $\alpha$ clusters are  condensed only partly.
  Even under the condensation rate as small  20--30\%, which may be likely 
in the actual nuclei, we have the zero mode, consisting of 
a series of zero mode states associated with the zero mode operators,
in the same way as in the previous subsections. 
It is also noted that  the  first excited  zero mode state
  appears less than 2 MeV above the threshold systematically 
for all the nuclei investigated here.  The $0^+$ states, identified
as members states of the zero mode states, with low excitation energy
can be regarded as a soft mode. This is  a new kind of soft mode 
due to the BEC of $\alpha$ clusters and  has never been known in 
nuclei. 
The systematic appearance of this soft mode in a wide range of nuclei 
including light and heavy  mass region   is  natural, since
it originates from the locking  of the global phase in gauge space that
 violates the number conservation.
 
The emergence of  a soft mode in physical systems  has been  well-known
  \cite{Anderson1984}. The soft modes, by its 
definition \cite{Anderson1984},  appear at low excitation energy from the ground state. 
In nuclei a soft mode of quadrupole collective motion appears in heavy 
nuclei under the quadrupole force. 
Then the NG zero mode connected to 
the spontaneous breaking of rotational symmetry plays a crucial role 
in the rotational motion \cite{Ring1980}.
Also pairing vibration and pairing rotation \cite{Broglia1973,Brink2005,Hinohara2016} 
are a soft mode in heavy superfluid nuclei in gauge space (number space) 
due to the spontaneous  breaking of particle number symmetry caused 
by the  paring interaction, i.e. condensation of the Cooper pairs. 
The soft modes in hadronic systems have been studied extensively, 
see for example, see Ref.\cite{Hatsuda1984}.

The vibrational and rotational collective  motion due to $\alpha$ clustering, 
accompanied by SSB in Euclidean space, have been known. 
The vibrational motion is a radial (higher nodal) excitation of the relative motion 
and the rotational motion is due to the deformation caused by clustering. They have been
  observed in experiment, for example, typically as the ${\mathcal N}=10$
 higher nodal band and the parity doublet bands with the $\alpha$+$^{16}$O cluster
 structure in $^{20}$Ne \cite{Fujiwara1980}, and the ${\mathcal N}=14$ 
higher nodal band and the parity doublet bands with the $\alpha$+$^{40}$Ca 
cluster structure in $^{44}$Ti \cite{Michel1998}. Here ${\mathcal N}=2n+\ell$ 
with $n$ and $\ell$ being the number of the nodes in the relative wave function 
between the clusters and the orbital angular momentum, respectively. However, 
no collective motion with  soft mode nature 
related to $\alpha$ cluster condensation and 
SSB  has been known so far. 
The collective states with low excitation energies
as the zero mode states in the present study definitely possess soft mode nature.
It is noted that the zero mode states emerge as $0^+$  levels
at very high excitation energies as well as at low excitation energy above the 
threshold. 

The emergence of the well-developed  $0_3^+$ and $0_4^+$ $\alpha$ 
cluster states in $^{12}$C have been  understood based on the geometrical 
configuration of $\alpha$ clusters  in the traditional microscopic 
cluster models without an order parameter. For example, the $0_3^+$ state 
has been interpreted  to have an $\alpha$+$^{8}$Be geometrical configuration 
 with its  relative motion being excited (higher nodal radial excitation) 
\cite{Kurokawa2007}. The $0_4^+$ state has been interpreted to have 
a geometrical configuration of a  linear chain  \cite{Suhara2014,Funaki2016}.  
The origin of  the emergence  of the two $0_3^+$ and $0_4^+$  states 
just above the threshold has been ascribed  to the specific structure 
of the $0_2^+$ state (the Hoyle state) in $^{12}$C.
It is not clear if such  two $0^+$ states with low excitation energy 
 also appear, for example, in  $^{16}$O with four alphas,  
 $^{20}$Ne with five alphas,   $^{24}$Mg with six alphas and  $^{28}$Si 
with seven alphas, {\it etc.}, before performing  microscopic many $\alpha$ 
 cluster model  calculations and/or   {\it ab initio} calculations, 
which are formidably difficult nowadays for $N>4$ or $5$. 
 Our viewpoint is that there may be a profound  {\it raison d'\^etre} 
for the  very specific structure in $^{12}$C that the collective 
three $0^+$ states with a very developed  dilute $\alpha$ cluster 
structure  appear nearby   within a couple of MeV just above the threshold.
 In other word, 
 there may be an underlying fundamental principle  related to symmetry that 
causes this specific structure in $^{12}$C because emergence of a   
{\it collective motion with low excitation energy} is often related 
to an underlying symmetry of the system 
\cite{Anderson1984,Nambu1961,Watanabe2012,Hidaka2013}. 
In the present systematic calculations from $A=12$--52, it is 
found that the two $0^+$ states appear just above the vacuum $0^+$ 
state (analog of the Hoyle  state) inevitably and universally as the 
zero mode states of the NG operators
  \cite{Nambu1961,Watanabe2012,Hidaka2013}. The present calculations have  shown 
that  even in the three $\alpha$ cluster system of $^{12}$C, 
these gas-like $0_3^+$ and $0_4^+$  states can be understood as a collective 
motion that originates from the fundamental principle of SSB of the global 
phase. This is important because it enables us to expect the 
existence of such a gas-like $\alpha$ cluster states of BEC 
in the NG phase universally in light and heavy nuclei, 
which has never expected in the Ikeda diagram \cite{Ikeda1968} 
 in the configuration space considered in the Wigner phase. 
The emergence of the low-lying  two $0^+$ states above the Hoyle state 
is not accidental to $^{12}$C but has a profound physical meaning of 
SSB of the global phase, which may persists as a emergence of a new kind 
of soft mode in light and heavy nuclei. 
It is interesting to search for  experimentally  not only the first two    soft mode $0^+$  states, which are observed  in $^{12}$C,  but also  higher members of   low-lying soft mode  states of the Nambu-Goldstone mode  in light  and  heavy nuclei.

\subsection{Characteristic of the zero mode spectrum}

The spectrum of the zero mode states or collective states of the 
NG operators is derived from Eq.~(\ref{eq:HuQPeigen}). There the Hamiltonian
$\hat H_u^{QP}$ in Eq.~(\ref{eq:HuQP}) is specified by the six
 coefficients $C_{iji'j'}$ defined in Eq.~(\ref{eq:Cijij}) and the parameter $I$
in Eq.~(\ref{eq:etaI}).
In the most naive estimation, which is valid for large $N_0$
and in which $\xi(r)$ and $\eta(r)$ behave
as $\sqrt{N_0}$ and $1/\sqrt{N_0}$, respectively, 
the $N$-dependences are 
$C_{2020} \propto N_0^2$\,,\,$C_{2011}\propto N_0^1$\,,\,$C_{2002}\,,\,C_{1111}
\propto N_0^0$\,,\,$C_{1102}\,,\, I\propto N_0^{-1}$\,,\,$C_{0202}\propto N_0^{-2}$,
respectively. The results of actual numerical calculations for the coefficients 
$C_{iji'j'}$ and $I$, 
ranging from $N_0=2$ to $13$, are given in Fig.~\ref{fig:N0CI},
indicating that the hierarchical order of the magnitudes is 
$C_{2020}\, > \, C_{2011}\, > \, I \, > \, C_{2002}\,,\, C_{1111} 
\, > \, C_{1102} \,> \,C_{0202}$ for the whole range of $N_0$ and
that $C_{1102}$ and $C_{0202}$ are negligible.

\begin{figure}[tbh!]
\begin{center}
		\includegraphics[width=8cm]{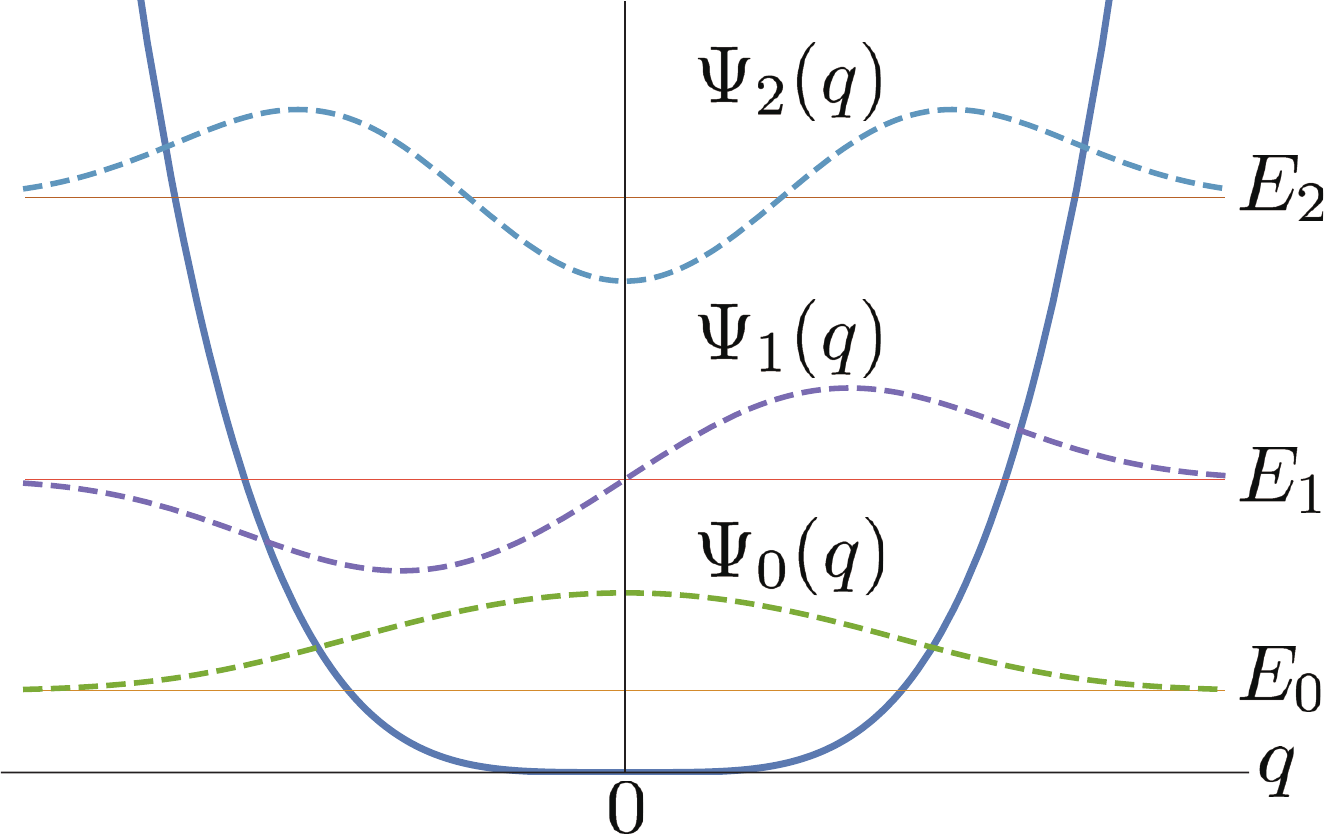}
\caption{(Color online)
The $q^4$ potential and
the eigenfunctions $\Psi_\nu(q)$ belonging to the eigenvalues
$E_\nu$ $(\nu=0,1,2)$ for $\hat H_0^{QP}$.}
\label{fig:ZMoutline}
\end{center}
\end{figure}

\begin{figure}[tbh!]
\begin{center}
		\includegraphics[width=8cm]{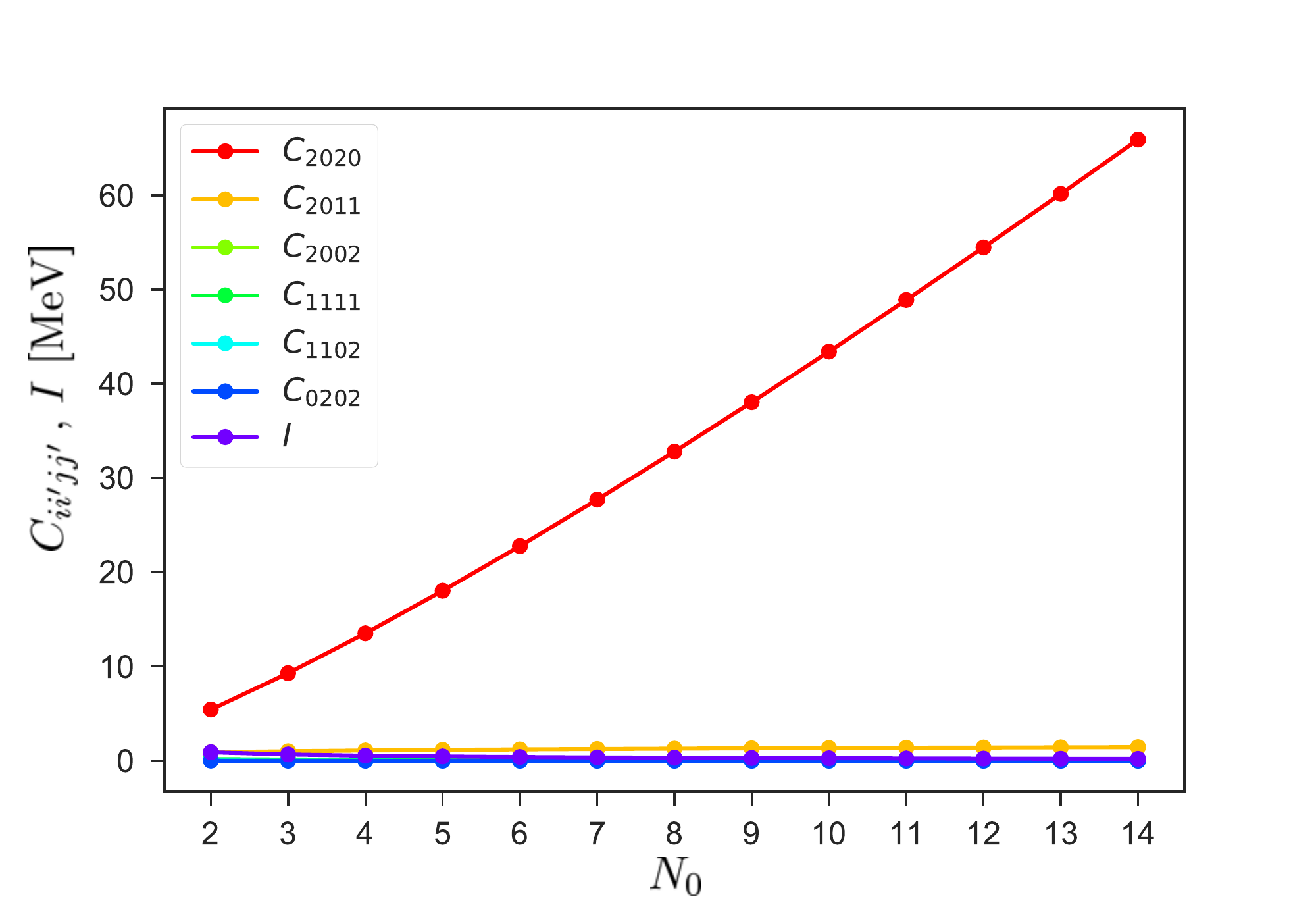}
		\includegraphics[width=8cm]{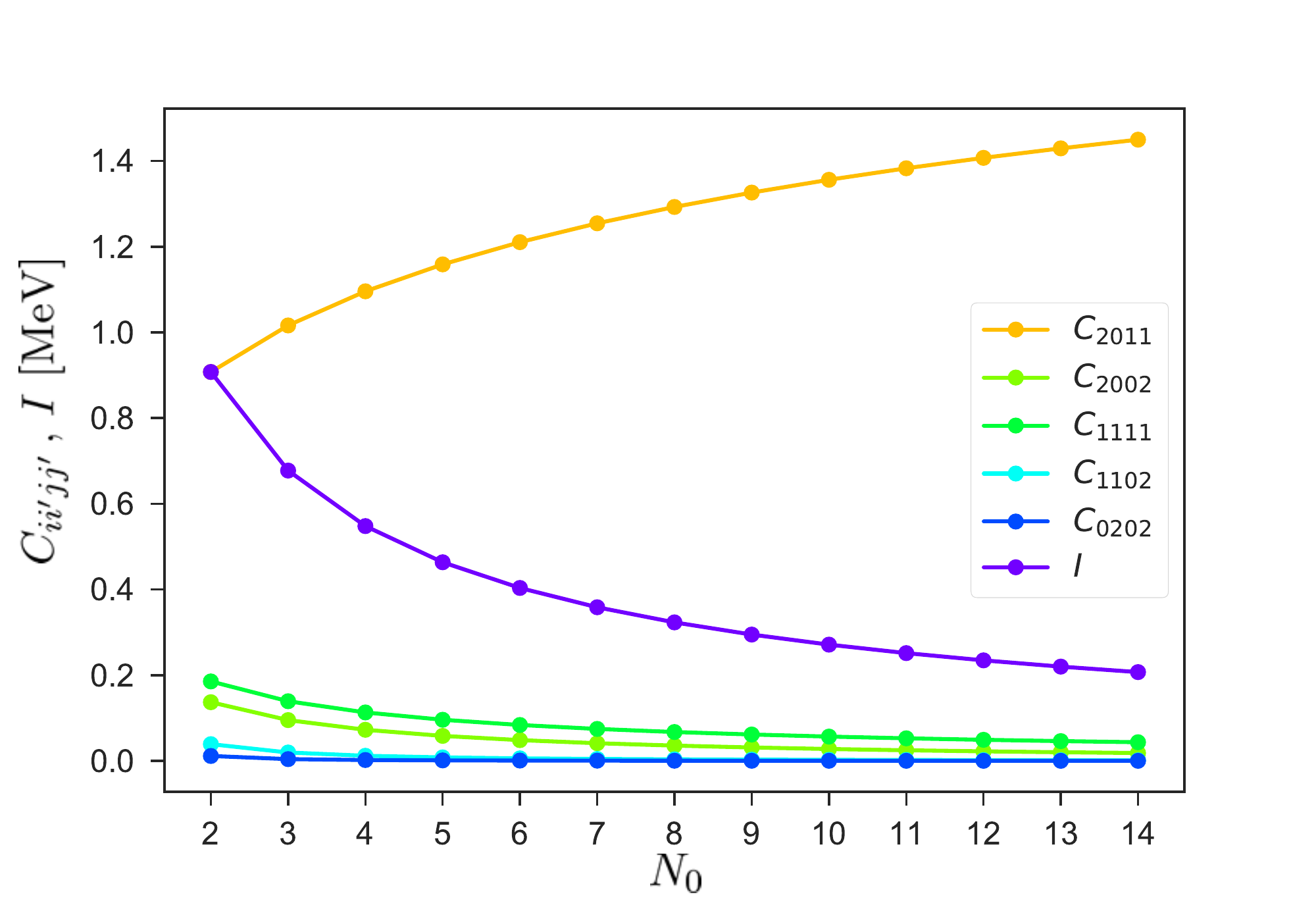}
\caption{(Color online) 
The $N_0$-dependences of $C_{iji'j'}$ and $I$ with the parameters 
$V_r=403$[MeV]\,,\, $\Omega=2.62$[MeV] and for $N_0=2$--13
are shown. (b) is a magnified graph of the small coefficients.}
\label{fig:N0CI}
\end{center}
\end{figure}

\begin{figure}[tbh!]
	\begin{center}
		\includegraphics[width=8cm]{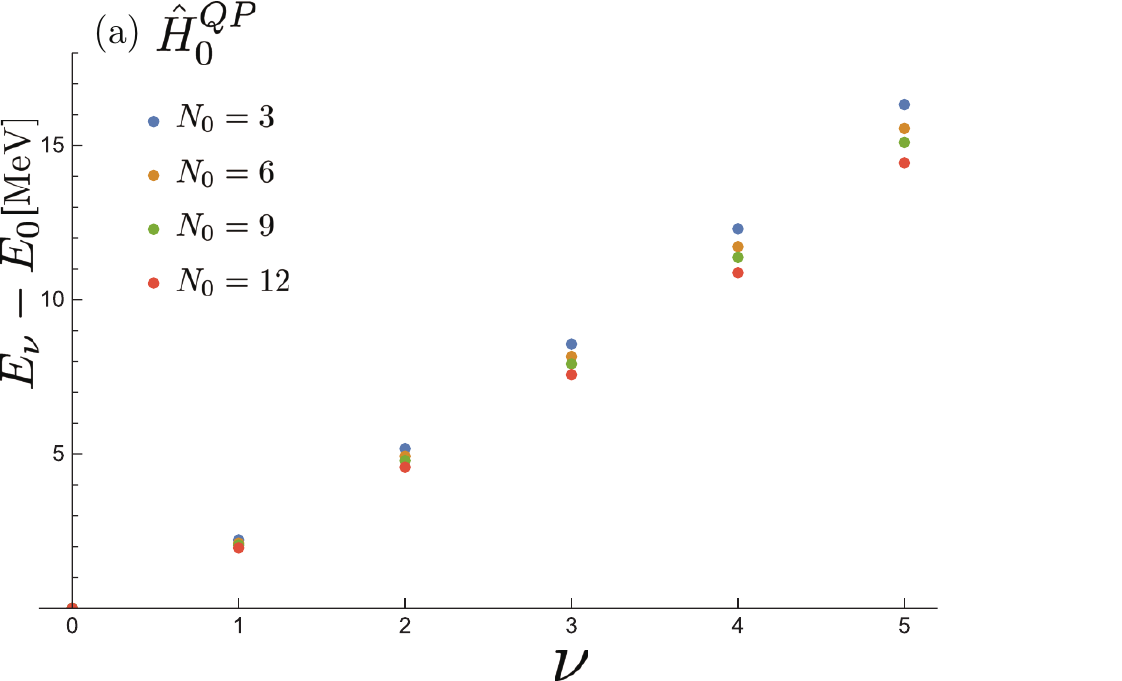}
		\bigskip

		\includegraphics[width=8cm]{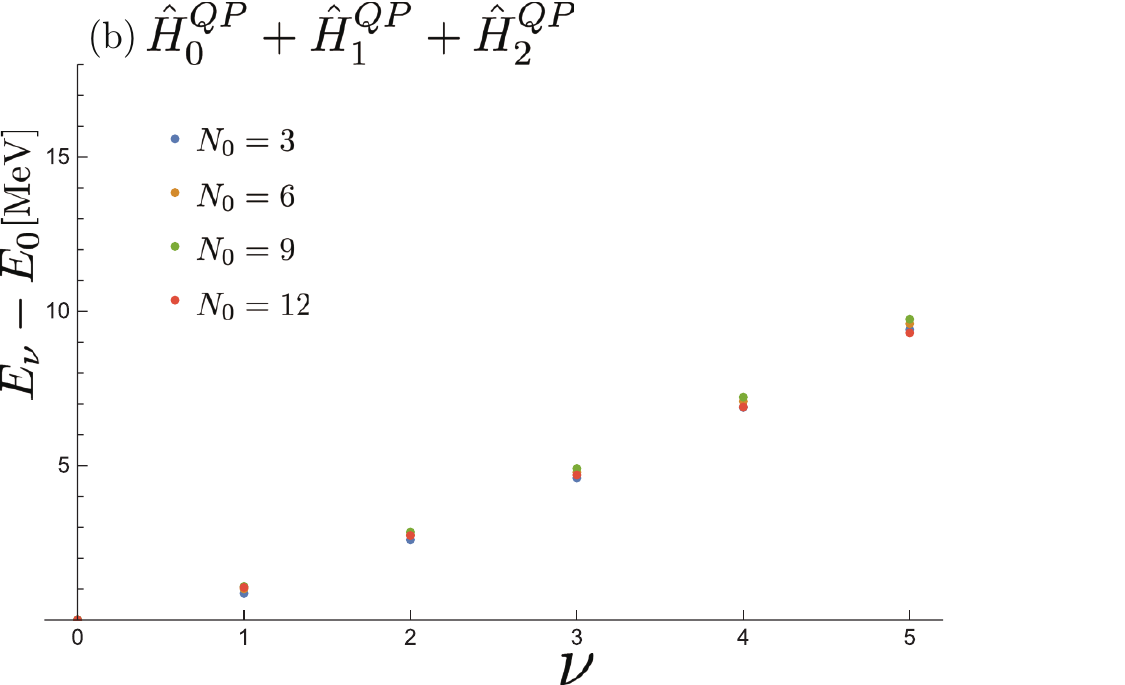}
		\caption{(Color online) Numerically 
		calculated excitation energies 
		$E_\nu-E_0$
		of (a) ${\hat H}_0^{QP}$ and (b)  ${\hat H}_0^{QP}+{\hat H}_1^{QP}
		+{\hat H}_2^{QP}$ are plotted for $\nu=1$--5 and
		$N_0=3,6,9,12$ with the parameters  $V_r=403$[MeV] and 
		$\Omega=2.62$[MeV].} 
		\label{fig:ESall36912}
	\end{center}
\end{figure}

The dominant contribution in $\hat H_u^{QP}$ comes from the ${\hat Q}^4$
term with $C_{2020}$, and the ${\hat P}^2$ term can not be neglected due to
the virial theorem. Thus the leading order zero mode Hamiltonian 
is the Hamiltonian for a one-dimensional quantum mechanical system under 
 the ${\hat Q}^4$ potential, 
\begin{align}
\hat H_0^{QP} = \frac{I}{2}\hat P^2  +
\frac{1}{2}C_{2020}\hat Q^4  \,.
\label{eq:H0QP}
\end{align}
Figure~\ref{fig:ZMoutline} outlines the $q^4$ potential and
the eigenfunctions $\Psi_\nu(q)$ belonging to the eigenvalues
$E_\nu$ $(\nu=0,1,2)$ for $\hat H_0^{QP}$. 
The eigenvalues for $\hat H_0^{QP}$ depends only on a single parameter
of a dimension of energy $w=(I^2 C_{2020})^{1/3}$. This is because 
$\hat H_0^{QP}$ becomes
\begin{align}
\hat H_0^{QP} = \frac{ I'}{2}({\hat P}')^2  +
\frac{1}{2}C'_{2020}\hat Q^4
\end{align} 
with $I'= s^2 I$ and $C'_{2020}=C_{2020}/s^4$ when 
the scale transformation with a dimensionless parameter $s$,
${\hat Q}' = s{\hat Q}$ and ${\hat P}' = {\hat P}/s$, keeping
$[{\hat Q}',{\hat P}']=1$, is performed. According to numerical calculations,
the parameter $w$ shows a very weak $N_0$ dependence as $N_0^{-0.08}$ and 
is almost constant. This implies that the spectrum of the zero mode states
are almost the same regardless to a value of $N_0$ at this level of approximation,
as in Fig.~\ref{fig:ESall36912} (a).

The terms of the next orders in $\hat H_u^{QP}$ are
\begin{align} 
& {\hat H}_1^{QP} =  2C_{2011}\hat Q\hat P\hat Q 
 -2C_{2011} \hat Q^2 \,, 
\label{eq:H1QP} \\
&\hat H_2^{QP} = - 2 \left(C_{2002} + C_{1111} \right) \hat P+
 + C_{2002}\hat Q\hat P^2\hat Q \,.
\label{eq:H2QP}
\end{align}
We see that ${\hat H}_1^{QP}$ pushes down $E_\nu$, while ${\hat H}_2^{QP}$
pushes up $E_\nu$, both larger for larger $\nu$. The calculated spectrum
of ${\hat H}_0^{QP}+{\hat H}_1^{QP} +{\hat H}_2^{QP}$ 
that approximates $\hat H_u^{QP}$ well is plotted in 
Fig.~~\ref{fig:ESall36912} (b). It is seen there that the spectrum is almost
independent of $N_0$ and that the corrections by  ${\hat H}_1^{QP} +{\hat H}_2^{QP}$
shift the spectrum of  ${\hat H}_0^{QP}$ downwards for all the $\nu$ and $N_0$.

We remark on the interpretations of ${\hat Q}$ and ${\hat P}$,
 and the excitations in $\Psi_\nu(q)$. 
When the quantum fluctuation of $\hat{Q}$ is small, 
it may be interpreted 
as the phase operator, as $(1-i\hat{Q }) \xi \simeq e^{-i\hat{Q}} \xi$, 
and the localization of $\Psi_0(q)$ around $q=0$
 corresponds to the phase locking. 
The excited wavefunctions $\Psi_\nu(q)$ $(\nu=1,2,…)$ are extended more widely as $\nu$
 goes up,  which implies that the excitation in the zero mode sector 
loosens the phase locking.
However, the interpretation of $\hat{Q}$ as the phase operator is valid 
only for small quantum fluctuations. As quantum fluctuations of 
$\hat{Q}$ and $\hat{P}$ become large due to a finite size of the 
system, the simple interpretation that $\hat{Q}$ and $\hat{P}$ are 
the phase operator and the generator of the phase transformation, respectively, 
breaks down. 
The operators $\hat{Q}$ and $\hat{P}$ should be treated as the
canonical coordinate and momentum in a fictitious one-dimensional space, 
as is indicated by the field expansion in Eq.~(\ref{eq:PhiExp})
 and the commutation relation Eq.~(\ref{eq:CCRQPa}).

The analysis of this subsection serves as a general grasp of
 the energy levels of 
the $0^+$ states, identified as the zero mode states, 
in Figs.~\ref{fig:ex_energy_C_N70p_32_38}, \ref{fig:energyN0}, 
\ref{fig:energy}, \ref{fig:energy_12C_p}, \ref{fig:Nalpha_fit:Omega}
 and \ref{energy_nalpha_Omega_70p}.

\section{SUMMARY}\label{sec-Summary}

 Bose--Einstein condensation (BEC) of $\alpha$ clusters in light and 
medium-heavy  $4N$ nuclei is studied in the frame of the field theoretical 
superfluid cluster model.
There it is crucial that the order parameter is expressed as the vacuum expectation
of the field operator explicitly.
The order parameter is a superfluid amplitude that satisfies Gross--Pitaevskii 
equation and characterizes  the phase transition from the normal $\alpha$ 
cluster state  in the Wigner phase to the Nambu--Goldstone phase with 
Bose--Einstein condensate of $\alpha$ clusters. The Nambu--Goldstone operators 
(zero mode operators) due to  spontaneous symmetry breaking 
of the global phase in the {\it finite} 
number of $\alpha$ clusters is rigorously treated.  
 We have analyzed 
 the $\alpha$  cluster structure in $^{12}$C,  assuming a realistic condensation rate.
 It is found that the energy levels of the $\alpha$ cluster 
structure above the Hoyle state are well reproduced.  
The well-developed $2^+$ and $4^+$ states are understood to be  
Bogoliubov--de Gennes vibrational modes built on the Hoyle state rather 
than a rotational band.  The calculations at various condensation rates have revealed
 that the similar energy level structures are appear repeatedly 
once the BEC is realized, even if the  condensation rate is not very large.

  It is found in the superfluid cluster model with the order parameter that 
the two gas-like collective $0_3^+$ and  $0_4^+$ states above the Hoyle state 
in $^{12}$C, which have been understood in the traditional cluster 
models based on  a geometrical configuration space, can be understood 
as  a manifestation of the emergence of the zero mode that has a 
profound field theoretical physical meaning of the locking of the global 
phase in the gauge space.  
   The present theory of $^{12}$C was extended to $N\alpha$ nuclei,  
$^{16}$O--$^{52}$Fe,  
in light and medium-heavy mass region, assuming different condensation rates of 
$\alpha$ clusters.  Then, due to spontaneous symmetry breaking, we similarly 
have the zero mode operators and the associated zero mode states 
 in these nuclei, even if the condensation rates are not very large.
The energy level structure of the zero mode  $0^+$ states change 
little, depending on the $N$ and the condensation rates. This means that  
the collective $0^+$ states with  low excitation energy appear systematically 
above the threshold energy in light and heavy nuclei. They are new kind 
of soft mode states due to the BEC of $\alpha$ clusters.  It is highly 
expected to search for such a soft mode in experiment.
The field theoretical superfluid cluster model 
 may be applicable to much heavier nuclei beyond the present study.  It is 
expected that 
the new zero mode states, originating from the spontaneous symmetry breaking 
of the global phase of the superfluid $\alpha$ clusters in nuclei may persist 
throughout the periodic table from light nuclei to heavy nuclei. 
The application of the present field theoretical superfluid cluster 
model study  to the heavy mass region around $^{212}$Po is a future challenging subject.
 
 \par
\begin{acknowledgements}
This work is supported by JSPS KAKENHI Grant No.~16K05488.
 The authors thank Yasuhiro Nagai and Ryo Yoshioka for 
their numerical calculations at an early stage of this work, and
the Yukawa Institute for Theoretical Physics, Kyoto University for  
making possible  the present collaborative work. One of the authors (SO) 
is also grateful to the Research Center for Nuclear Physics, Osaka University for supporting the present work. Part of this work was done during the authors'
 stay in YITP in 2018.
\end{acknowledgements}

\end{document}